\PassOptionsToPackage{dvipsnames}{xcolor}
\documentclass[aps,prb,reprint,groupedaddress]{revtex4-2}

\usepackage{mathtools, nccmath}
\usepackage{amssymb}
\usepackage{dsfont}
\usepackage{afterpage}
\usepackage[T1]{fontenc}
\usepackage{float}
\usepackage{pgfplots}
\usepackage{tikz}
\usetikzlibrary{decorations.pathmorphing}
\usetikzlibrary{arrows, backgrounds}
\usepackage{ifthen}
\usepackage[caption=false]{subfig}
\usepackage{placeins}
\usepackage{graphicx}
\graphicspath{ {fig-paper-1/} }
\definecolor{almond}{rgb}{0.94, 0.87, 0.8}

\newcommand{\be}{\begin{equation}}
\newcommand{\ee}{\end{equation}}
\newcommand{\bea}{\begin{eqnarray}}
\newcommand{\eea}{\end{eqnarray}}
\newcommand{\Eq}[1]{Eq.\,(\ref{#1})}
\newcommand{\Eqs}[2]{Eqs.\,(\ref{#1}) and (\ref{#2})}
\newcommand{\Eqsss}[3]{Eqs.\,(\ref{#1}), (\ref{#2}), and (\ref{#3})}

\newcommand{\Fig}[1]{Fig.\,\ref{#1}}
\newcommand{\Figs}[2]{Figs.\,\ref{#1} and \ref{#2}}

\newcommand{\Sec}[1]{Sec.\,\ref{#1}}
\newcommand{\Secs}[2]{Secs.\,\ref{#1} and \ref{#2}}
\newcommand{\App}[1]{Appendix\,\ref{#1}}

\newcommand{\q}{{\bf q}}

\newcommand{\Ks}[6]{
\draw[fill=#6,fill opacity=0.5] (#1,#2) rectangle (#1+#3,#2+#4);
\draw node at (#1+#3/2,#2+#4/2) {#5};
}
\newcommand{\Lshape}[4]{
\begin{pgfonlayer}{Llayer}
\draw[-,line width=0.8mm,#1] (#2*#3,#2*#3) -- (#2*#3+#2*#4,#2*#3) -- (#2*#3+#2*#4,#2*#3+#2) -- (#2*#3+#2,#2*#3+#2) -- (#2*#3+#2,#2*#3+#2*#4) -- (#2*#3,#2*#3+#2*#4) -- (#2*#3,#2*#3);
\draw[-,step=10mm,dashed,line width=0.5mm,white] (#2*#3,#2*#3) -- (#2*#3+#2*#4,#2*#3) -- (#2*#3+#2*#4,#2*#3+#2) -- (#2*#3+#2,#2*#3+#2) -- (#2*#3+#2,#2*#3+#2*#4) -- (#2*#3,#2*#3+#2*#4) -- (#2*#3,#2*#3);
 \end{pgfonlayer}
}
\makeatletter
\def\p@subsubsection{}
\makeatother

\begin{document}

\title{Impact of the phonon environment on the nonlinear quantum-dot-cavity QED. \\ I. Path-integral approach}
\author{L.S. Sirkina}
\affiliation{School of Physics and Astronomy, Cardiff University, The Parade, Cardiff CF24 3AA, United Kingdom}
\author{E.A. Muljarov}
\affiliation{School of Physics and Astronomy, Cardiff University, The Parade, Cardiff CF24 3AA, United Kingdom}


\begin{abstract}

We demonstrate a strong influence of the phonon environment on the coherent dynamics of the quantum dot (QD)-cavity system in the quantum strong coupling regime. This regime is implemented in the nonlinear QD-cavity QED and can be reliably measured by heterodyne spectral interferometry. We present
a semi-analytic asymptotically exact path integral-based approach to the nonlinear optical response of this system, which includes two key ingredients: Trotter's decomposition and linked-cluster expansion. Applied to the four-wave-mixing optical polarization, this approach provides access to different excitation and measurement channels, as well as to higher-order optical nonlinearities and quantum correlators. Furthermore, it allows us to extract useful analytic approximations and analyze the nonlinear optical response in terms of quantum transitions between phonon-dressed states of the anharmonic Jaynes-Cummings (JC) ladder. Being well described by these approximations at low temperatures and small exciton-cavity coupling, the exact solution deviates from them for stronger couplings and higher temperatures, demonstrating remarkable non-Markovian effects, spectral asymmetry, and strong phonon renormalization of the JC ladder.
\end{abstract}

\maketitle

\section{Introduction}
\label{section:intro}

A quantum dot (QD) embedded in an optical microcavity offers a technologically accessible platform for the study of quantum-optics phenomena in solid state. This system can be experimentally realized, for example, in self-assembled QDs inside semiconductor micropillar or photonic-crystal cavities~\cite{gerard1996quantum,painter1999two}.  Some properties of the system, including generation of single photons~\cite{pelton2002efficient,gazzano2013bright,press2007photon}, are important for development of quantum devices with applications in the field of quantum information.

The repeated mutual conversion of excitation between the QD exciton and the cavity mode, well known as strong light-matter coupling, is typically described by the Jaynes-Cummings (JC) model~\cite{jaynes1963comparison}. The strong coupling itself is a linear classical effect leading to polariton formation~\cite{HopfieldPR58} and observation of the vacuum Rabi splitting in various systems, such as quantum wells inside planar microcavities~\cite{WeisbuchPRL92} and atoms inside optical cavities~\cite{HoodPRL98}, as well as QDs coupled to cavity modes~\cite{reithmaier2004strong,yoshie2004vacuum}. However, owing to the fermionic nature of QD excitons, their coupling to bosonic cavity modes introduces a quantum nonlinearity, which manifests itself in anharmonic ladder-like structure of hybrid QD-cavity states. This is know in the literature as the quantum strong coupling regime recently observed QD-cavity systems~\cite{Faraon2008,kasprzak2010up} and  superconducting circuits~\cite{Fink2008}. This quantum nonlinearity can naturally be measured by means of nonlinear optical spectroscopy~\cite{langbein2006heterodyne,kasprzak2010up}, in which  higher rungs of the JC ladder are accessed through an external excitation in a form of a sequence of optical pulses and subsequent selection of different channels of optical nonlinearity~\cite{allcock2022quantum}.

In practice, due to the solid-state environment, the QD can be subject to significant non-Markovian behavior~\cite{KrummheuerPRB02,hohenester2009phonon,madsen2011observation,stock2011acoustic}. This is mainly a result of the coupling of QD excitons to longitudinal acoustic (LA) phonons via the deformation potential~\cite{takagahara1993electron, KrummheuerPRB02,Ramsay2010}. In specific parameter regimes, when a QD exciton is not very strongly coupled to a cavity mode, the effect of phonons may be addressed perturbatively~\cite{kaer2010non}, or via the polaron transformation combined with a perturbation theory ~\cite{RamseyPRL10,NazirJPCM16}. However, for a stronger QD-cavity coupling, phonons play a more significant role that requires developing non-perturbative techniques~\cite{Hughes2011}. This has been recently demonstrated by an exact theoretical approach to the linear optical response~\cite{morreau2019phonon,morreau2020phonon}. 
The linear response is, however, harder to observe  experimentally, as compared to a nonlinear optical polarization that can be reliably measured by means of a heterodyne spectral interferometry~\cite{langbein2006heterodyne}, and is also a suitable tool for the study of quantum nonlinearities and the quantum strong coupling regime. Alternatively, quantum nonlinearities can be studied by analyzing photon counting statistics in photoluminiscence measurements~\cite{Laussy2012}.

The four-wave mixing (FWM) polarization of a QD-cavity system has been the focus of many experimental and theoretical works but the exciton-phonon interaction has either been neglected or treated perturbatively~\cite{kasprzak2010up,kasprzak2013coherence,albert2013microcavity,groll2020four}. In the absence of a cavity, the effect of the phonon dephasing measured in the FWM response~\cite{borri2007four} has been addressed theoretically for ensembles of isotropic~\cite{vagov2004nonmonotonous}  and anisotropic QDs, also accounting for real phonon-assisted transitions~\cite{muljarov2006nonlinear}. 
One important advantage of using a FWM scheme is a possibility to use a two-dimensional hyperspectral imaging that reveals the information about coherences in the system~\cite{kasprzak2011coherent}. The effect of phonons on two-dimensional spectra for a nonlinear optical response has been studied in the absence of a cavity using a path integral (PI)  based approach~\cite{sahrapour2010multitime,duan2015efficient,liang2014simulating} in the context of energy harvesting. In addition to the FWM, other quantities of experimental relevance, including photoluminescence and photon indistinguishability~\cite{laussy2008strong,grange2015cavity}, have been recently studied in QD-cavity systems~\cite{bundgaard2021non,denning2020optical}, using a PI-based approach to include a phonon contribution. With an increasing interest in understanding of non-Markovian effects and the need to study non-perturbative regimes of more complex models, there has been a rise in popularity of such treatments, including applications in the context of QDs~\cite{vagov2011real,barth2016path,cygorek2017nonlinear}.

PI-based approaches offer a possibility to achieve numerically exact solutions for a QD-cavity system in a phonon environment and are capable of addressing arbitrary parameter regimes, where approximations fail. Such solutions typically require Hilbert spaces, which rapidly grow in size. However, the environment degrees of freedom can be reduced dramatically by means of introducing a memory kernel~\cite{makri1995methods}. The memory kernel takes into account all the necessary information introduced by temporal correlations in the system evolution as a result of its interaction with the environment~\cite{breuer2002theory}.
The complexity of the problems one can address is generally limited by the computational memory and speed.
The approaches based on the early real-time PI methods developed in Ref.~\cite{makri1995theory,makri1995methods} typically
depend on the combined use of Trotter's decomposition and Feynman-Vernon influence functional~\cite{feynman2000theory}. Instead of the latter, Ref.~\cite{morreau2019phonon} uses a linked-cluster, or cumulant expansion~\cite{breuer2004time}, that formed the basis of this work, as it is a suitable approach for treating a linear coupling betweeen the system and a continuum of bath modes.

In order to tackle the computational costs and improve efficiency, a number of further PI optimization schemes have been proposed. Early approaches use basic selective filtering methods, such as those which filter out path segments with negligible weight by Monte-Carlo importance sampling ~\cite{sim1997filtered,sim2001quantum}.
More recently, an idea to spread the temporal correlations over path segments of increasing length, whilst their contribution is reduced, was formulated~\cite{makri2020small}.
Other recent works have developed an optimization scheme, which is based on rewriting the PI as a tensor network (TN), where the propagator is represented by a matrix product state (MPS). Compression of large tensors can be achieved by a selective truncation that relies on singular value decomposition~\cite{strathearn2018efficient}. The propagation can then be done very efficiently using existing tensor network algorithms. This approach has seen further development~\cite{jorgensen2019exploiting,cygorek2021numerically} and has been recently combined with mean-field theory~\cite{fowler2021efficient} to address systems with large Hilbert spaces. 
Another work has combined the approach developed in Ref.~\cite{strathearn2018efficient} with a different MPS-based algorithm, introducing a time-discrete quantum memory as a second non-Markovian reservoir, resulting in a quasi-two dimensional TN~\cite{Kaestle2021}.

In addition to PI-based approaches, there are other numerically exact methods. These include solving hierarchical equations of motion~\cite{tanimura1989time,KnorrPRL03,ishizaki2009unified} and numerical renormalization group approaches~\cite{wilson1975renormalization,bulla2003numerical}. An implementation of the density matrix (DM) renormalization group~\cite{white1992density} is based on the optimization of matrix product states~\cite{schollwock2011density}, which also uses singular value decomposition.
One further optimization implements a mapping of the the environment onto a one-dimensional chain with effective nearest-neighbor interactions~\cite{chin2010exact}, which makes a subsequent application of time-adaptive DM renormalization group algorithm~\cite{daley2004time} very efficient.

In the present work, we focus on the FWM polarization of a QD-cavity system linearly coupled to LA phonons. We develop an asymptotically exact semi-analytic approach which is a generalization of the method presented in Ref.~\cite{morreau2019phonon}, with key ingredients being the Trotter decomposition~\cite{trotter1959product,suzuki1976generalized} and linked-cluster expansion~\cite{muljarov2004dephasing,muljarov2005phonon}. This method is an explicit and physically intuitive PI-based approach allowing a number of useful analytic approximations~\cite{morreau2019phonon,morreau2020phonon}, also presented for the case of the FWM in this and the follow-up paper~\cite{paper1}. Since the approach is non-perturbative, it allows us to explore regimes of comparable exciton-cavity and exciton-phonon coupling strengths, when the similarity of the system and environment timescales opens up a possibility of phonon-assisted transitions between different polariton states~\cite{MullerPRX15,dory2016complete,morreau2019phonon}.  Although the treatment of optical nonlinearities increases the complexity of the approach, it remains physically intuitive and computationally straightforward. Furthermore, it allows us to address all possible channels of optical excitation and measurement and provides perspectives to a further application of the method to higher-order optical nonlinearities, arbitrary elements of the density matrix, and other quantum correlators and physical observables.

We demonstrate that the exciton coherent dynamics in semiconductor quantum dots can be significantly modified by the phonon environment, showing a remarkable non-Markovian behavior. Based on the developed technique, we propose two computationally simple approaches to the regime of small exciton-cavity coupling. One of these approaches has a fully analytic form while the other reduces the general method to a simple matrix multiplication. Being not limited to perturbative regimes, our microscopic approach can deal with arbitrary temperatures, as well as with situations where the interactions of the exciton with the cavity mode and the phonon environment are comparable. In the latter case, the phonon cloud around the quantum dot is unable to adiabatically adapt to a varying optical state, resulting in a non-Markovian dynamics and phonon-assisted transitions between states of the Jaynes-Cummings ladder. We observe a spectral asymmetry, a modification in anharmonicity of the ladder, and even deviation from the ladder-like structure, which becomes more pronounced with increasing temperature. By extracting complex fit parameters we analytically characterize the long-time behavior of individual transitions which contribute to the FWM signal and demonstrate the  non-Markovian nature of the latter.

\section{PI-based approach to the FWM dynamics using Trotter's decomposition and cumulant expansion}
\label{section:theory}

In this section, we describe the formalism of our asymptotically exact semi analytic approach to the FWM dynamics of the QD-cavity system surrounded by a phonon environment. We first introduce in \Sec{theory:system} the system Hamiltonian and the master equation containing Lindblad dissipators for both the QD exciton and the cavity mode. We also discuss there in detail optical excitation and measurement channels for observation of the FWM response of the system and comparing them, where appropriate, to those of the linear response. We then focus in \Secs{theory:Trotter}{theory:linkedcluster} on two key elements of our approach: Trotter's decomposition and linked-cluster expansion. Finally, in Secs.\,\ref{theory:NN}, \ref{theory:longtime}, and \ref{theory:LN} we show how these elements work together to accurately describe the FWM dynamics of the system. This includes the full rigorous $L$-neighbor approach presented in \Sec{theory:LN} and a number of useful approximations following from it, such as nearest-neighbor and polaron approximation, presented in \Secs{theory:NN}{theory:longtime}, respectively.

\subsection{System, its excitation, and measurement of the linear and  FWM polarizations}
\label{theory:system}

We consider a composite system, which presents a combination of two analytically solvable models, the JC model describing the exciton-photon coupling and the independent boson (IB) model~\cite{mahan2013many} taking care of the exciton-phonon interaction. The composite system is, however, not solvable analytically, to the best of our knowledge. Its full Hamiltonian is given by
\begin{align}
    H=H_{\rm JC}+H_{\rm IB}\,,
    \label{hamilt}
\end{align}
where
\begin{align}
    H_{\rm JC}&=\Omega_x d^\dagger d+\Omega_c a^{\dagger} a+g(a^{\dagger} d +d^{\dagger} a)\,,\\
\label{IB}
    H_{\rm IB}&=H_{\rm ph}+d^\dagger d V\,,\\
    H_{\rm ph}&=\sum_\q \omega_q b^\dagger_\q b_\q\,, \quad V=\sum_\q \lambda_\q(b^\dagger_{-\q}+b_\q)\,.
\label{V}
\end{align}	
Here, the fermionic (bosonic) creation operator $d^\dagger (a^\dagger$) corresponds to the  exciton (cavity) mode with real frequency $\Omega_x$ ($\Omega_c$), and $g$ is the exciton-cavity coupling strength. We work in the units of $\hbar=1$ throughout this paper.
The bosonic operator $b_\q^\dagger$ creates a phonon with momentum $\q$ and frequency $\omega_q$, where $q=|\q|$.
The coupling of the exciton to the LA phonon mode $\q$ is given by the matrix element $\lambda_\q$, which has a general symmetry property: $\lambda_{\q}^\ast=\lambda_{-\q}$. The explicit form of $\lambda_\q$ in case of a linear phonon dispersion is given by \Eq{ex-ph-coupling} of \App{appendix:Propagator}.

The exciton-phonon Hamiltonian \Eq{IB} contains a single exciton mode, so we assume that other exciton states of the QD do not couple to this state via phonons.
Such a diagonal form of the exciton-phonon interaction is known in the literature as the IB model which leads in the absence of the cavity to exciton pure dephasing~\cite{KrummheuerPRB02}. While this model can adequately describe a measured quick relaxation of the FWM polarization~\cite{vagov2004nonmonotonous}, it does not include any mechanisms of experimentally observed~\cite{borri2001ultralong} density or polarization relaxation at longer times. To rectify on this, we add to the model a phenomenological exponential decay rate $\gamma_x$ of the exciton state, which reffers to the population decay time $T_2$. This can be further refined in a rigorous way within the single-mode (i.e. diagonal) exciton Hamiltonian by mapping the off-diagonal phonon coupling to other exciton states onto a diagonal quadratic interaction~\cite{muljarov2004dephasing,grosse2008phonons} that accounts for the pure dephasing $T_2^*$. We also introduce a decay rate $\gamma_c$ of the electromagnetic eigenmode of the cavity, which is strictly the negative of the imaginary part of the complex frequency of the cavity mode. Both $\gamma_x$ and $\gamma_c$ are introduced consistently via the Lindblad dissipation formalism and do not include any phenomenological pure dephasing.

The evolution of the composite system between the pulses is modeled by a master equation which has a Lindblad form:
\be
i\dot{\rho}= \mathcal{L} \rho
\equiv[ H_{\rm JC},\rho]+[ H_{\rm IB},\rho]+i\gamma_c \mathcal{D}[a]+i \gamma_x \mathcal{D}[d]\,,
\label{master}
\ee
where $\mathcal{L}$ is the Liouvillian superoperator of the composite system.
Here, the dephasing of the cavity and the exciton modes is represented by a Lindblad dissipator defined as
$\mathcal{D}[c] \rho  =2c\rho c^\dagger-c^\dagger c\rho -\rho c^\dagger c$.

In the dipole approximation, the excitation of the system by a sequence of ultrashort pulses is given by an interaction operator
\be
 \mathcal{Q}(t)= -\sum_j \delta(t-t_j) \left[ \mathcal{E}_j \mu_{c_j} c_j^\dagger + \mathcal{E}_j^\ast \mu_{c_j}^\ast c_j\right]\,,
     \label{interactExcitationField}
\ee
in which the external classical field due to each pulse is represented by a delta function peaked at $t=t_j$ and a pulse area $\mu_{c_j}\mathcal{E}_j$, with the index $j={\rm I,\,II,\dots}$ labeling the pulses. This corresponds to a situation in which each pulse duration is much shorter than the characteristic timescales of the system. The effective dipole moment $\mu_{c_j}$ describes the coupling strength of the mode $c_j^\dagger$ to the pulse field $\mathcal{E}_j$. The excitation can take place via the cavity mode ($c_j^\dagger=a^\dagger$) or the exciton mode ($c_j^\dagger=d^\dagger$), or a combination of both~\cite{morreau2020phonon}, and is further referred to as the excitation channel.
The time evolution of the system due to each pulse is then given by a product of exponential operators~\cite{kasprzak2010up,allcock2022quantum}.
The DM under the action of the interaction $\mathcal{Q}(t)$ can then be expanded in all orders of the excitation field as a series of nested commutators. The phase of $\mathcal{E}_j$ permits selection of a particular channel in the optical polarization, which is linear or nonlinear in terms of the powers of $\mathcal{E}_j$.

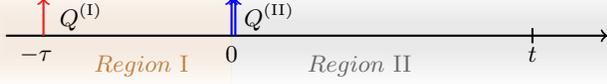
\begin{figure}[t]
\begin{tikzpicture}
\foreach \x in {0,...,10} {
      \draw[fill=lightgray,fill opacity=0.03,draw=none] (3.5,0) rectangle ++(5+0.05*\x,0.2+0.05*\x);
      \draw[fill=almond,fill opacity=0.07,draw=none] (3.5,0) rectangle ++(-3-0.05*\x,0.2+0.05*\x);
      \draw[fill=lightgray,fill opacity=0.03,draw=none] (3.5,0) rectangle ++(5+0.05*\x,-0.2-0.05*\x);
      \draw[fill=almond,fill opacity=0.07,draw=none] (3.5,0) rectangle ++(-3-0.05*\x,-0.2-0.05*\x);
}

\draw [thick,->,black](0.5,0) -- (8.5,0);

\draw[thick,->,Red] (1,0) -- (1,0.5);
\draw[thick,->,blue] (3.5,0) -- (3.5,0.5);
\draw[thick,->,blue] (3.55,0) -- (3.55,0.5);
\draw[thick,-,black] (7.5,-0.1) -- (7.5,0.1);

\draw node at (1.5,0.25) {$ Q^{(\mathrm{I})} $};
\draw node at (4,0.25) {$ Q^{(\mathrm{II})} $};
\draw[brown] node at (2.3,-0.4) {$Region$  $\mathrm{I} $};
\draw[darkgray!80!white] node at (5.2,-0.4) {$Region$  $\mathrm{II} $};

\draw node at (0.9,-0.25) {$ -\tau $};
\draw node at (3.5,-0.25) {$ 0 $};
\draw node at (7.5,-0.25) {$ t $};
\end{tikzpicture}
\caption{Pulsed excitation scheme showing a sequence of ultrashort laser pulses with a positive delay time $\tau$. The first pulse with the pulse area $\mu_{c_{\rm I}}\mathcal{E}_{\rm I}$ and action ${Q}^{\mathrm{(I)}}$ on the DM arrives at $t_{\rm I}=-\tau$. After delay $\tau$, two identical pulses with the resulting pulse area $\mu_{c_{\rm II}}\mathcal{E}_{\rm II}$ and the total action ${Q}^{\mathrm{(II)}}$ arrive together  at $t_{\rm II}=0$. The observation time is denoted by $t$. }
\label{excitation}
\end{figure}

For the linear polarization, we need to consider only a first-order effect on the DM by a single pulse, which is proportional to $\mathcal{E}_{\rm I}^\ast$,
\begin{align}
\rho^\mathrm{(I)}(-\tau)&=i \mu_{c_{\rm I}}^\ast\mathcal{E}_{\rm I}^\ast[c_{\rm I},\rho^\mathrm{(0)}(-\tau)]\,,
\label{select-rho1}
\end{align}
which is due to a single pulse at $t_{\rm I}=-\tau$.
For a nonlinear polarization,
we focus on the well-known case of degenerate FWM in which three-pulse excitation is effectively reduced to two pulses, at $t_{\rm I}=-\tau$ and $t_{\rm II}=0$, with $\tau$ being the delay time between the pulses.
Figure \ref{excitation} schematically illustrates this pulsed excitation scheme. The action of the first pulse, denoted by ${Q}^{\mathrm{(I)}}$, is described by \Eq{select-rho1}, the same way as in the linear polarization. The second and the third pulses, both arriving at $t=0$, are considered as a single pulse, with the action denoted by ${Q}^{\mathrm{(II)}}$. The FWM signal is proportional to $\mathcal{E}_{\rm I}^\ast \mathcal{E}_{\rm II}^2$ in the lowest excitation order, and
a third-order correction to the DM  is given by
\be
\rho^\mathrm{(II)}(0)=\frac{(i \mu_{c_{\rm II}}\mathcal{E}_{\rm II})^2}{2} \Big[c_{\rm II}^\dagger,\big[c_{\rm II}^\dagger,\rho^\mathrm{(I)}(0) \big]\Big]\,,
\label{select-rho3}
\ee
describing the action ${Q}^{\mathrm{(II)}}$ of the second pulse. We emphasize that the excitation channels for pulses I and II need not be the same, so one can have $c_{\rm I} \neq c_{\rm II}$. Having discussed the selection of the desired channels for both the linear and FWM signals, in what follows, we will ignore the constant prefactors $-i \mu_{c_{\rm I}}^\ast\mathcal{E}_{\rm I}^\ast$ and  $(i \mu_{c_{\rm II}}\mathcal{E}_{\rm II})^2$, respectively, in \Eqs{select-rho1}{select-rho3}, consistent with~\cite{allcock2022quantum}.

The total optical polarization is given by:
\begin{align}
P(t)= \langle \langle  {\rho}(t) {O}\rangle \rangle\,,
\label{Ptrace-quantum-thermal}
\end{align}
where the expectation value is taken with respect to the states of the exciton-cavity system, as well as phonon degrees of freedom, which is expressed by double brackets. Similar to the excitation, the response is detected through the coupling to an external field,  and ${O}$ represents an annihilation operator of the observation channel: either cavity ${O}=a$, or exciton ${O}=d$, or a combination (i.e. a linear superposition) of both~\cite{morreau2020phonon}.

It is convenient to expand the DM into the basis of the JC model,
\begin{align}
\rho(t)=\sum_{\eta \xi} \rho_{\eta \xi}(t) |\eta \rangle \langle \xi|\,,
\label{dens-mat-decomp}
\end{align}
where $\rho_{\eta \xi}(t)$ includes the phonon degrees of freedom.
In order to describe the linear and FWM polarizations, in addition to the ground state $|0\rangle$, the following reduced basis~\cite{kasprzak2010up} is used in \Eq{dens-mat-decomp}:
\begin{align}
        |1\rangle &= d^\dagger |0\rangle\,,  \qquad 
    |3\rangle = a^\dagger d^\dagger |0\rangle\,,  \nonumber \\ 
          |2\rangle &= a^\dagger |0\rangle\,,  \qquad 
    |4\rangle = \frac{1}{\sqrt{2}}(a^\dagger)^2 |0\rangle\,. 
    \label{basis}
\end{align}
For the linear polarization, only the ground state $|0\rangle$, the exciton state $|1\rangle$ and the single photon state $|2\rangle$ are required. In order to describe a third-order nonlinearity, two states of the second rung, are also needed. These are state $|3\rangle$ containing single excitations both in the exciton and cavity modes and the two-photon state $|4\rangle$. As we are dealing with a linearly polarized light, the system also has  in principle  a biexciton state involved in the FWM dynamics. However, owing to the biexciton binding energy (which is typically much larger than $g$), this state is detuned out of the energy range in focus (which is of the order of $g$) and can be ignored. The effect of phonons on a nonlinear response of an excitonic system with a possibility of exciton-biexciton transitions was explored in~\cite{krugel2006coherent}, in the absence of a cavity.

Using the expansion \Eq{dens-mat-decomp}, the master equation (\ref{master}) can be expressed in terms of the Liouvillian written as a matrix and the DM written as a vector~\cite{kasprzak2010up}:
\begin{align}
i\dot{\vec{\rho}}=&\mathcal{L} \vec{\rho}\,.
\label{master-matrix}
\end{align}
When expressed as a vector in the basis of states given by \Eq{basis}, the DM takes the following form~\cite{allcock2022quantum}:
\begin{align}
\vec{\rho}^{\mathrm{(0)}}=
\left(
  \begin{array}{cccc}
  \rho_{00}
  \end{array}
\right),
\quad
\vec{\rho}^{\mathrm{(I)}}=
\left(
  \begin{array}{cccc}
  \rho_{01} \\
  \rho_{02}\\
  \end{array}
\right),
\quad
\vec{\rho}^{\mathrm{(II)}}=
\left(
  \begin{array}{cccc}
  \rho_{10} \\
  \rho_{20}\\
  \rho_{31}\\
  \rho_{32}\\
  \rho_{41}\\
  \rho_{42}\\
  \end{array}
\right) .
\label{rho-vector}
\end{align}
The superscripts 0 and I\,(II) are used to distinguish quantities describing the system, respectively, before the pulses and after the application of the first (second) pulse. They refer to a specific size of the reduced basis needed to describe the signal.
The initial DM has only one component $\rho_{00}=\rho_\mathrm{\rm ph}$ with phonons in thermal equilibrium,
\begin{align}
\rho_\mathrm{\rm ph}=\frac{\exp(-\frac{H_{\rm ph}}{k_B T})}{\mathrm{tr}_{\rm ph}\{ \exp(-\frac{H_{\rm ph}}{k_B T} )\}}\,,
\label{rho-ph-equilib}
\end{align}
where $k_B$ is the Boltzmann constant and $T$ is the temperature. Before the first pulse (at $t_{\rm I}=-\tau$) and immediately after it (or after both pulses if $\tau=0$), the DM is factorizable into the JC and phonon parts. In fact, the excitation pulse acts only on the JC component of the DM. Since the pulse is infinitely short, the phonon component of the DM remains unaffected at the time of excitation. At any later time these parts of the DM are entangled, and the phonon bath surrounding the QD is no longer in equilibrium.

When the DM is factorizable or any effects of phonons are neglected, the Liouvillian $\mathcal{L}_{\rm JC}$ of the exciton-cavity system can be diagonalized analytically -- a detailed procedure can be found in the supplement of Ref.\,\cite{allcock2022quantum} where the FWM and higher-order responses were studied with inclusion of many rungs of the JC ladder, though in the absence of exciton-phonon interaction. Here, we exploit this procedure in order to solve the full problem, in which the exciton-phonon interaction is taken into account.

\subsubsection{Linear response}

A linear polarization can be produced by exciting the system with a single pulse.
With $\vec{\rho}^{\mathrm{(0)}}$ being a $1\times1$ vector [see \Eq{rho-vector}], the action on the system of an ultrashort pulse at $t_{\rm I}=-\tau$ can be represented by a vector $\vec{Q}^{\mathrm{(I)}}$. Then \Eq{select-rho1} takes the form:
\begin{align}
\vec{\rho}^{\mathrm{(I)}}(-\tau)={\vec{Q}^{\mathrm{(I)}}} \rho_{00}\,,
\label{firstpulse}
\end{align}
with
\begin{align}
{\vec{Q}^{\mathrm{(I)}}}_x=
\begin{pmatrix}
   {1}  \\
	{0}  \\
\end{pmatrix}\,,
\hspace{2cm}
{\vec{Q}^{\mathrm{(I)}}}_c=
\begin{pmatrix}
   {0}  \\
	{1}  \\
\end{pmatrix},
\label{Q1-x-c}
\end{align}
where the index $x$ ($c$) refers to the exciton (cavity) excitation channel.
The effect of $\vec{Q}^{\mathrm{(I)}}$ is to produce a single-particle excitation (in either the exciton or the cavity mode), converting the element $|0 \rangle \langle 0|$ of the DM into $|0 \rangle \langle 1|$ or $ |0 \rangle \langle 2|$  in the decomposition \Eq{dens-mat-decomp}.
The Liouvillian for the exciton-cavity system after the first pulse has the following form:
\begin{align}
\small
 \mathcal{L}^{\mathrm{(I)}}_{\rm JC}=
\begin{pmatrix}
   {-\Omega_x-i \gamma_x} & {-g}\\
   {-g} & {-\Omega_c-i \gamma_c}
\end{pmatrix}.
\label{L1-JC}
\end{align}
Following the evolution of the DM, from $\vec{\rho}^{\mathrm{(I)}}(-\tau)$ to $\vec{\rho}^{\mathrm{(I)}}(t)$, a linear response is measured in a specific observation channel ${O}$.
With ${\rho}(t)$ and $ {O}$ represented as vectors, \Eq{Ptrace-quantum-thermal} takes the form of a dot product
\begin{align}
P_{\mathrm{Lin}}(t)=\vec{O}^{\mathrm{(I)}} \cdot  \langle \vec{\rho}^{\mathrm{(I)}}(t) \rangle\,,
\label{Plin}
\end{align}
where, for the linear polarization,
\begin{align}
\vec{O}_x^{\mathrm{(I)}}=
\begin{pmatrix}
   {1} \\ {0}
\end{pmatrix} , \qquad \vec{O}_c^{\mathrm{(I)}}=
\begin{pmatrix}
   {0} \\ {1}
\end{pmatrix},
\label{O1-x-c}
\end{align}
and the thermal expectation value in \Eq{Plin} is taken over the phonon bath only.

\subsubsection{FWM response}

Starting with the initial excitation, given by \Eq{firstpulse}, followed by a delay dynamics, in which the system evolves from $t=-\tau$ to $t=0$, the FWM response is produced by applying another pulse (consisting of two identical simultaneous pulses) at $t_{\rm II}=0$.
With $\vec{\rho}^{\mathrm{(I)}}$ given by a $2\times1$ vector in  \Eq{rho-vector}, the action of the second pulse can be represented by a matrix ${Q}^{\mathrm{(II)}}$, so \Eq{select-rho3} becomes
\begin{align}
\vec{\rho}^{\mathrm{(II)}}(0)={Q}^{\mathrm{(II)}} \vec{\rho}^{\mathrm{(I)}}(0)\,,
\label{secondpulse}
\end{align}
where
\begin{align}
{Q}^{\mathrm{(II)}}_x=\frac{1}{2}
\begin{pmatrix}
   {-2} & {0} \\
	{0} & {0} \\
	{0} & {0} \\
   {0} & {0} \\
	{0} & {0} \\
	{0} & {0} \\	
\end{pmatrix},
\hspace{1cm}
{Q}^{\mathrm{(II)}}_c=\frac{1}{2}
\begin{pmatrix}
   {0} & {0} \\
	{0} & {-2} \\
	{0} & {0} \\
   {0} & {0} \\
	{\sqrt{2}} & {0} \\
	{0} & {\sqrt{2}} \\	
\end{pmatrix}.
\label{Q2-x-c}
\end{align}
In terms of the decomposition \Eq{dens-mat-decomp}, $Q^{\mathrm{(II)}}_x$ converts the element $|0\rangle \langle 1|$ into $-2 |1\rangle \langle 0|$, while  $|0\rangle \langle 2|$ does not contribute.
In the case of the cavity excitation, $Q^{\mathrm{(II)}}_c$ converts $|0\rangle \langle 1|$  and $|0\rangle \langle 2|$ , respectively, into $|4\rangle \langle 1|/\sqrt{2} $ and $-|2\rangle \langle 0|$ + $|4\rangle \langle 2|/\sqrt{2} $. At any time after the second pulse ($t>0$), there are six components making up the DM: $|1 \rangle \langle 0|$, $ |2 \rangle \langle 0|$, $|3 \rangle \langle 1|$, $ |3 \rangle \langle 2|$, $|4 \rangle \langle 1|$, and $ |4 \rangle \langle 2|$, in accordance with \Eq{rho-vector}. The Liouvillian matrix then takes the form:
\begin{widetext}
\begin{align}
\mathcal{L}^{\mathrm{(II)}}_{\rm JC}=
\begin{pmatrix}
   {\Omega_x-i \gamma_x} & {g} & {0} & {2i \gamma_c} & {0} & {0} \\
   {g} & {\Omega_c-i \gamma_c} & {2 i\gamma_x} & {0} & {0} & {2\sqrt{2} i \gamma_c} \\
   {0} & {0} & {\Omega_c- i \gamma_c-2 i \gamma_x} & {-g} & {\sqrt{2} g} & {0} \\
   {0} & {0} & {-g} & {\Omega_x- i \gamma_x- 2 i \gamma_c} & {0} & {\sqrt{2} g} \\
   {0} & {0} & {\sqrt{2} g} & {0} & {2\Omega_c-\Omega_x- i\gamma_x- 2 i \gamma_c} & {-g} \\
   {0} & {0} & {0} & {\sqrt{2} g} & {-g} & {\Omega_c- 3 i \gamma_c} \\
\end{pmatrix}.
\label{L2-JC}
\end{align}
\end{widetext}
The FWM polarization is given by
\begin{align}
P_{\mathrm{FWM}}(t)=\vec{O}^{\mathrm{(II)}}  \cdot \langle \vec{\rho}^{\mathrm{(II)}}(t) \rangle
\label{Pfwm}
\end{align}
and is measured (at $t>0$) in an observation channel $\vec{O}^{\mathrm{(II)}}$, which can be either exciton or cavity,
\begin{align}
\vec{O}^{\mathrm{(II)}}_x=
\begin{pmatrix}
   {1} \\ {0}  \\ {0}  \\ {1} \\ {0}  \\ {0}
\end{pmatrix},
\hspace{0.5cm}
\vec{O}^{\mathrm{(II)}}_c=
\begin{pmatrix}
   {0} &\\ {1}  \\ {1}  \\ {0} \\ {0}  \\ {\sqrt{2}}
\end{pmatrix},
\label{O2-x-c}
\end{align}
or a linear combination of the two.

\subsection{Trotter's decomposition}
\label{theory:Trotter}

We now consider the evolution of the system between and after the pulses which is described by the master equation (\ref{master}), or its matrix analogue \Eq{master-matrix}. The first key element of our asymptotically exact approach to the system dynamics is the Trotter decomposition. It is based on Trotter's theorem~\cite{trotter1959product} for two non-commuting operators $A$ and $B$:
\begin{align}
e^{(A+B)t}=\lim_{\Delta t \rightarrow 0} \Big( e^{A \Delta t} e^{B \Delta t}\Big)^N\,,
\label{Trotter}
\end{align}
where $\Delta t=t/N$ and $N$ is an integer. In our case, the IB and JC models are two exactly solvable parts of the system, which are described by non-commuting operators. We therefore set $A=-i\mathcal{L}_{\rm IB}$ and $B=-i\mathcal{L}_{\rm JC}$, the two non-commuting operators of the full composite system, described by the Liouvillian
\begin{equation}
\mathcal{L}=\mathcal{L}_{\rm IB}+\mathcal{L}_{\rm JC}\,.
\label{Lfull}
\end{equation}
We then use \Eq{Trotter}, in order to separate the evolution of the full system  into discrete time intervals, and assume independent evolution of the IB and JC parts within each time interval, exploiting the exact analytic solvability of each part.

The time interval between the pulse time $t_0$ (where $t_0=-\tau$ or  $t_0=0$) and the observation time $t$ is split into $N$ discrete time steps, which  need not be equidistant. Concentrating on the $n$-th time step, between $t_{n-1}$ and $t_n$, with $\Delta t=t_n-t_{n-1}$, and assuming that  $\Delta t$ is small, the evolution of the DM over this time interval can be approximated, in line with \Eq{Trotter}, as
\be
\vec{\rho}(t_n)=  e^{-i \mathcal{L}\Delta t} \vec{\rho}(t_{n-1}) \approx e^{-i \mathcal{L}_{\rm IB}\Delta t} e^{-i \mathcal{L}_{\rm JC}\Delta t} \vec{\rho}(t_{n-1})\,.
\label{evolve-original0}
\ee
To account for the effects of $ \mathcal{L}_{\rm IB}$ and $ \mathcal{L}_{\rm JC}$, we introduce, respectively,  a matrix operator ${W}$ and a matrix ${M}$ (both defined below), such that \Eq{evolve-original0} can be written as
\begin{align}
\vec{\rho}(t_n)  \approx
W(t_n,t_{n-1})   {M}  \vec{\rho}(t_{n-1})  {W}^\dagger (t_n,t_{n-1}) \,.
 \label{evolve-original}
\end{align}

The exciton-phonon dynamics, represented by ${W}$, follows a unitary evolution that can be fully described by the Hermitian Hamiltonians $H_{\rm IB}$ and $H_{\rm ph}$:
\begin{align}
W(t_n,t_{n-1}) &= e^{i H_{\rm ph}t_n} e^{-i H_{\rm IB}(\Delta t)} e^{-i H_{\rm ph}t_{n-1}}  \nonumber\\
&= T \exp{\left\{ -i \int_{t_{n-1}}^{t_n}  dt' V(t') d^\dagger d \right\}}\,,
\label{W-matrix}
\end{align}
where
\begin{align}
     V(t')= e^{i H_{\rm ph} t'} V e^{-i H_{\rm ph} t'}
     \label{V-interaction}
\end{align}
is the interaction representation of the phonon coupling $V$ defined in \Eq{V}. Due to the diagonal form of the exciton-phonon interaction, $W$ and $W^\dagger$ are diagonal matrices, and owing to the Hermiticity of the IB model, $W(t_n,t_{n-1})= W^\dagger (t_{n-1},t_n)$. The non-Hermitian dynamics of the exciton-cavity system is in turn described by the matrix ${M}$ defined as
\begin{align}
    {M}= e^{-i \mathcal{L}_{\rm JC} \Delta t},
\label{M-matrix}
\end{align}
where matrix $\mathcal{L}_{\rm JC}$ depends on the basis used and for the linear and FWM polarizations is given, respectively, by \Eq{L1-JC} and \Eq{L2-JC}.

Introducing the components of the DM vector, the approximate \Eq{evolve-original} can be written as
\begin{align}
{\rho}_{i_n}  &= \sum_{i_{n-1}}
W_{i_{n} i_{n}}  {M}_{i_n i_{n-1}}{\rho}_{i_{n-1}} {W}^\dagger_{i_{n} i_{n}} \,,
 \label{evolve-component}
\end{align}
where the phonon operators  ${W}_{i_{n} i_{n}}$ and ${W}^\dagger_{i_{n} i_{n}}$ are the matrix elements of the diagonal matrix operators $W(t_n,t_{n-1})$ and $W^\dagger (t_{n},t_{n-1})$, respectively, the numbers ${M}_{i_n i_{n-1}}$ are the matrix elements of ${M}$, and ${\rho}_{i_n}$ is the $i_n$-th component of the vector $\vec{\rho}(t_n)$ representing the DM at the time moment $t_n$. Note that the subscript $n$ is included in the index $i_n$ labeling matrix and vector components, in order to keep the information about this time moment and the selected time interval (between $t_{n-1}$ and $t_n$) of the system evolution described by \Eq{evolve-component}. Physically, $i_n$ labels the quantum state of the full system at time $t_n$ on the selected path of its evolution.
In the case of non-equidistant time steps, $W$ and $M$ depend also on $n$, which is omitted for the brevity of notations.

Now, we introduce a specific  time-ordering operator $\tilde{T}$, which allows us to move the operator ${W}^\dagger_{i_{n} i_{n}}$ next to ${W}_{i_{n} i_{n}}$:
\begin{align}
{\rho}_{i_n} &= \sum_{i_{n-1}} \tilde{T}
W_{i_{n} i_{n}} {W}^\dagger_{i_{n} i_{n}}   {M}_{i_n i_{n-1}}{\rho}_{i_{n-1}} \,.
 \label{evolve-Tordering}
\end{align}
In doing so, the order of phonon operators should be preserved. The operator $\tilde{T}$ ensures that all phonon operators in $W$ (${W}^\dagger$) stand to the left (right) of $\vec{\rho}$ in normal (inverse) order. We further introduce an operator
\be
 {Y}_{i_n} =\tilde{T} \exp{\left\{ -i \int_{t_{n-1}}^{t_n} \tilde{V}_{i_n}(t')
 d t' \right\}}
  \label{Y-vector}
\ee
which includes the effects of both $W$ and $W^\dagger$, with the help of $\tilde{T}$ and the interaction
\be
\tilde{V}_{i}(t')=\alpha_{i} V^{(+)}(t')-\beta_{i} V^{(-)}(t')\,.
     \label{Vtilde-left-right-vec}
\ee
This allows us to write \Eq{evolve-Tordering} as
\begin{align}
{\rho}_{i_n} &= \sum_{i_{n-1}} \tilde{T}
{Y}_{i_n}   {M}_{i_n i_{n-1}}{\rho}_{i_{n-1}} \,.
 \label{evolve-Yvec}
\end{align}
Introduced in \Eq{Vtilde-left-right-vec},  operators $V^{(\pm)}(t')$ are the same as the operator $V(t')$, defined by \Eq{V-interaction}, with the only difference that under the action of $\tilde{T}$ all phonon operators in $V^{(+)}$ ($V^{(-)}$) stand to the left (right) of $\vec{\rho}$ in the normal (inverse) order. Vectors $\vec{\alpha}$ and $\vec{\beta}$ are related, respectively, to the left and right sides of the DM and have Boolean components $\alpha_i$ and $\beta_i$ indicating whether or not the exciton-phonon coupling is affecting the given element ${\rho}_{i}$ of the DM from either side. Within \Eq{evolve-Yvec}, ${Y}_{i_n}$ has effect on ${M}_{i_n i_{n-1}}{\rho}_{i_{n-1}}$ from the left (right) only if $\alpha_{i_n} =1$ ($\beta_{i_n} =1$), which takes place when the corresponding component of the DM contains exciton. Otherwise $\alpha_{i_n} =0$ ($\beta_{i_n} =0$), and there is no effect of ${Y}_{i_n}$. Note that $\vec{\alpha}$ and $\vec{\beta}$  are constant vectors, independent of the time step $n$, but are different for different orders of optical nonlinearity. Their explicit form for the linear and FWM polarizations is provided in this section below.

Applying \Eq{evolve-Yvec} to  the full dynamics of the DM, from the initial time $t_0$ of pulsed excitation to the final time $t=t_N$ of signal observation, and using the excitation and measurement conditions, as detailed in \Sec{theory:system}, the optical polarization takes the form
\begin{align}
P (t)= \sum_{i_{{N}}...i_1 i_0} &{O}_{i_{{N}}} {M}_{{i_{{N}}} i_{{N}-1}}\dots {M}_{i_2 i_1}{M}_{i_1 i_0}{Q}_{i_0} \nonumber \\
     & \times \left\langle
    \tilde{T}{Y}_{i_{{N}}} \dots{Y}_{i_2} {Y}_{i_1}
     \right\rangle ,
\label{P-full-before-LCE-zerotau}
\end{align}
where in the last line, a thermal expectation value is taken over the phonon ensemble.
Clearly, the phonon contribution is present only in the time-ordered product of the elements ${Y_{i_n}}$  consisting of $V^{(\pm)}$ operators. This  expectation value will be evaluated in \Sec{theory:linkedcluster} with the help of the linked-cluster theorem~\cite{mahan2013many}.

Equation~(\ref{P-full-before-LCE-zerotau}) is valid for any channel of optical nonlinearity, including linear and FWM response, provided that the system is excited by ultrashort optical pulses. Below we provide details of how to use \Eq{P-full-before-LCE-zerotau} for calculating the linear and FWM polarization at  zero delay ($\tau=0$). The case of a non-zero delay between the pulses ($\tau \neq 0$) is considered in detail in Appendix~\ref{appendix:delay} and also in \Sec{theory:NN}.

\subsubsection{Linear polarization}
\label{Trotter-Plin}
The linear polarization $P_\mathrm{Lin} (t)$ can be extracted from \Eq{P-full-before-LCE-zerotau}, by using  ${M}={M}^{\mathrm{(I)}}$ with $\mathcal{L}_{\rm JC}=\mathcal{L}_{\rm JC}^{\mathrm{(I)}}$ given by \Eq{L1-JC}, $\vec{O}=\vec{O}^{\mathrm{(I)}}$ given by \Eq{O1-x-c}, and  $\vec{Q}=\vec{Q}^{\mathrm{(I)}}$ given by \Eq{Q1-x-c}. The phonon contribution is taken care of by $Y_{i_n}$, defined by \Eq{Y-vector}, with $\vec{\alpha}=\vec{\alpha}^{\mathrm{(I)}}$ and $\vec{\beta}=\vec{\beta}^{\mathrm{(I)}}$, where
\begin{align}
\vec{\alpha}^{\mathrm{(I)}}=&
\begin{pmatrix}
   {0}\\
   {0}\\
\end{pmatrix},
\qquad
\vec{\beta}^{\mathrm{(I)}}=
\begin{pmatrix}
   {1}\\
   {0}\\
\end{pmatrix},
\label{alphabeta1}
\end{align}
as follows from the DM components involved, which are given by \Eq{rho-vector}.

\subsubsection{FWM polarization ($\tau=0$)}
\label{Trotter-Pfwm}
Similarly, the FWM polarization at zero delay, $P_\mathrm{FWM} (t,\tau=0)$, can be obtained from \Eq{P-full-before-LCE-zerotau} by using ${M}={M}^{\mathrm{(II)}}$ with $\mathcal{L}_{\rm JC}=\mathcal{L}_{\rm JC}^{\mathrm{(II)}}$ given by \Eq{L2-JC},
$\vec{O}=\vec{O}^{\mathrm{(II)}}$ given by \Eq{O2-x-c},
and
\begin{align}
\vec{Q}= Q^{\mathrm{(II)}}
\vec{Q}^{\mathrm{(I)}}
\label{O-Q-fwm}
\end{align}
with $\vec{Q}^{\mathrm{(I)}}$ and ${Q}^{\mathrm{(II)}}$ given, respectively, by \Eqs{Q1-x-c}{Q2-x-c}.
For the phonon part, $Y_{i_n}$ is defined by the same \Eq{Y-vector}, with $\vec{\alpha}=\vec{\alpha}^{\mathrm{(II)}}$ and $\vec{\beta}=\vec{\beta}^{\mathrm{(II)}}$, where
\begin{align}
\vec{\alpha}^{\mathrm{(II)}}=&
\begin{pmatrix}
   {1}\\
   {0}\\
   {1}\\
   {1}\\
   {0}\\
   {0}\\
\end{pmatrix},
\qquad
\vec{\beta}^{\mathrm{(II)}}=
\begin{pmatrix}
   {0}\\
   {0}\\
   {1}\\
   {0}\\
   {1}\\
   {0}\\
\end{pmatrix}.
\label{alphabeta2}
\end{align}
This follows from the components of the DM involved in the FWM dynamics, as given by \Eq{rho-vector}.
\subsection{Linked-cluster expansion}
\label{theory:linkedcluster}

The second element of our approach to the system dynamics is the cumulant, or linked-cluster expansion. It allows us
to address the exciton-phonon interaction exactly, by providing an analytic evaluation of the expectation value of the time-ordered product of operators in \Eq{P-full-before-LCE-zerotau}.
As a result, any explicit phonon dependence is removed at a cost of introducing temporal correlations between the states of the system at different time steps.

The linked-cluster theorem~\cite{mahan2013many} is valid for arbitrary operators and a wide class of time orderings, including the normal ($T$) and the inverse ($T_{\mathrm{inv}}$) time orderings.
The specific form of time ordering $\tilde{T}$ introduced above can be considered as a combination of $T$ and $T_{\mathrm{inv}}$, which are distinctly separated: In fact, differently ordered operators stand on either side of the DM and do not mix.
As the two classes of time ordering do not mix in $\tilde{T}$, the linked-cluster theorem holds for this time ordering, resulting in cumulants similar to those which appear in the IB model.

Within the IB model, only second-order connected diagrams contribute to the cumulant.
Applying the linked-cluster theorem, the expectation value of the product of $Y_{i}$ operators in \Eq{P-full-before-LCE-zerotau} takes the form
\be
\Xi(t)\equiv\left\langle
    \tilde{T}{Y}_{i_{{N}}} \dots{Y}_{i_2} {Y}_{i_1}
     \right\rangle =\exp{\left(\sum_{m=1}^N \sum_{n=1}^N \tilde{\mathcal{K}}_{{i}_{m} {i}_{n}}(m,n)\right)},
\label{linked-cluster-expansion}
\ee
where
\begin{align}
   \tilde{\mathcal{K}}_{i_m i_n}(m,n) =-\frac{1}{2} \int_{t_{m-1}}^{t_m} d\tau_1 \int_{t_{n-1}}^{t_n} d\tau_2  \left\langle \Tilde{T} \tilde{V}_{i_m}(\tau_1) \tilde{V}_{i_n}(\tau_2) \right\rangle
    \label{cumulant-Vtilde}
\end{align}
and $\tilde{V}_{i}(\tau)$ is defined by \Eq{Vtilde-left-right-vec}.
Shown above is a general result for an arbitrary channel of nonlinearity. Depending on which channel is considered,  vectors $\vec{\alpha}$ and $\vec{\beta}$ take a particular form, see e.g. \Eqs{alphabeta1}{alphabeta2}.

Owing to the linked-cluster expansion, the phonon contribution to the system dynamics is thus expressed in a form of two-time correlations in system variables determining the path, indexed by $i_n$. The possible realizations (or paths) are indicated via combinations of the basis state labels $i_n$ for all values of $n$ on the time grid. In the optical polarization \Eq{P-full-before-LCE-zerotau}, all possible realizations are summed over.

With a general derivation given in Appendix~\ref{appendix:delay}, Eq.(\ref{cumulant-Vtilde}) becomes
\begin{align}
\tilde{\mathcal{K}}_{i_m i_n}(m,n) =    (\alpha_{i_m}-\beta_{i_m})( \alpha_{i_n} {K}_{mn} - \beta_{i_n} {K}_{mn}^\ast )
    \label{cumulant-general}
\end{align}
for $m \geqslant n$, and is symmetric with respect to the interchange of indices:
\be
\tilde{\mathcal{K}}_{i_n i_m}(n,m) =\tilde{\mathcal{K}}_{i_m i_n} (m,n)\,.
\label{Ksym}
\ee
Within \Eq{cumulant-general}, each cumulant element is given by:
\be
   {K}_{mn} =-\frac{1}{2} \int_{t_{m-1}}^{t_m} d\tau_1 \int_{t_{n-1}}^{t_n} d\tau_2  D(\tau_1-\tau_2)\,,
    \label{cumulant-element-defn}
\ee
where
\be
   D(t) =\sum_\q | \lambda_\q |^2 \left[ (N_q+1) e^{-i \omega_q |t|} + N_q e^{i \omega_q |t|}  \right]
    \label{D}
\ee
is the phonon propagator and
\be
N_q=\frac{1}{\exp[\omega_q/(k_b T)]-1}
\label{Bose}
\ee
is the Bose occupation number. Substituting \Eqsss{linked-cluster-expansion}{cumulant-general}{cumulant-element-defn} into \Eq{P-full-before-LCE-zerotau} allows us to find the optical polarization.

Note that we have used so far an arbitrary time grid, which is not necessary equidistant. Below, for clarity of presentation, we use an equidistant time grid, with a constant time step $\Delta t$.
As we would like the time evolution to be composed of identical memory kernels (visualized by L-shapes, see below), an equidistant grid is also more relevant because the cumulant function is a function of the difference between the two times $t_1$ and $t_2$.
In this case, all the cumulant elements depend on the difference $l=|n-m|$ only and  can be written as $K_{nm}=K_{mn}=R_l$, with $R_l$ found recursively via
\be
2R_{l-1}= K(l\Delta t)-lR_0 -2 \sum_{k=1}^{l-2} (l-k) R_k\,,
\label{Rl}
\ee
starting from $l=2$ and using $R_0=K(\Delta t)$. Here the cumulant function of the IB model is defined as
\be
K(t)=-\dfrac{1}{2} \int_0^{t} d\tau_1 \int_0^{t} d\tau_2 {D}(\tau_1-\tau_2)
\label{diag-cumulant-element-fn}
\ee
[cf. \Eq{cumulant-element-defn}]. Then for $m\geqslant n$, \Eq{cumulant-general} simplifies to
\begin{align}
\tilde{\mathcal{K}}_{i_{m}i_n}(m,n) =& \mathcal{K}_{i_{n+l} i_n}(l)
\nonumber\\
\equiv&    (\alpha_{i_{n+l}}-\beta_{i_{n+l}})( \alpha_{i_n} {R}_{l} - \beta_{i_n} {R}_{l}^\ast )\,,
    \label{cumulant-equidistant}
\end{align}
where $l=m-n\geqslant0$. In the opposite case of $m\leqslant n$, one can use the symmetry \Eq{Ksym} to obtain $\tilde{\mathcal{K}}_{i_{m}i_n}(m,n) = \mathcal{K}_{i_{m+l} i_m}(l)$ with $l=n-m$ and the same $\mathcal{K}_{ij}(l)$ defined in \Eq{cumulant-equidistant}.

If the phonon bath  has a memory time $\tau_{\rm IB}$ (introduced in Ref.~\cite{morreau2019phonon}, see also \App{appendix:timescales}), it is not necessary to take into account all $R_l$ on the time grid up to $l=N$, but it is  sufficient to cover only a portion of the two-dimensional time grid including a distance of about $\tau_{\rm IB}/2$ from the main diagonal (see \Sec{theory:LN}). In this case the number of cumulant elements $R_l$ which are taken into account is limited to some finite value $l\leqslant L$. The number of time steps in the memory segment of the system evolution, or the number of ``neighbors'' $L$ is chosen in our calculation as to satisfy a criterion $\Delta t \ll \tau_{\rm JC},\,\tau_{\rm IB}$, which is imposed by the Trotter decomposition.  Here $\tau_{\rm JC}$ is the timescale for the JC dynamics, see \App{appendix:timescales}.
Below we first consider in \Sec{theory:NN} the $L=1$ case called a nearest-neighbor approximation and then proceed in \Sec{theory:LN} with a general case of $L\geqslant1$ which we call an $L$-neighbor approach.

\subsection{Nearest-neighbor approximation}
\label{theory:NN}

The theory presented so far has focused on the case of zero delay between optical excitation pulses.
For the purpose of illustration of the full method, we consider here both cases of zero and non-zero delay, illustrating and comparing them in the nearest-neighbor (NN) approximation. The latter is achieved by setting $L=1$ so that only $|n-m| \leqslant1$ are kept in the double sum in \Eq{linked-cluster-expansion}, and the DM or the optical polarization can be expressed at any time simply as a product of matrices. This approach is valid for a relatively small exciton-cavity coupling strength $g$, for which $\tau_{\rm JC} \gg \tau_{\rm IB}$, and breaks down as the two timescales become comparable. Owing to its simplicity, the NN approximation serves as a clear illustration of the Trotter decomposition method in general  and provides a basic start point for the full $L$-neighbor ($L$N) solution presented in \Sec{theory:LN} below. It also allows us to clearly demonstrate how one can treat an arbitrary sequence of excitation pulses, at the same time taking into account any memory effects and the phonon contribution to the evolution of the system between the pulses.

In the limit $\tau_{\rm JC} \gg \tau_{\rm IB}$, the full cumulant in Eq.(\ref{linked-cluster-expansion}) can be approximated by a reduced cumulant which includes only the diagonal ($m=n$) and the NN ($m=n\pm1$) terms. For the linear or FWM polarization with $\tau=0$, the cumulant for a given realization (or a path) then takes the form:
\be
\ln{\Xi(t)} \approx \sum_{n=1}^N \mathcal{K}_{i_n i_n}(0)+ 2\sum_{n=1}^{N-1} \mathcal{K}_{i_{n+1}i_n }(1)\,,
\label{NN-zerotau-cumulant}
\ee
where $\mathcal{K}_{ij}(l)$ is defined in \Eq{cumulant-equidistant}.
For the linear polarization, using the explicit form of $\vec{\alpha}^{\mathrm{(I)}}$ and $\vec{\beta}^{\mathrm{(I)}}$ given by \Eq{alphabeta1}, this reduces to
\be
\ln{\Xi(t)} \approx \delta_{i_N,1} R_0^\ast +\sum_{n=1}^{N-1} \delta_{i_n,1} \left({R}^\ast_{0} + 2
\delta_{i_{n+1},1} {R}^\ast_{1}\right)\,,
\label{NN-Plin-cumulant}
\ee
where $\delta_{nm}$ is the Kronecker delta. Note that \Eq{NN-Plin-cumulant} is the complex conjugate of the expression given by Eq.\,(19) of Ref.~\cite{morreau2019phonon}, which is obtained here in accordance with the phase selection determined by \Eq{select-rho1}.

The optical polarization can then be written as a product of matrices,
\be
P_{\rm NN}(t)=
\vec{O} \cdot E {\mathcal{G}}^{N-1} {M} \vec{Q}\,,
\label{P-NN-zerotau}
\ee
where
\begin{align}
\label{G-matrix}
\mathcal{G}_{ij}=& {M}_{ij} e^{\mathcal{K}_{jj}(0) +2\mathcal{K}_{ij}(1)}\,, \\
   E_{ij}= & \delta_{ij} e^{\mathcal{K}_{jj}(0)}\,,
\label{E-matrix}
\end{align}
and $t=N\Delta t$. The time step $\Delta t$ satisfies the condition $\Delta t\leq \tau_{\rm IB}$, and for short times ($0<t<\tau_{\rm IB}$) is chosen as $\Delta t= t/2$, so that the NN approximation takes all cumulant elements into account. \Eqsss{P-NN-zerotau}{G-matrix}{E-matrix} are valid both for linear and FWM polarization at $\tau=0$. For the FWM (linear) polarization, one should use $\vec{O}$, $\vec{Q}$, $\vec{\alpha}$, and $\vec{\beta}$ with index ${\mathrm{II}}$ (${\mathrm{I}}$), except $\vec{Q}$ which is given by \Eq{O-Q-fwm} in the case of the FWM with $\tau=0$. For the linear polarization, \Eq{P-NN-zerotau} reproduces the result obtained in Ref.\,\cite{morreau2019phonon}.

Physically, $ {M} \vec{Q}$ in \Eq{P-NN-zerotau} represents the state of the system following a pulsed excitation and a subsequent propagation by one time step under the influence of $\mathcal{L}_{\rm JC}$. The full evolution of the system by a single time step is carried out through a multiplication with the matrix $\mathcal{G}$, which is repeated $N-1$ times. The phonon contribution at the final step of evolution is included in the matrix $E$, and the optical polarization is then obtained by taking the dot product with $\vec{O}$, in this way projecting the DM onto a selected observation channel.

For the FWM polarization with $\tau > 0$, the cumulant \Eq{NN-zerotau-cumulant} modifies to include in the dynamics both  the delay time $\tau=N_\mathrm{I}\Delta \tau$ and the observation time $t=N_\mathrm{II}\Delta t$, with the number of discrete steps $N_\mathrm{I}$ and $N_\mathrm{II}$, respectively.
The time steps satisfy the condition $\Delta t, \Delta \tau \leq \tau_{\rm IB}$. Equation (\ref{P-NN-zerotau}) then modifies as follows
\begin{align}
P_{\rm NN}(t,\tau)=&
\vec{O}^{\mathrm{(II)}} \cdot E^{\mathrm{(II)}} \left[{\mathcal{G}^{\mathrm{(II)}}}\right]^{N_\mathrm{II}-1} \nonumber \\
&\times \mathcal{G}^{\mathrm{(I-II)}} \left[{\mathcal{G}^{\mathrm{(I)}}}\right]^{N_\mathrm{I}-1} {M}^{\mathrm{(I)}} \vec{Q}^{\mathrm{(I)}}\,, 
\label{P-NN-delay}
\end{align}
where
\begin{align}
\mathcal{G}^{\mathrm{(I-II)}}_{ij}=& [{M}^{\mathrm{(II)}} Q^{\mathrm{(II)}}]_{ij} \exp\left({\mathcal{K}^\mathrm{(I)}_{jj}(0)  +2\mathcal{K}^{\mathrm{(I-II)}}_{ij}(1)}\right),
\label{G1-2-mat}
\\
\mathcal{K}^{\mathrm{(I-II)}}_{ij}(1)&=(\alpha_{i}^{\mathrm{(II)}}-\beta_{i}^{\mathrm{(II)}})
\left(\alpha_{j}^{\mathrm{(I)}} R_1^{\mathrm{(I-II)}}-\beta_{j}^{\mathrm{(I)}} {R_1^{\mathrm{(I-II)}}}^\ast\right),
\label{cumulant-1-2-mat}
\end{align}
and matrices $\mathcal{G}^{\mathrm{(\zeta)}}$ and $\mathcal{K}^{\mathrm{(\zeta)}}(l)$ are defined as before, respectively, by \Eqs{G-matrix}{cumulant-equidistant}, with the upper index $\zeta$ (taking the values I or II) added to the matrix ${M}$ and vectors $\vec{\alpha}$ and $\vec{\beta}$, see \Eqsss{L1-JC}{L2-JC}{M-matrix}, as well as \Eqs{alphabeta1}{alphabeta2} for their definitions. Matrix
$E_{ij}^{\mathrm{(II)}}= \delta_{ij} e^{\mathcal{K}^\mathrm{(II)}_{jj}(0)}$ is also defined as in \Eq{E-matrix}, now with the upper index II added. The range of the matrix indices is different in different regions:  $\mathcal{G}^{\mathrm{(I)}}$  and ${M}^{\mathrm{(I)}}$ are $2\times2$ matrices; $\mathcal{G}^{\mathrm{(II)}}$, ${M}^{\mathrm{(II)}}$, and $E^{\mathrm{(II)}}$ are $6\times6$ matrices, while $\mathcal{G}^{\mathrm{(I-II)}}$ is a $6\times2$ matrix.

The specific cumulant elements, which are required for the NN approximation and appear in \Eqs{cumulant-equidistant}{cumulant-1-2-mat} are obtained in the same way as $R_l$ in \Eq{Rl}, and take the form
\begin{align}
&R^{\mathrm{(I)}}_0=K(\Delta \tau)\,,  \qquad  \qquad  \qquad
R^{\mathrm{(II)}}_0=K(\Delta t)\,, \nonumber \\
&R^{\mathrm{(I)}}_1=K(2\Delta \tau)/2-R^{\mathrm{(I)}}_0, \ \ \ \ \
R^{\mathrm{(II)}}_1=K(2\Delta t)/2-R^{\mathrm{(II)}}_0,
\nonumber \\
&2R^{\mathrm{(I-II)}}_1=K(\Delta \tau +\Delta t)-R^{\mathrm{(I)}}_0-R^{\mathrm{(II)}}_0\,.
\label{NN-cumulant-elements}
\end{align}
Note that they are all expressed in terms of the cumulant function $K(t)$ given by \Eq{diag-cumulant-element-fn}. Clearly, within the NN approximation there are only five distinct cumulant elements given by \Eq{NN-cumulant-elements} which must be considered.

\begin{figure}[t]
\resizebox{9cm}{9cm}{%
 \begin{tikzpicture}
\pgfdeclarelayer{Llayer}
\pgfsetlayers{main,Llayer}
\definecolor{lightyellow}{rgb}{1.0, 0.86, 0.35}

\foreach \x in {0,...,10} {
      \draw[fill=lightgray,fill opacity=0.03,draw=none] (0,0) rectangle ++(4.5+0.05*\x,4.5+0.05*\x);
      \draw[fill=almond,fill opacity=0.08,draw=none] (0,0) rectangle ++(-4-0.05*\x,-4-0.05*\x);
      \draw[fill=lightyellow,fill opacity=0.01,draw=none] (0,0) rectangle ++(-4-0.05*\x,4.4+0.05*\x);
      \draw[fill=lightyellow,fill opacity=0.01,draw=none] (0,0) rectangle ++(4.5+0.05*\x,-4-0.05*\x);
}
\draw node at (-3.5,-4.3) {(I)};
\draw node at (0.7,4.8) {(II)};
\draw node at (-3.5,4.8) {(I-II)};
\draw node at (0.7,-4.3) {(I-II)};

\foreach \x [
evaluate=\x as \xb using int(4-\x),
evaluate=\x as \xp using int(3-\x),
evaluate=\x as \xy using int(1+\x),
evaluate=\x as \xr using int(2+\x)
] in {0,...,3} {
 \Ks{-\x-1}{-\x-1}{1}{1}{ }{cyan};
\ifthenelse{\x<3}{
 \Ks{-\x-2}{-\x-1}{1}{1}{}{Purple};	
 \Ks{-\x-1}{-\x-2}{1}{1}{}{Purple};	
 \Ks{\x*1.5}{\x*1.5}{1.5}{1.5}{}{yellow}
 \Lshape{black}{1}{\x-4}{2};
}{}
\ifthenelse{\x<2}{
 \Ks{1.5*\x}{1.5*\x+1.5}{1.5}{1.5}{}{red};
 \Ks{1.5*\x+1.5}{1.5*\x}{1.5}{1.5}{}{red};
\Lshape{black}{1.5}{\x}{2};
}{}
\ifthenelse{\x<1}{
 \Ks{\x-1}{1.5*\x}{1}{1.5}{}{Gray}
 \Ks{1.5*\x}{\x-1}{1.5}{1}{}{Gray}
 \Lshape{black}{1}{\x-1}{2.5};
 \Lshape{black}{1.5}{\x+3}{1};
}{}
}

\draw[<->] (3,1.3) -- (4.5,1.3) node at (3.75,1) {$\Delta t$};
\draw[<->] (-4,-1.7) -- (-3,-1.7) node at (-3.5,-1.5) {$\Delta \tau$};

\draw[style=help lines,step=1.5] (0,0) grid (5,5);
\draw[style=help lines,step=1] (-4.5,-4.5) grid (0,0);
\draw[help lines,xstep=1,ystep=1.5] (-4.5,0)  grid (0,5);
\draw[help lines,xstep=1.5,ystep=1] (0,-4.5)  grid (5,0);

\draw [black,thick] (0,0) circle [radius=0.1];
\draw[->] (-4.5,0) -- (5,0) node[right] {};
\draw[->] (0,-4.5) -- (0,5) node[above] {};
\draw node at (5.25,0.0) {$t_1$};
\draw node at (0.0,5.25) {$t_2$};
\draw node at (-4,-0.2) {$-\tau$};
\draw node at (-0.3,-4) {$-\tau$};
\draw node at (4.5,-0.2) {$t$};
\draw node at (-0.3,4.5) {$t$};

\foreach [count=\i] \x in {$i_4$, $i_3$, $i_2$, $i_1$} {\node (\i) at (-\i+0.5,-4.7) {\x};}
\foreach [count=\i] \x in {$i_5$, $i_6$, $i_7$} {\node (\i) at (1.5*\i-0.75,-4.7) {\x};}
\foreach [count=\i] \x in {$i_4$, $i_3$, $i_2$, $i_1$} {\node (\i) at (-4.7,-\i+0.5) {\x};}
\foreach [count=\i] \x in {$i_5$, $i_6$, $i_7$} {\node (\i) at (-4.7,1.5*\i-0.75) {\x};}
\end{tikzpicture}
}
\caption{
An example of the time grid for the FWM polarization with non-zero delay in the NN approach with $N_\mathrm{I}=4$ and $N_\mathrm{II}=3$, showing the delay time ($\tau$) region (I), the observation time ($t$) region (II) and the mixed region (I-II). Only the self interaction (main diagonal) and the NN interaction (nearest off-diagonal elements) are included.
The evolution along this grid consists of L-shaped regions (these are bounded by a dashed line), which correspond to an element of a matrix $\mathcal{G}$.
Grid squares are colored to distinguish different cumulant elements [see \Eq{NN-cumulant-elements}].
}
\label{NNkr}
\end{figure}
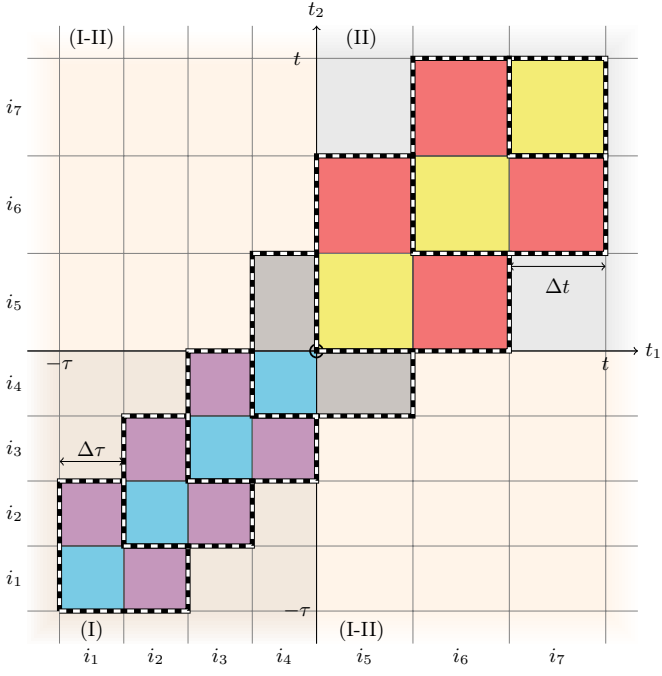

The NN approximation \Eq{P-NN-delay} for the FWM polarization with $\tau > 0$ is illustrated in \Fig{NNkr}, indicating three different regions in the time domain, labeled with (I), (II), and (I--II).
Different cumulant elements ${R}_j$ defined in \Eq{NN-cumulant-elements} are distinguished by color.
The correlations in memory, arising from the phonon contribution, are limited to the indicated L-shaped regions. Note in particular that the (I-II) region, where the reduced basis size changes, includes path segments connecting states of the system evolution both in the delay and the observation time regions.

\subsection{Long-time asymptotics}
\label{theory:longtime}
A much better understanding of system's behavior can be achieved by separating the dynamics which occurs on different timescales, $\tau_{\rm IB}\ll \tau_{\rm JC}$. We provide two different ways to describe the long-time behavior analytically: (i) applying a multi-exponential fit to our numerical solution and (ii) using the polaron approximation (PA).
This section is devoted to the PA, while the multi-exponential fit approach is introduced at the end of \Sec{theory:LN} below. For the review of the PA in greater depth and its behavior with the delay time, see Ref.~\cite{paper1}.

The PA is obtained by taking the long-time limit of \Eq{diag-cumulant-element-fn}, $K(t)\to -S+i\Omega_p t$. For the linear and $\tau = 0$ FWM polarization the PA is given by (see Ref.~\cite{paper1} for the derivation)
\begin{align}
P_\mathrm{PA}(t) = \vec{O} \cdot e^{-\mathcal{S}/2} e^{-i \tilde{\mathcal{L}}_{\rm JC}  t} e^{-\mathcal{S}/2} \vec{Q}\,,
\label{polaron-approx-zerotau}
\end{align}
where $\mathcal{S}$ is a diagonal matrix, given by $\mathcal{S}^{\mathrm{(I)}}=\mathrm{diag}(S,0)$ for the linear and $\mathcal{S}^{\mathrm{(II)}}= \mathrm{diag}(S,0,0,S,S,0)$ for the FWM polarization, and
$\tilde{\mathcal{L}}_{\rm JC}$ is the polaron transformed JC Liouvillian, obtained from $\mathcal{L}_{\rm JC}$ provided in \Eqs{L1-JC}{L2-JC} by replacing the exciton transition energy $\Omega_x$ and the coupling strength $g$ with, respectively,  a polaron-shifted exciton energy $\tilde{\Omega}_x$ and an effective coupling strength $\tilde{g}$, which are given by
\begin{align}
\tilde{\Omega}_x&=\Omega_x+\Omega_p\,, \nonumber\\
\tilde{g}&=g e^{-S/2}\,.
\label{Polaron}
\end{align}
The polaron shift $\Omega_p$ and Huang-Rhys~\cite{huang1950theory} factor $S$ have the following explicit form:
\begin{align}
\Omega_p&=-\int_{0}^{\infty}d\omega \frac{J(\omega)}{\omega}\,,
\nonumber \\
S&=\int_{0}^{\infty}d\omega \frac{J(\omega)}{\omega^2} \coth{\Big(\frac{\omega}{2k_BT}\Big)}\,,
\label{Huang-Rhys}
\end{align}
and are determined by the phonon spectral density $J(\omega)=\sum_\q |\lambda_\q|^2 \delta(\omega-\omega_q)$. For QD systems, considered in this work,
the phonons spectral density has a superohmic form $J(\omega)= J_0 \omega^3 e^{-\omega^2/\omega_0^2}$, where $J_0$ and $\omega_0$ are constants determined by the material parameters of the QD (see Appendix~F of Ref.~\cite{morreau2019phonon} for details and derivation).

For $\tau > \tau_{\rm IB}$, the PA for the FWM polarization can be obtained from \Eq{P-NN-delay} (see Ref.~\cite{paper1} for the derivation) and becomes
\begin{align}
P_{\mathrm{PA}}(t,\tau)= \vec{O}^{\mathrm{(II)}} \cdot e^{-\frac{\mathcal{S}^{\mathrm{(II)}}}{2}} e^{-i \tilde{\mathcal{L}}^{\mathrm{(II)}}_{\rm JC} t} \tilde{Q}^{\mathrm{(II)}}  e^{-i \tilde{\mathcal{L}}^{\mathrm{(I)}}_{\rm JC} \tau} e^{-\frac{\mathcal{S}^{\mathrm{(I)}}}{2}} \vec{Q}^{\mathrm{(I)}},
\label{polaron-approx-delay}
\end{align}
where
\be
\tilde{Q}^{\mathrm{(II)}}_x= {Q}^{\mathrm{(II)}}_x e^{-2S}\,,\quad \tilde{Q}^{\mathrm{(II)}}_c={Q}^{\mathrm{(II)}}_c\,,
\ee
with  ${Q}^{\mathrm{(II)}}_x$ and ${Q}^{\mathrm{(II)}}_c$ defined in \Eq{Q2-x-c}.

\subsection{Full numerical $L$-neighbor approach}
\label{theory:LN}

Here we present a full $L$N approach, already mentioned at the end of \Sec{theory:linkedcluster}, which is based on the scheme proposed in Ref.~\cite{morreau2019phonon} and can be understood as a ``forward- memory'' approach (see below for details). For clarity of presentation, we return back to linear and zero-delay FWM polarizations -- the general case of non-zero delay is derived in Appendix~\ref{appendix:delay} which also provides details of the formalism presented here.

The $L$N approach is a generalization of the NN approximation, given by \Eq{P-NN-zerotau}, and consists of a recursive generation of tensors $\mathcal{F}^{(n)}$, starting from $n=1$, and ending up with $n=N$, the latter determining the optical polarization:
\begin{align}
\mathcal{F}_{i_{L}...i_{1}}^{(1)}=&\sum_{j=1}^h {M}_{i_1 j} Q_{j}\,,
 \label{F-initial}
 \\
\mathcal{F}^{(n+1)}_{i_{L}...i_{1}}=&\sum_{j=1}^h \mathcal{G}_{i_{L}...i_{1}j} \mathcal{F}^{(n)}_{i_{L-1}...i_{1}j} , \label{F-evolve}
\\
\mathcal{G}_{i_{L}...i_{1}j}=& {M}_{i_{1} j}
e^{\mathcal{K}_{jj}(0) +2\mathcal{K}_{i_1 j}(1)+\dots +2\mathcal{K}_{i_L j}(L)}\,,
\label{G-tensor}
\\
P_{L{\rm N}}(t)=& \sum_{j=1}^h e^{\mathcal{K}_{jj}(0)} O_j \mathcal{F}^{(N)}_{p\dots p j}\,.
\label{F-remove-tails}
\end{align}
Tensors $\mathcal{F}^{(n)}$ of rank $L$ are generated with the help of a constant tensor $\mathcal{G}$ of rank $L+1$ which is the $L$N analogue of the matrix $\mathcal{G}$, given by \Eq{G-matrix}, that appears in the NN approach. Calculation of the tensor $\mathcal{G}$ requires only $L+1$ distinct cumulant elements $\mathcal{K}_{i j}(l)$ defined in \Eq{cumulant-equidistant}.   As in \Eq{P-NN-zerotau}, for the FWM (linear) polarization, one should use vectors, tensors, and matrices with the upper index ${\mathrm{II}}$ (${\mathrm{I}}$), except $\vec{Q}$ which is given by \Eq{O-Q-fwm} in the case of the FWM with $\tau=0$. 
We shall introduce a short-hand notation, where we first specify the excitation channel and then the measurement channel (e.g. c-x refers to the situation where the system is initially excited in the cavity mode and the response is measured via the excitonic mode).
In \Eq{F-remove-tails} $h=p=6$ ($h=p=2$) for the zero-delay c-c and c-x FWM (x-x, x-c FWM and linear) polarization. The time step is set to $\Delta t = \tau_{\rm IB}/L$ for $t\geqslant \tau_{\rm IB}$ and at earlier times ($t<\tau_{\rm IB}$) to $\Delta t = t/(L+1)$, in the latter case taking all possible cumulant elements into account. This allows one to resolve the initial dynamics on finer timescales and capture the fast initial decay of the polarization.

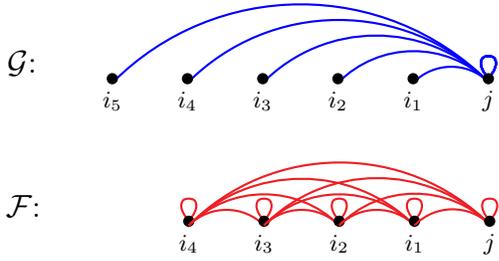
\begin{figure}[t]
\begin{tikzpicture}[thick]

\node at (-0.2,0.5) {\large $\mathcal{G}$:};

  \foreach [count=\i] \x in {$i_5$, $i_4$, $i_3$, $i_2$, $i_1$, $j$} {
      \node (\i) at (\i,0) {\x};

  }
  \foreach [count=\i] \x in {$i_5$, $i_4$, $i_3$, $i_2$, $i_1$, $j$} {
 \draw [blue] (\i.north) to[bend left=45] (6.north);
 \draw [blue] plot [smooth, tension=1] coordinates {(6,0.3) (6-0.1,0.5) (6,0.6) (6+0.1,0.5) (6,0.3)};
 \node at (\i,0.3)[circle,fill,inner sep=1.5pt]{};}
 \node (7) at (7,0) {};
\end{tikzpicture}

\begin{tikzpicture}[thick]

\node at (-0.2,0.5) {\large $\mathcal{F}$:};

  \foreach [count=\i] \x in {$ $, $i_4$, $i_3$, $i_2$, $i_1$, $j$, $ $} {
      \node (\i) at (\i,0) {\x};
  }
  \foreach [count=\i] \x in {$ $, $i_4$, $i_3$, $i_2$, $i_1$, $j$, $ $} {
  	\foreach [count=\j] \x in {$ $, $i_4$, $i_3$, $i_2$, $i_1$, $j$, $ $} {
  		\ifthenelse{\i>\j \AND \j>1 \AND \i<7}
           {\draw [Red] (\j.north) to[bend left=45] (\i.north);}	
  	}
\ifthenelse{\i>1 \AND \i<7 }
           {\draw [Red] plot [smooth, tension=1] coordinates {(\i,0.3) (\i-0.1,0.5) (\i,0.6) (\i+0.1,0.5) (\i,0.3)};
           \node at (\i,0.3)[circle,fill,inner sep=1.5pt]{};}
 }
  \draw [opacity=0] (1.north) to[bend left=45] (6.north);
  \node [text opacity=0](1) at (1,0) {$i_5$};
  \end{tikzpicture}
\caption{
Diagrams showing how the system variables determining the path are linked within  $\mathcal{G}$ (blue) and $\mathcal{F}$ (red) of \Eq{F-evolve}  for $L=5$. 
$\mathcal{G}$ contains path segments connecting the time of observation (indexed by $j$) to itself, as well as to five other nearest time intervals (indexed by $i_n$). These links are weighted by appropriate cumulant elements if the interaction is present or by zero if there is no interaction and form a memory kernel.
In $\mathcal{F}$, all system variables that describe $L$-step segment along some path, are linked together. This tensor contains a future `history' of exciton-phonon interactions that extends over five nearest timesteps. 
}
\label{connections}
\end{figure}

\begin{figure}[t]
\resizebox{8cm}{8cm}{%
 \begin{tikzpicture}
\pgfdeclarelayer{Llayer}
\pgfsetlayers{main,Llayer}

\foreach \x in {0,...,10} {
      \draw[fill=lightgray,fill opacity=0.03,draw=none] (0,0) rectangle ++(4.5+0.05*\x,4.5+0.05*\x);
}
\draw[darkgray!80!white] node at (0.7,4.5) {(II)};

\foreach \x in {0,...,7} {
 \Ks{\x*0.4}{\x*0.4}{0.4}{0.4}{}{yellow}
 \Lshape{black}{0.4}{\x}{6};

\ifthenelse{\x<7}{
 \Ks{0.4*\x}{0.4*\x+0.4}{0.4}{0.4}{}{red};
 \Ks{0.4*\x+0.4}{0.4*\x}{0.4}{0.4}{}{red};
}{
 \Ks{0.4*\x}{0.4*\x+0.4}{0.4}{0.4}{$\times$}{red};
 \Ks{0.4*\x+0.4}{0.4*\x}{0.4}{0.4}{$\times$}{red}; }

\ifthenelse{\x<6}{
 \Ks{0.4*\x}{0.4*\x+0.8}{0.4}{0.4}{}{Blue};
 \Ks{0.4*\x+0.8}{0.4*\x}{0.4}{0.4}{}{Blue};
}{
 \Ks{0.4*\x}{0.4*\x+0.8}{0.4}{0.4}{$\times$}{Blue};
 \Ks{0.4*\x+0.8}{0.4*\x}{0.4}{0.4}{$\times$}{Blue}; }

\ifthenelse{\x<5}{
 \Ks{0.4*\x}{0.4*\x+1.2}{0.4}{0.4}{}{pink};
 \Ks{0.4*\x+1.2}{0.4*\x}{0.4}{0.4}{}{pink};
}{
 \Ks{0.4*\x}{0.4*\x+1.2}{0.4}{0.4}{$\times$}{pink};
 \Ks{0.4*\x+1.2}{0.4*\x}{0.4}{0.4}{$\times$}{pink}; }

\ifthenelse{\x<4}{
 \Ks{0.4*\x}{0.4*\x+1.6}{0.4}{0.4}{}{gray};
 \Ks{0.4*\x+1.6}{0.4*\x}{0.4}{0.4}{}{gray};
}{
 \Ks{0.4*\x}{0.4*\x+1.6}{0.4}{0.4}{$\times$}{gray};
 \Ks{0.4*\x+1.6}{0.4*\x}{0.4}{0.4}{$\times$}{gray}; }

\ifthenelse{\x<3}{
 \Ks{0.4*\x}{0.4*\x+2}{0.4}{0.4}{}{Green};
 \Ks{0.4*\x+2}{0.4*\x}{0.4}{0.4}{}{Green};
}{
 \Ks{0.4*\x}{0.4*\x+2}{0.4}{0.4}{$\times$}{Green};
 \Ks{0.4*\x+2}{0.4*\x}{0.4}{0.4}{$\times$}{Green}; }

}

\draw[->] (0,0) -- (5,0) node[right] {};
\draw[->] (0,0) -- (0,5) node[above] {};

\draw[-] (0.1,3.2) -- (-0.1,3.2) node[left] {$t$};
\draw[-] (3.2,0.1) -- (3.2,-0.1) node[below] {$t$};

\draw[style=help lines,step=1.6] (0,0) grid (5,5);

\draw[->] (0,0) -- (5,0) node[right] {};
\draw[->] (0,0) -- (0,5) node[above] {};
\draw node at (5.25,0.0) {$t_1$};
\draw node at (0.0,5.25) {$t_2$};
\end{tikzpicture}
}
\caption{
A time grid for $L=5$ corresponding to the evolution of the system between the pulse time and the observation time $t$.
The evolution along this grid consists of L-shaped regions, which are bounded by the dashed lines. The time steps within each L-shape are linked by
an element of the tensor $\mathcal{G}$. Grid squares are colored to distinguish different cumulant elements \Eq{cumulant-equidistant}, which characterize the strength of non-local interaction between system variables for a pair of time steps within the memory. Cumulants (elements of the grid) not included in the path sum (as they are beyond $t$) are labeled by $\times$.
}
\label{LNkrr}
\end{figure}
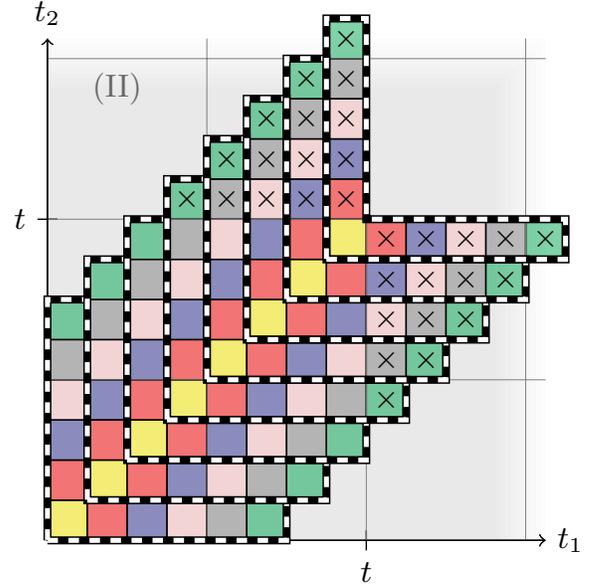

Physically, \Eq{F-initial} forms the initial DM, by applying a pulsed excitation to the ground state of the JC system and adding a propagation by one time step under the influence of $\mathcal{L}_{\rm JC}$ [see \Eq{M-matrix}], without
contribution from the phonon dynamics yet. The subsequent evolution of the DM is described by \Eq{F-evolve}, representing each single time step in the form of a tensor product, expressed as a summation over $j$.
The final result expressed by \Eq{F-remove-tails} is obtained by taking a dot product $\vec{O} \cdot \langle \vec{\rho}(t) \rangle$, in accordance with \Eqs{Plin}{Pfwm}, projecting the DM onto a selected observation channel.

The tensor $\mathcal{G}$, given by \Eq{G-tensor}, is a memory kernel responsible for phonon-induced correlations which are schematically demonstrated in \Fig{connections} for both $\mathcal{G}$ and $\mathcal{F}$. The tensor $\mathcal{G}$ contains the full information required to propagate the system by a single time step and includes path segments connecting the current time interval to $L$ nearest intervals and to itself (\Fig{connections}, blue). The tensor $\mathcal{F}^{(n)}$ contains the information about the state of the system and its interactions with the phonon environment in the form of two-time correlations in system variables determining the path (\Fig{connections}, red). It includes all possible path segments within the memory kernel. In the literature, $\mathcal{G}$ and $\mathcal{F}$, or their analogues,  are sometimes referred to as the propagator tensor and the augmented density tensor, respectively  (see e.g. \cite{makri1995methods,makri1995theory,strathearn2018efficient}).

As already mentioned above, we follow the forward memory $L$N approach introduced in  Ref.~\cite{morreau2019phonon}. Figure~\ref{LNkrr} illustrates the method for $L=5$, showing a time grid representing temporal correlations in system variables.  The L-shaped regions (bounded by dashed lines) contain contributions within the memory kernel. A product of terms in each such region forms a rank-6 tensor $\mathcal{G}_{i_5 ... i_1 j}$. As it is clear from direct L shapes, the memory kernel includes correlations with the future time steps, rather than the past, and any cumulant contributions after the observation time (labeled by crosses) are discarded.
In \Eq{F-remove-tails}, representing the physical observable extracted from the tensor $\mathcal{F}$, the index $p$ is chosen in such a way that any cumulant contributions which fall outside the range of the propagation are removed. In other words, the choice of the index $p$ is dictated by the fact that $L$ last L-shape regions are truncated. Choosing $p=2$ or $p=6$ for the FWM polarization (both corresponding to the cavity components of the DM) ensures that there is no phonon contribution beyond the observation time $t$, in accordance with the vanishing second and sixth elements of both vectors $\vec{\alpha}^{\mathrm{(II)}}$ and $\vec{\beta}^{\mathrm{(II)}}$, see \Eq{alphabeta2}. In fact, this corresponds to paths beyond time $t$ with no change of the DM, since phonons do not interact with the cavity.

It has been demonstrated in Ref.~\cite{morreau2019phonon} that the linear polarization shows a bi-exponential behaviour at long times. Expecting a similar property of the FWM polarization, we introduce
a multi-exponential form of a fit function,
\begin{align}
P_\mathrm{fit}(t) = \sum_{j=1}^{h} A_j e^{-i\omega_j t}\,,
\label{fit}
\end{align}
which we use in \Sec{section:results} below for approximating the full numerical data at long times,  $t\gg \tau_{\rm IB}$. Here $h=6$ ($h=2$)  for the FWM (linear) polarization, corresponding to six (two) optical transitions involved in the coherence dynamics. The amplitudes $A_j$ and the corresponding frequencies $\omega_j$  are complex-valued and found numerically by fitting the numerical data at long times. Note that when excited via the cavity channel, the system reaches second rung of the JC ladder, so in fact six transitions take part in the FWM dynamics, and $h=6$. Excitonic excitation can only produce ground state to first rung coherences in the FWM, therefore only two possible transition frequencies contribute, so $h=2$, the same as in the linear polarization.

\section{Results}
\label{section:results}

In this work we use the exciton-phonon parameters of InGaAs QDs, same as in Ref.~\cite{morreau2019phonon}.
These parameters are in turn taken from Refs.~\cite{muljarov2004dephasing,muljarov2005phonon}:
$D_c-D_v=-6.5$\,eV, $\rho_m=5.65$\,g $\mathrm{cm}^{-3}$, $v_s=4.6$\,km $\mathrm{s}^{-1}$, $l=3.3$\,nm. They describe, respectively, the difference in deformation potentials of the conduction and valence bands, mass density, speed of sound in the QD material, and the electron and hole Gaussian confinement radius.
The cavity parameters are extracted from the FWM experiment~\cite{kasprzak2010up} with QDs embedded in a micropillar cavity: $g=50\,\mu$eV (corresponding to $\tilde{g}=39\,\mu$eV at $T=50\,$K), $\gamma_c=30\,\mu$eV, and $\gamma_x=2\,\mu$eV. We also use in all results a zero effective detuning, Re$(\tilde{\omega}_x-\omega_c)=0$, in which the polaron shift $\Omega_p$ is compensated by the energy difference between the bare exciton and the cavity modes, see \Eqs{Polaron}{Huang-Rhys}. As for the coupling strength $g$, in addition to the above value of $g=50\,\mu$eV used in \Sec{results:smallg} and  corresponding to the experiments~\cite{kasprzak2010up,albert2013microcavity}, we explore in \Secs{results:changeg}{results:changeT} and in Appendices \ref{appendix:channels} and \ref{appendix:convergence} also the regime of a much stronger coupling, with $g=0.3\,$meV and $g=0.8\,$meV. These large values of the coupling strength were shown to be achievable in modern experiments~\cite{peter2005,englund2007,Faraon2008,takamiya2013,dory2016complete}.

\subsection{Small g regime: Comparison of approaches}
\label{results:smallg}

In this section we explore the regime of relatively small exciton-cavity coupling strength $g=0.05$\,meV, for which $\tau_{\rm JC} \gg \tau_{\rm IB}$. Still, the exciton and the cavity mode are strongly coupled.
In this regime, both the NN (corresponding to $L=1$, see \Sec{theory:NN})  and the PA (see \Sec{theory:longtime}) provide good approximations to the exact optical polarization calculated in the full $L$N approach (see \Sec{theory:LN}). The latter is also analyzed in terms of the multi-exponential fit  described at the end of \Sec{theory:LN}. The convergence of the $L$N result to the exact solution is studied in detail in \App{appendix:convergence}. In this section we use only  $L=9$ (the numerical results are  referred below to as just $L$N). This value of $L$ is sufficient for the coupling strengths used in this paper (with root mean square deviation from the exact result of up to 0.03-0.3\% for the linear polarization), as demonstrated in \App{appendix:convergence}. 

While the main purpose of this paper is to study the FWM polarization, we also include here some results for the linear polarization, in order to provide a broader comparison. This allows us to see changes of the phonon contribution to the polarization, induced by the third-order optical nonlinearity. For a more detailed study of the linear polarization in this system see Ref.~\cite{morreau2019phonon}. As for the FWM, we focus here on the case of zero delay $\tau=0$ for clarity. A generalization of the theory for arbitrary delay times is available in Appendix~\ref{appendix:delay} and is studied in more depths in Ref.~\cite{paper1}.

To see how linear and nonlinear effects manifest themselves in different excitation and measurement channels, we focus here on (i) the FWM polarization in the x-x channel when the optical excitation and measurement are performed through the QD exciton state, and on (ii) both the linear and FWM polarizations in the c-c channel when the excitation and measurement are done via the cavity mode. Other combinations of excitation and measurement channels in the FWM and linear polarizations are explored, respectively, in \App{appendix:channels} and Ref.\,\cite{morreau2020phonon}. Note that the dynamics of the x-x FWM polarization is essentially the same as the x-x linear polarization, studied in Ref.\,\cite{morreau2019phonon}, since only the first rung of the JC ladder is involved, as it is clear from the form of the excitation operator ${Q}^{\mathrm{(II)}}_x$ given by \Eq{Q2-x-c}. However, this simpler coherent dynamics can still be measured~\cite{langbein2006heterodyne} as a nonlinear optical response of the system. As for the c-c channel, the linear and the FWM polarizations are entirely different, as they involve different quantum transitions, as discussed below in more depths.

\begin{figure}[t]
\centering
\raisebox{4cm}{}{\includegraphics[width=0.45\textwidth]{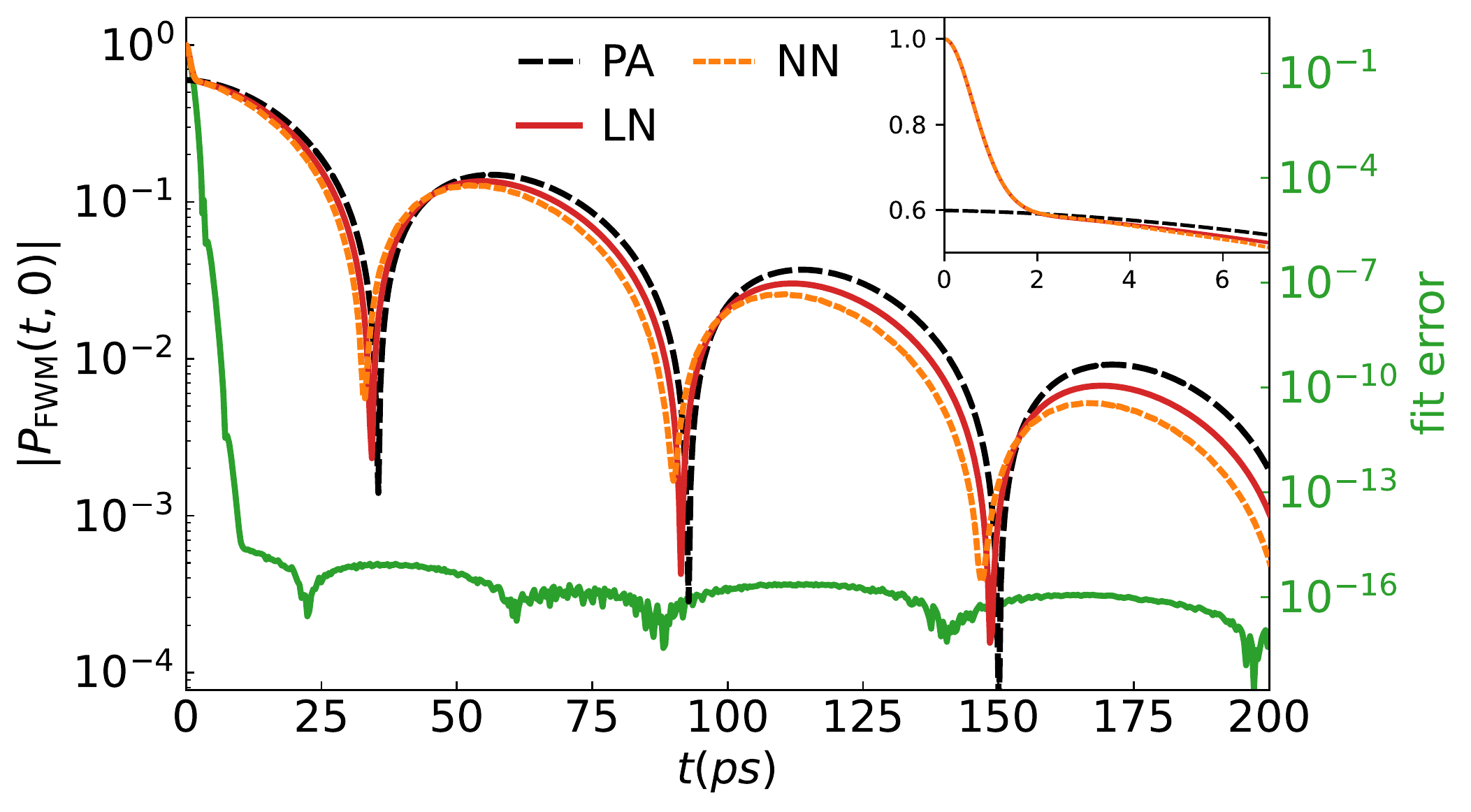}}
\caption{
x-x FWM polarization at $\tau=0$, $g=0.05$\,meV ($\tilde{g}=0.039$ meV), and $T=50$\,K, calculated in the $L$N ($L=9$, red solid), NN (red dotted), and PA approach (black dashed line). The $L$N result is fitted with two complex exponentials [see \Eq{fit}], to separate the long-time from the short-time behavior, here shown as a relative error of the fit (green curve). The inset shows the initial dynamics.
}
\label{res-xx-50}
\end{figure}

To begin with, we consider the situation when the system is excited and the response is measured via the excitonic mode (\Figs{res-xx-50}{spec-xx-50}). In this case the second-rung coherences are absent in the dynamics and the results reduce to the linear polarization, as mentioned above. 
Note that in the absence of the cavity, the zero-delay FWM and the linear response are identical up to a phase factor $P_\mathrm{FWM}(t,0)=-P_\mathrm{Lin}^*(t)$, and the full effect of the linear exciton-phonon coupling \Eq{V} can be taken into account analytically, which is know as the IB model. Moreover, even for arbitrary delay times, the FWM response  can be expressed entirely in terms of the linear response  using an analytic relation~\cite{muljarov2006nonlinear}
\be
P_\mathrm{FWM}(t,\tau)=-\frac{|P_\mathrm{Lin}(t)|^2 [P_\mathrm{Lin}(\tau)]^2}{P_\mathrm{Lin}(t+\tau)}\,.
\ee
In the presence of the cavity, the linear and the zero-delay FWM polarizations in x-x and x-c channels (where the second rung is not directly excited) are also identical up to a phase factor in the full calculation.

\begin{figure}[t]
\centering
\raisebox{4cm}{}{\includegraphics[width=0.45\textwidth]{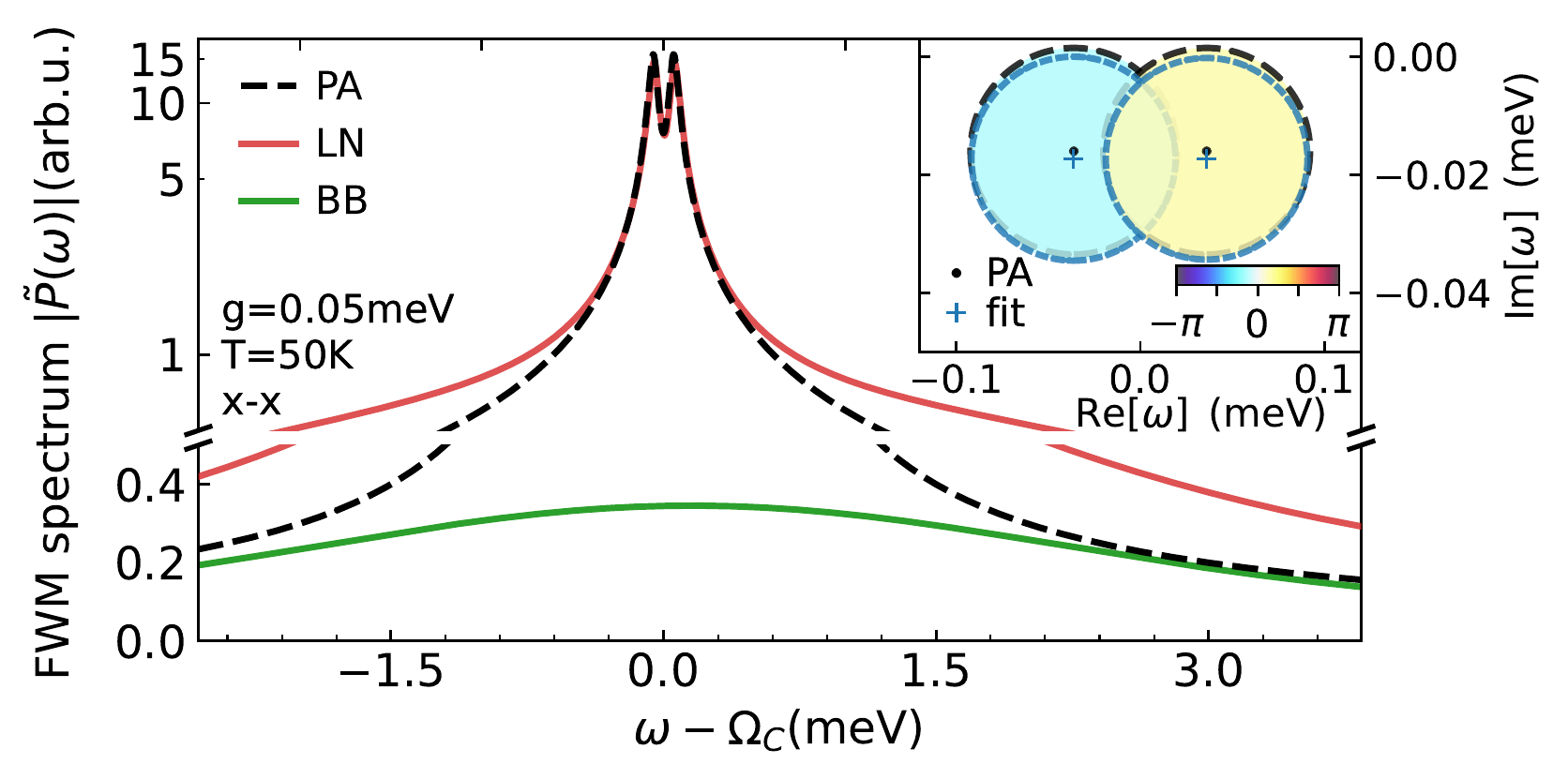}}
\caption{
x-x FWM spectrum, i.e. the Fourier transform of the FWM polarization, calculated in the $L$N and PA approaches shown in \Fig{res-xx-50}. The full numerical $L$N result (red), is separated into ZPLs (the inset) and a phonon broad band (green line), corresponding, respectively, to the long-time and short-time parts of the signal. The upper (lower) part of the spectrum has a logarithmic (linear) scale. The inset shows the complex frequencies of the bi-exponential fit \Eq{fit} (blue crosses) and  PA (black dots) along with the corresponding amplitudes $|A_i|$ given by the circle area 
and their phases (all shifted by $\pi$) color coded.
}
\label{spec-xx-50}
\end{figure}

Figures \ref{res-xx-50} and \ref{spec-xx-50} show, respectively, the time dependence of the FWM signal $|P_{\rm FWM}(t,0)|$ and the corresponding spectrum $|\tilde{P}_{\rm FWM}(\omega)|$, which is the Fourier transform of the former, $\tilde{P}_{\rm FWM}(\omega)=\int_0^\infty P_{\rm FWM}(t,0) e^{i\omega t} dt$. 
One can see a fast initial decay of the polarization (\Fig{res-xx-50}) followed by a multi-exponential (here, bi-exponential) behavior at later times. The former is well seen in the inset and the fit error (green line). 
The initial decay can be attributed to the rapid polaron formation as a result of the instantaneous excitation of the QD by a laser pulse. The polaron formation occurs on a  timescale of $\tau_{\rm IB}=3.25$\,ps and corresponds to a phonon broad band (BB) in the spectrum, represented by the Fourier transform of the complex absolute error of the fit (green line in \Fig{spec-xx-50}). The Fourier transform of the fit corresponds to the zero-phonon line (ZPL) in the spectrum. The ZPL is described in this case by four complex parameters, which are illustrated in the inset.  The fit error, i.e. the difference between the bi-exponential fit and the full calculation shown by the green line in \Fig{res-xx-50} is in the $10^{-4}-10^{-3}$ range for long times and gradually (exponentially) increases at earlier times, reflecting the non-Markovian dynamics of the polaron formation.

The fitted long-time behavior of the $L$N is captured well by the PA, although the decay rates are slightly underestimated, as can be seen from the bubble plot (the inset in \Fig{spec-xx-50}), representing the complex frequencies and the amplitudes of both exponentials (the circles are centered at the complex frequencies, with the circle area proportional to the modulus of the amplitude). In fact, the phonon BB is not taken into account in the PA, and any phonon-related effects are limited to reduction of the exciton-cavity coupling and a change in the detuning, see \Eq{Polaron}. With this renormalization, the quantum dynamics in the PA is entirely determined by the JC model. The NN approximation is capable of capturing correctly the BB and also describes the quantum dynamics at longer times quite well, which is expected in this regime of small $g$~\cite{morreau2019phonon,paper1}. Being very efficient and straightforward, the NN approximation is, however, a semianalytic approach. The fully analytic PA is derived from the NN approximation in the long-time limit, so the PA and NN naturally agree at long times and differ only in that PA underestimates and NN overestimates the decay compared to the $L$N.

\begin{figure}[t]
\centering
\raisebox{4cm}{(a)}{\includegraphics[width=0.45\textwidth]{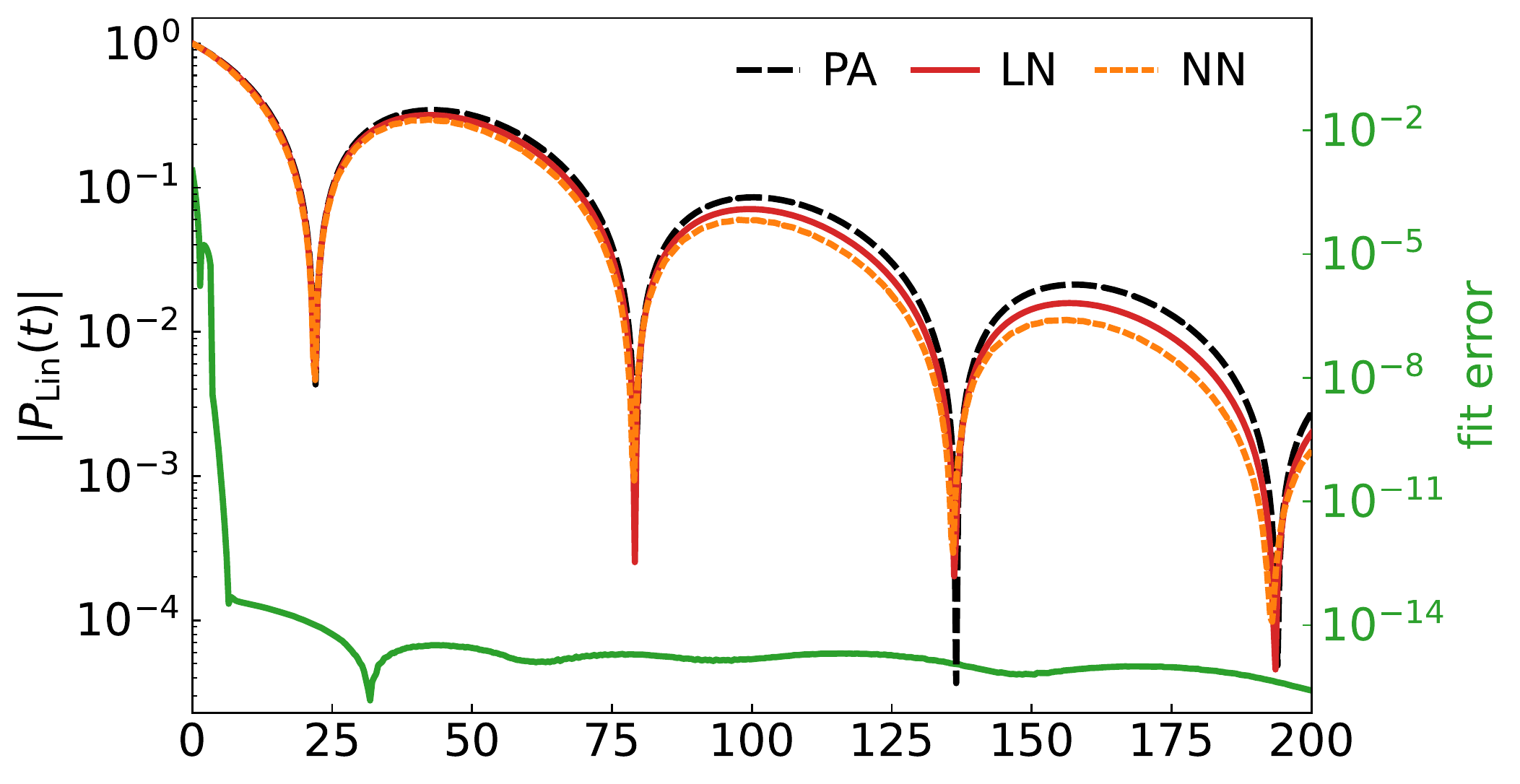}}
\raisebox{4cm}{(b)}{\includegraphics[width=0.45\textwidth]{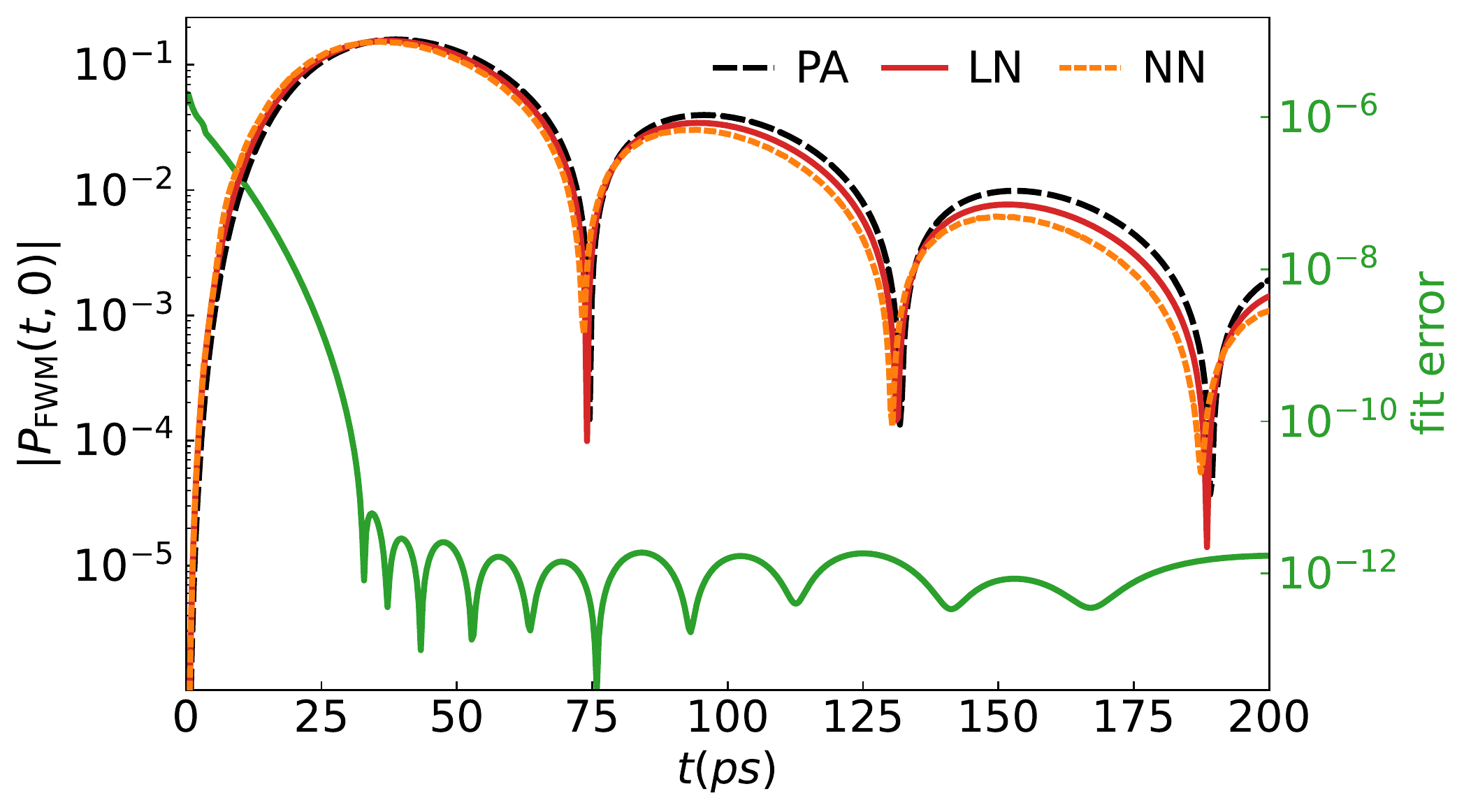}}
\caption{
As \Fig{res-xx-50} but for (a) linear and (b) FWM c-c polarizations. For the FWM polarization the multi-exponential fit \Eq{fit} consists of six exponentials, corresponding to the six transitions between the rungs of the JC ladder involved in the quantum dynamics.
}
\label{res-cc-50}
\end{figure}

\begin{figure}[t]
\centering
\raisebox{4cm}{(a)}{\includegraphics[width=0.45\textwidth]{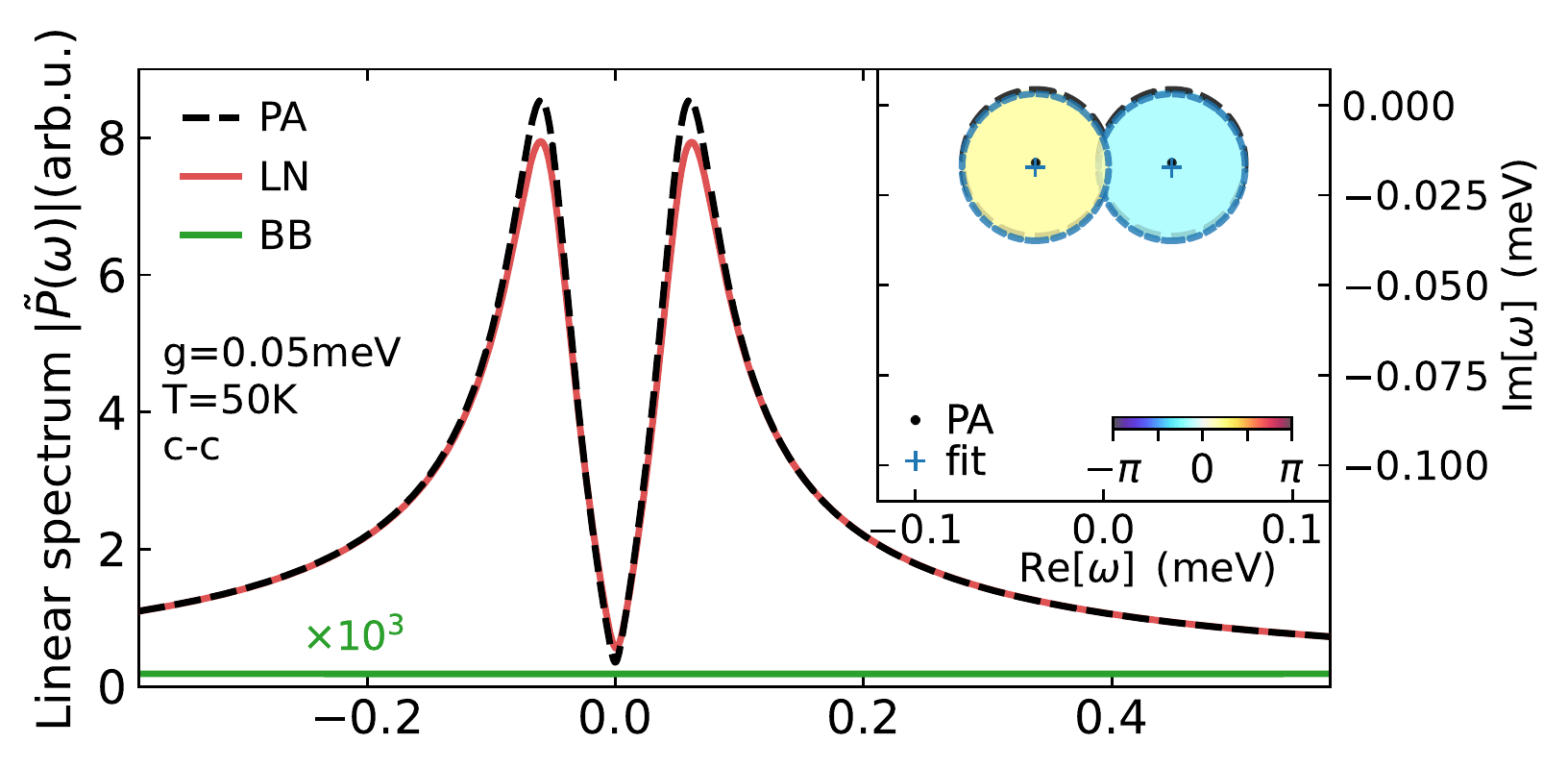}}
\raisebox{4cm}{(b)}{\includegraphics[width=0.45\textwidth]{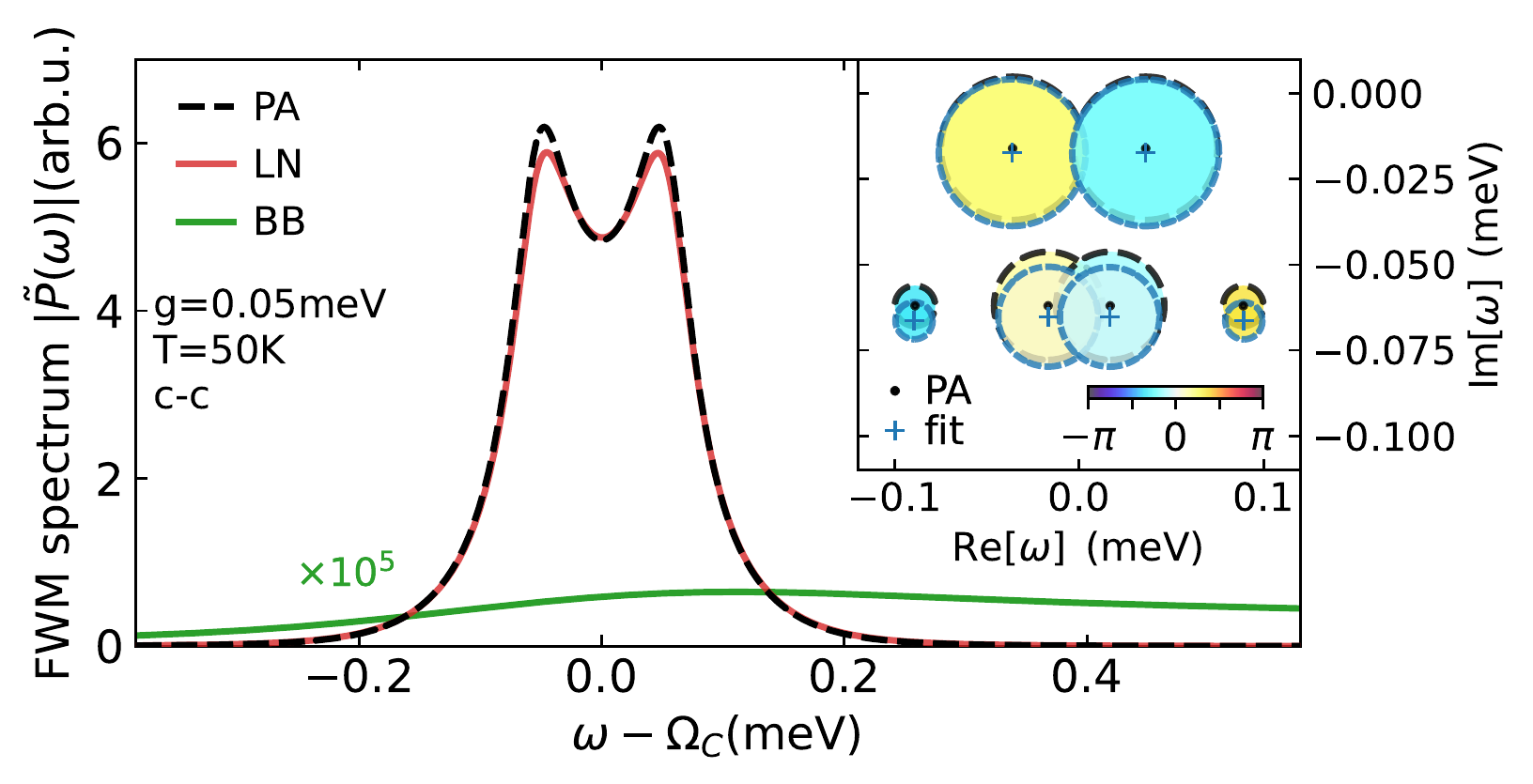}}
\caption{
As \Fig{spec-xx-50} but for (a) linear and (b) FWM c-c polarizations. The inset in (b) shows six exponentials of the fit (blue crosses and dotted circles) and  PA (black dots and dashed circles).
The amplitudes (circle areas) of the two transitions in the linear response in the inset in (a) are multiplied by a factor of 0.415 to match the amplitudes of the corresponding transitions in the FWM.
}
\label{spec-cc-50}
\end{figure}

Let us now focus on the case in which the initial excitation and the measurement of the response are both done via the cavity mode. Figures \ref{res-cc-50} and \ref{spec-cc-50} show, respectively, the time dependence and spectra for linear (upper panels) and FWM (lower panels) polarizations. In the linear response, the spectrum is dominated by two Lorentzian lines, corresponding to quantum transitions between the ground state and the first rung of the JC ladder, the same as in the x-x polarization, see \Fig{spec-cc-50}(a). In the time domain, the linear c-c polarization also starts from one but does not show any initial pure dephasing, see \Fig{res-cc-50}(a). As a result, the BB is practically absent in the spectrum. The c-c FWM response shows a few notable differences compared to the x-x FWM and c-c linear polarizations. One difference arises from the fact that there are six transitions involved in the quantum dynamics, including four transitions between the first and the second rungs, compare the insets in Figs.\,\ref{spec-cc-50}(a) and (b). The spectra in Figs.\,\ref{spec-cc-50}(a) and (b) show, respectively,  two and six Lorentzian lines, both in the fit and PA. As a result of destructive interference between the first- and second-rung transitions~\cite{kasprzak2010up,allcock2022quantum}, the FWM spectrum is shrunk significantly (i.e. the spectral tails are suppressed) compared to the linear spectra. The other difference is that in the time domain, the FWM signal starts from zero, see \Fig{res-cc-50}(b). This is because the excitation of the system via the cavity mode does not immediately lead to an optical nonlinearity, because the cavity itself is linear (photons do not interact directly with each other). In other words, the excitation has to be converted from photon to exciton and back to photon, in order to produce a nonlinearity measured in the c-c FWM, with the rise time of the signal proportional to $t^2$~\cite{allcock2022quantum}.

The absence of a visible phonon BB in the spectrum is a feature common to all c-c polarizations. Indeed, this is what we see in \Fig{spec-cc-50}, and there is no apparent fast decay in the $L$N data at short times, see \Fig{res-cc-50}. This is because the polaron formation is adiabatically slow in the case of cavity excitation and small $g$. In fact, the polaron cloud forms around the QD or disappears on the timescale $\tau_{\rm JC}$, and since $\tau_{\rm JC} \gg \tau_{\rm IB}$ the quick non-Markovian dynamics of the polaron formation is spread over the much slower JC evolution. As a result, there is a better overall agreement between $L$N,  PA, and NN, both in the linear and FWM signals. Also, the fit error is reduced by an order of magnitude compared to the x-x polarization (\Fig{res-cc-50}), reaching $10^{-5}$ for the FWM polarization. Still, there is a few times smaller (compared to x-x) leftover after the multi-exponential fit at short times and, as a consequence, a small BB is present in the spectra (green lines in \Fig{spec-cc-50}, magnified by a factor of $10^3$ and $10^5$ for the linear and the FWM response, respectively). Surprisingly, the BB in the FWM corresponds to a much longer non-Markovian time compared to the linear and even the x-x FWM polarizations (10--20\,ps versus 3\,ps). This is due to the quantum nonlinearity of the JC system, leading to a quicker polariton dynamics when the second-rung transitions are involved. These non-Markovian features enhance dramatically for higher-order nonlinearities, involving higher rungs of the JC ladder, and for larger values of $g$. The latter is demonstrated and discussed in more depths in \Sec{results:changeg} below.

\begin{figure}[t]
\centering
\raisebox{4cm}{(a)}{\includegraphics[width=0.45\textwidth]{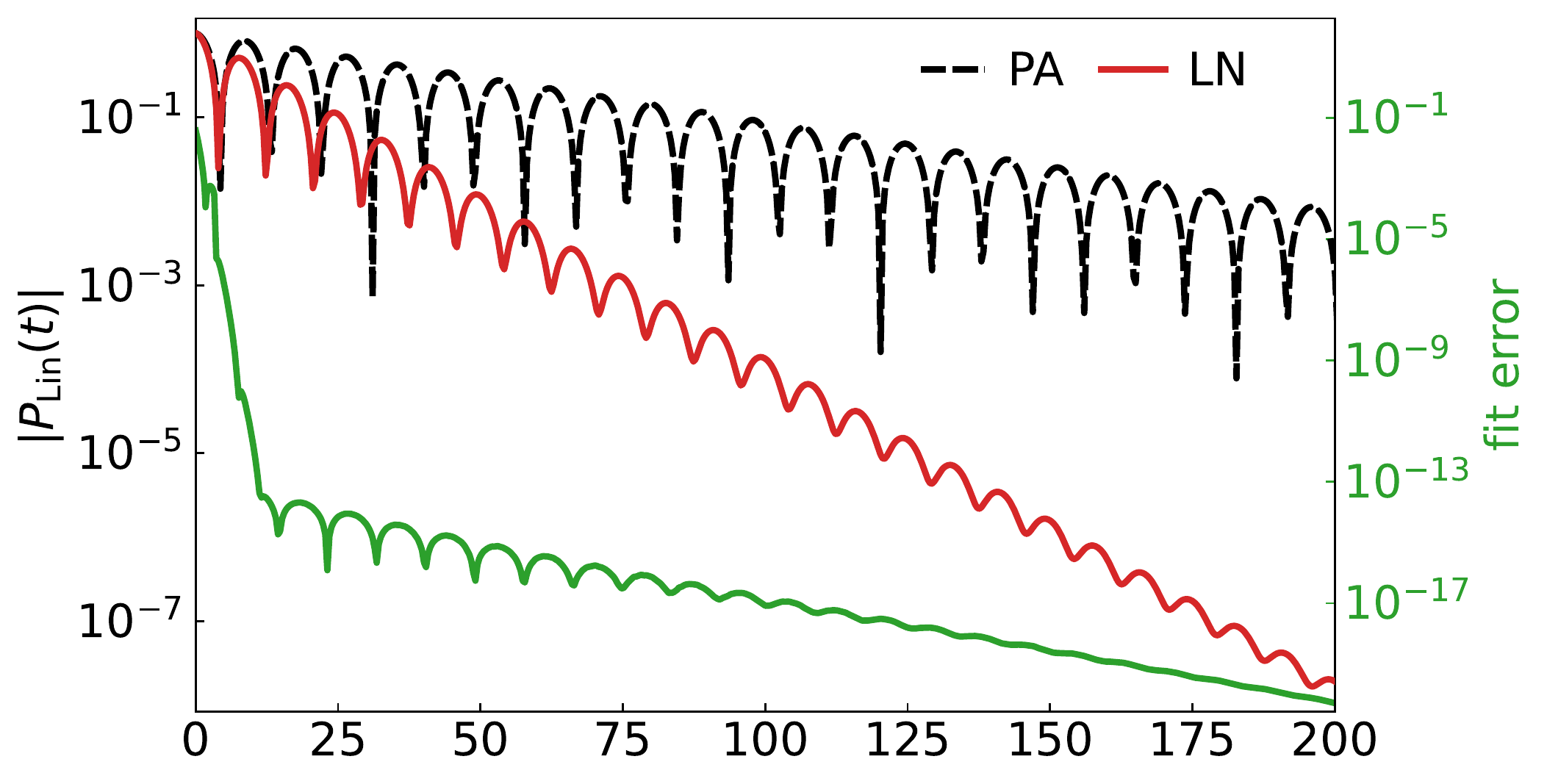}}
\raisebox{4cm}{(b)}{\includegraphics[width=0.45\textwidth]{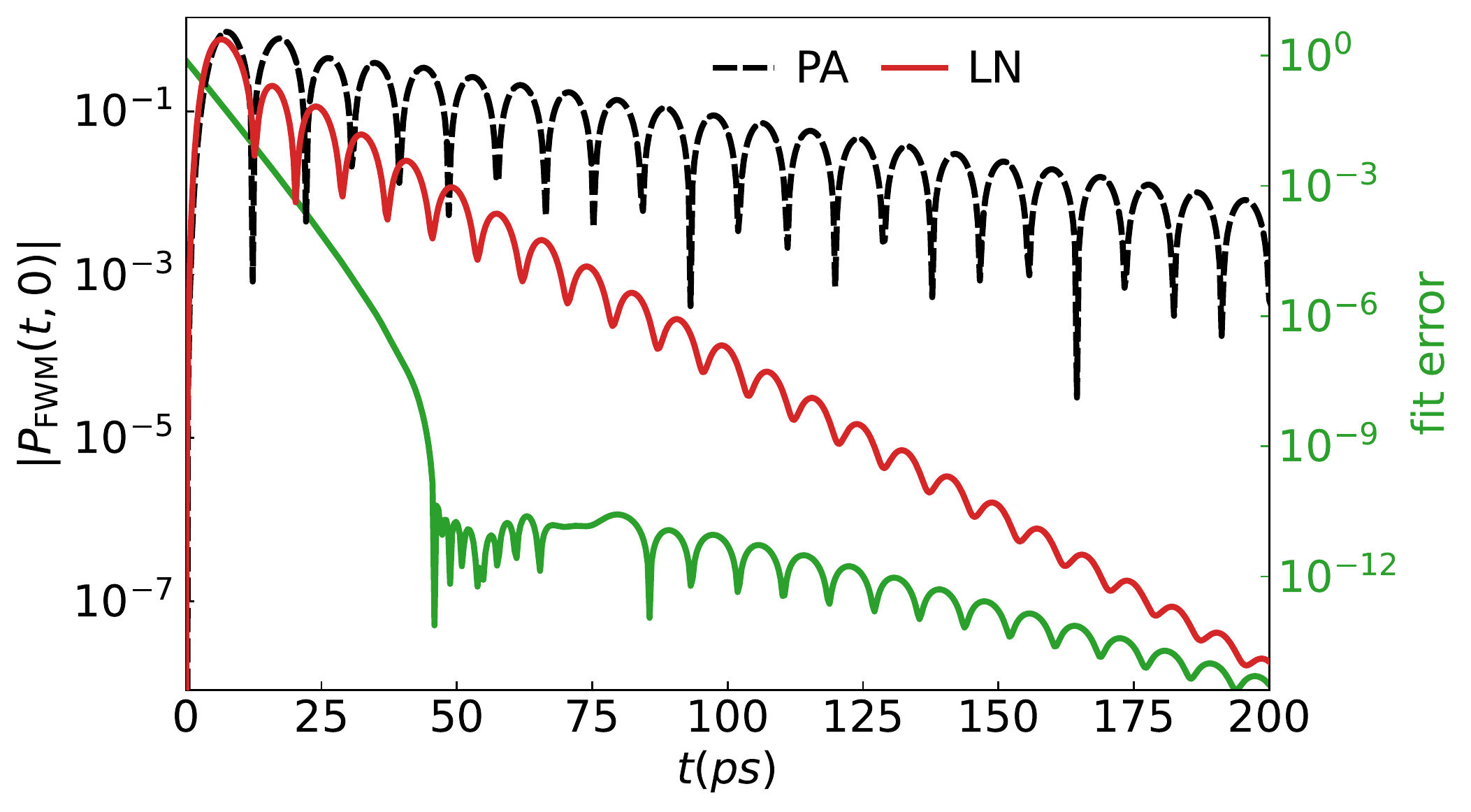}}
\caption{
As \Fig{res-cc-50} but for $g=0.3$\,meV ($\tilde{g}=0.23$ meV) and without the NN approximation.
}
\label{res-cc-300}
\end{figure}

\subsection{Non-Markovian dynamics with increasing $g$ }
\label{results:changeg}

While the two approximate solutions, NN and PA,  provide simple and convenient results, they show strong deviation from the correct behavior as the exciton-cavity coupling strength $g$  and the temperature $T$ are increased. For the linear polarization, increasing $g$ and $T$ are known to cause the breakdown of the PA and approaches based on the polaron master equation~\cite{morreau2019phonon,morreau2020phonon}. Here, we present results for the c-c FWM and linear response, focusing on the role of the coupling strength $g$ in the quantum dynamics. The temperature dependence is studied in \Sec{results:changeT} below.

As the timescales $\tau_{\rm JC}$ and $\tau_{\rm IB}$ become comparable, the $L$N shows a slower convergence with $L$, forcing us to use a finer time grid, with an increased memory kernel (which contains $6^{L+1}$ elements for the c-c FWM). In fact, with $g$ increasing, the time step must be reduced also to resolve a faster JC system dynamics. To understand the effect of $g$ on phonon induced dephasing, it is useful to first consider a simpler case of the linear polarization, where the kernel memory size is much smaller (contains $2^{L+1}$ elements). In this case we can store more time steps in memory and use a finer grid.
To make sure that the convergence was reached even in a regime of comparable timescales, $L=15$ was used in Ref.~\cite{morreau2019phonon}. In this work, we use $L=9$, which is also sufficient, with RMSD of 0.03-0.3\%, as demonstrated in \App{appendix:convergence}.  For the FWM response, we do not expect to see any significant change in convergence with $L$ (see \App{appendix:convergence}) and all relevant features of the physical behavior can be captured with $L=9$.

\begin{figure}[t]
\centering
\raisebox{4cm}{(a)}{\includegraphics[width=0.45\textwidth]{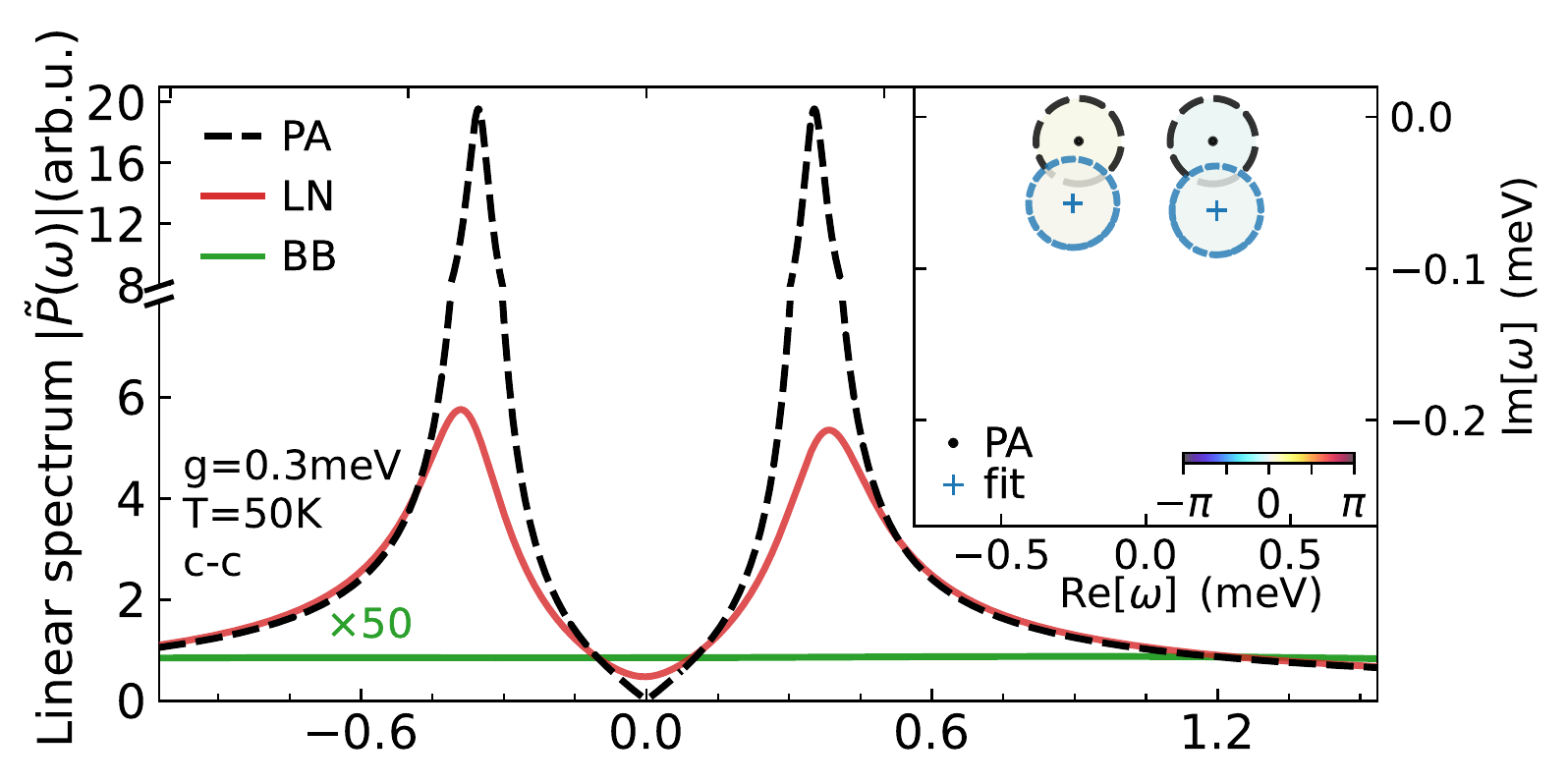}}
\raisebox{4cm}{(b)}{\includegraphics[width=0.45\textwidth]{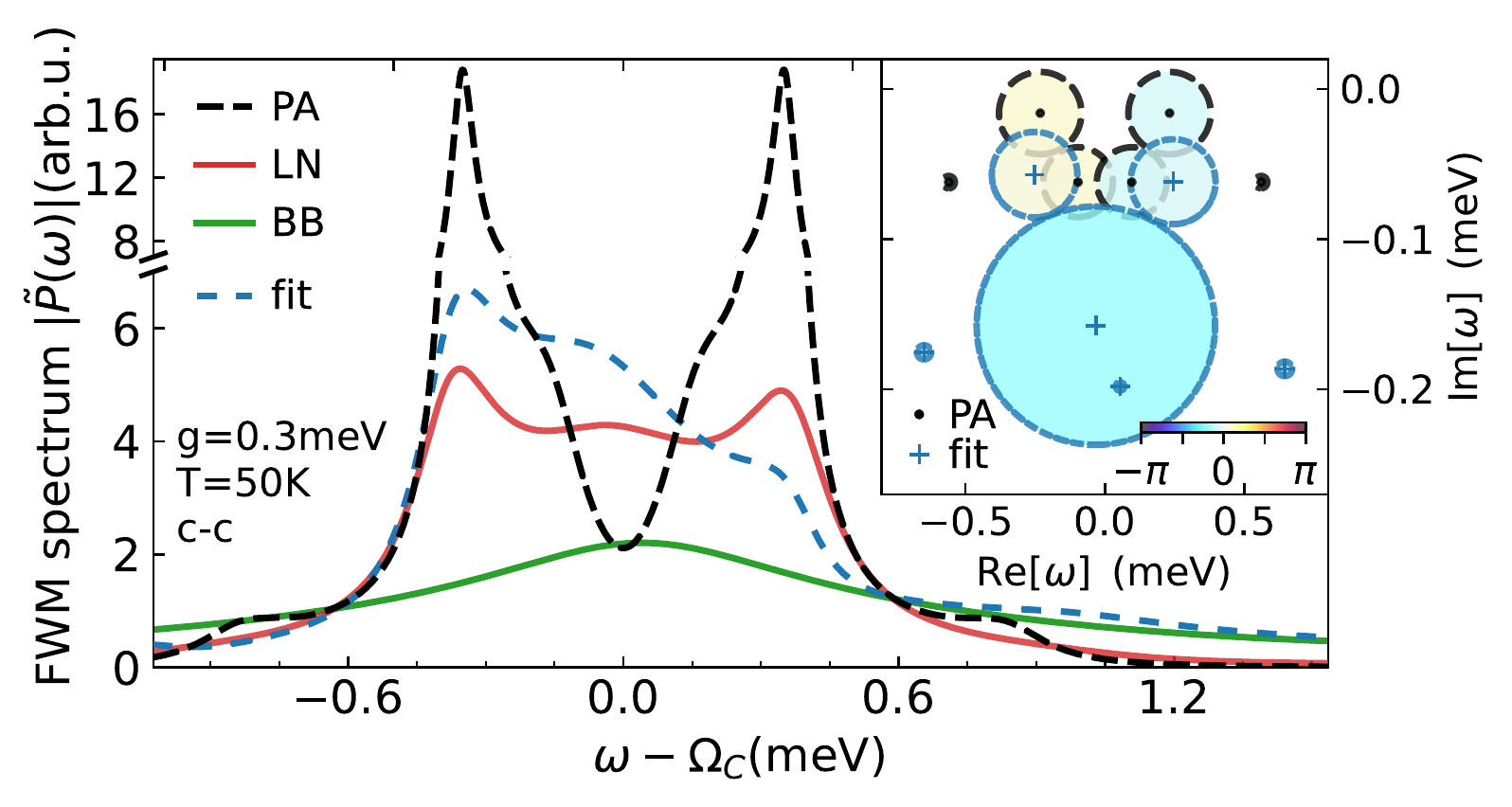}}
\caption{
As \Fig{spec-cc-50} but for $g=0.3$\,meV.
The amplitudes (circle areas) of the two transitions in the linear response in the inset of (a) are multiplied by a factor of 0.95 to match the amplitudes of the corresponding transitions in the FWM.
}
\label{spec-cc-300}
\end{figure}

\begin{figure}[t]
\centering
\raisebox{4cm}{(a)}{\includegraphics[width=0.45\textwidth]{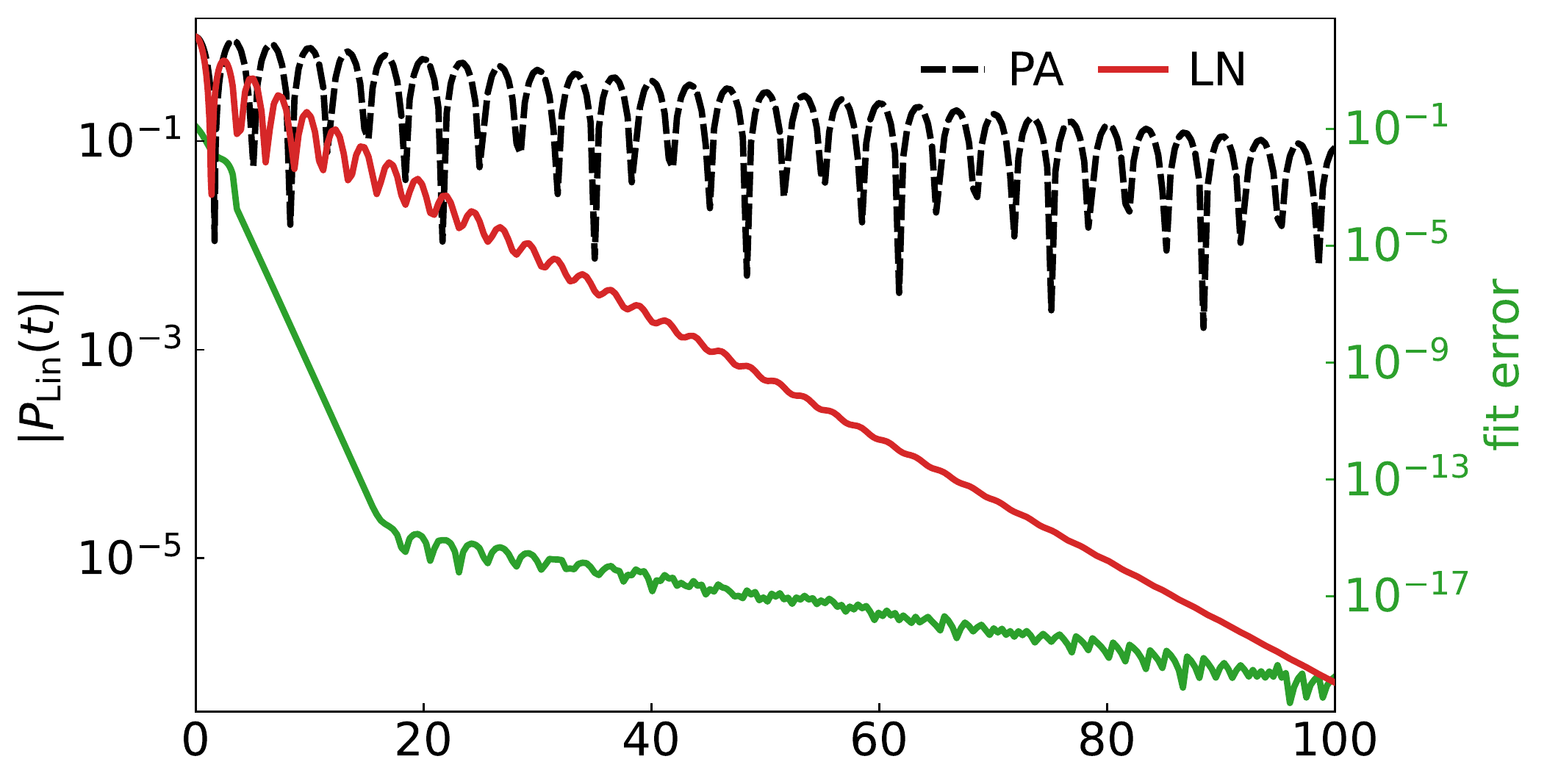}}
\raisebox{4cm}{(b)}{\includegraphics[width=0.45\textwidth]{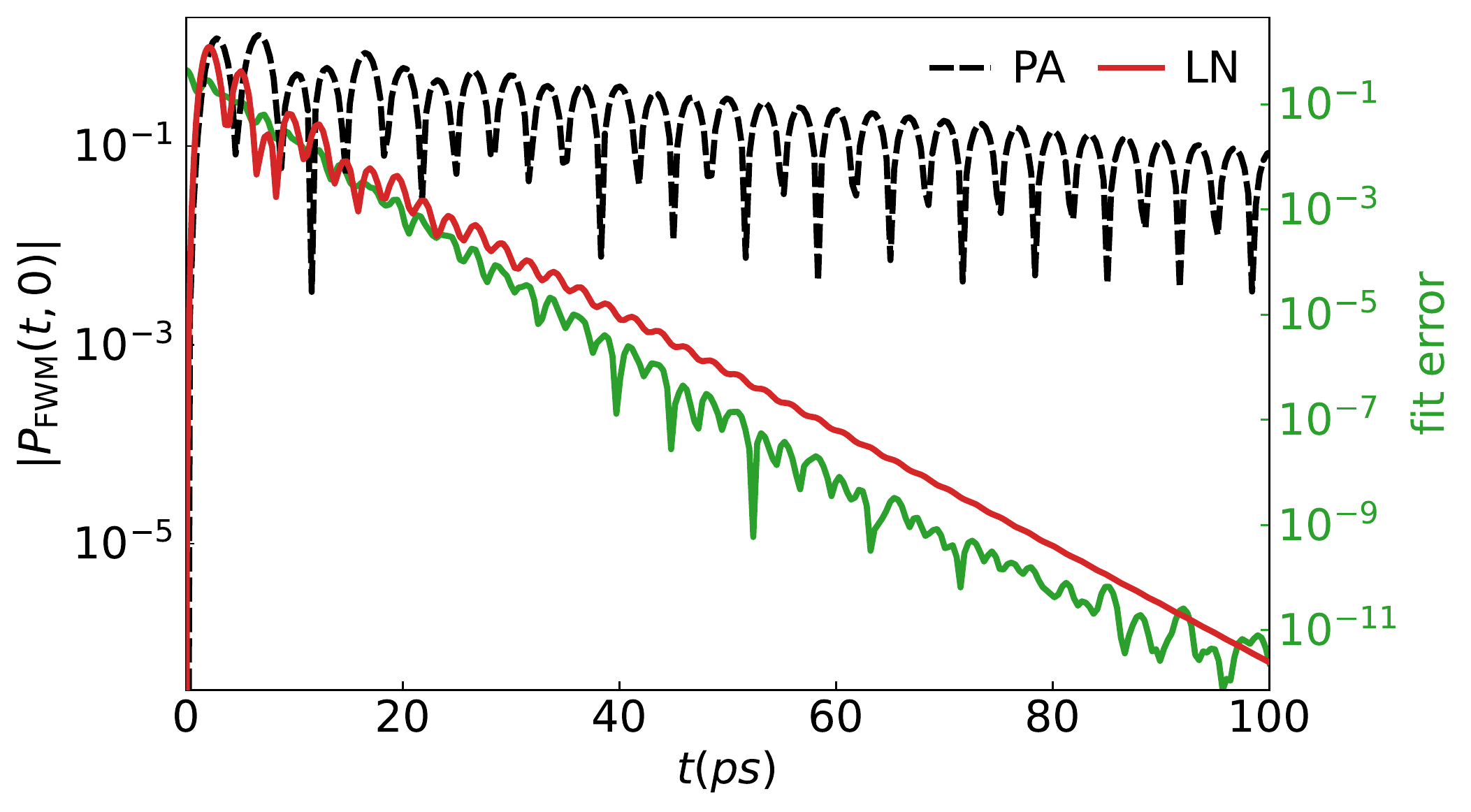}}
\caption{
As \Fig{res-cc-300} but for $g=0.8$\,meV ($\tilde{g}=0.62$ meV).
}
\label{res-cc-800}
\end{figure}

\begin{figure}[t]
\centering
\raisebox{4cm}{(a)}{\includegraphics[width=0.45\textwidth]{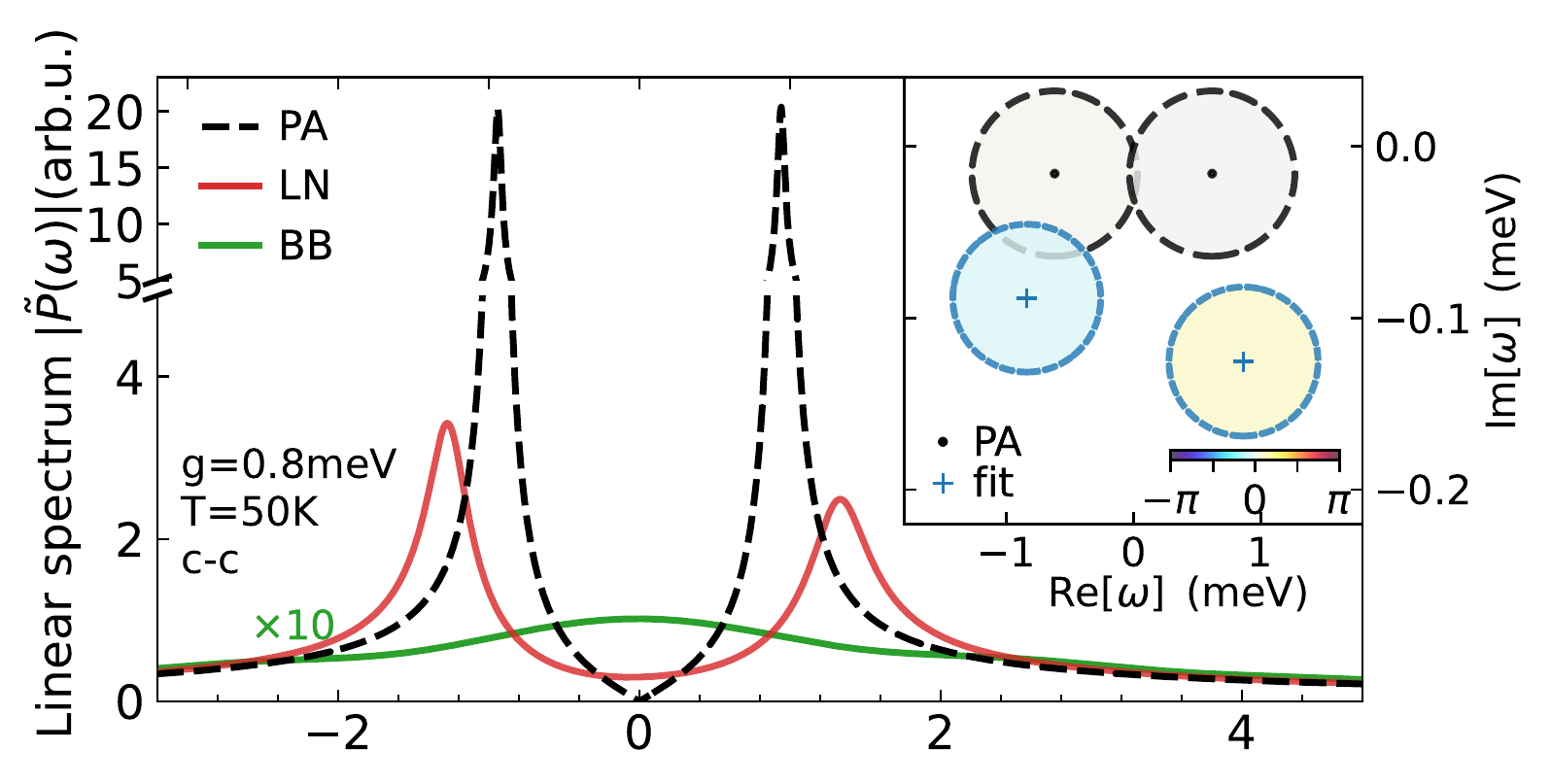}}
\raisebox{4cm}{(b)}{\includegraphics[width=0.45\textwidth]{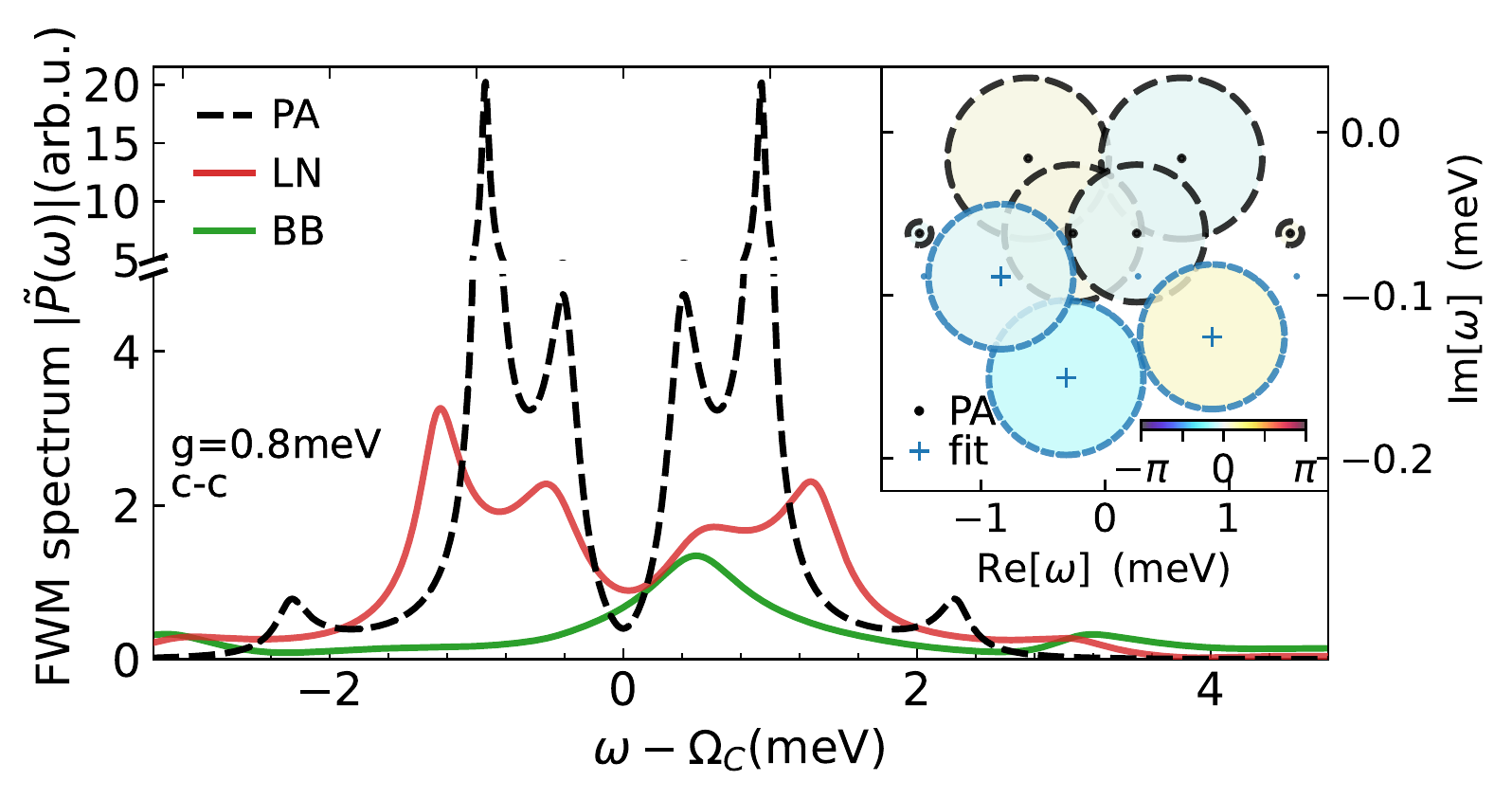}}
\caption{
As \Fig{spec-cc-300} but for $g=0.8$\,meV and the two amplitudes in the inset of (a) are scaled by a factor of 0.993.
}
\label{spec-cc-800}
\end{figure}

Results for the linear and FWM c-c polarizations are shown in Figs.~\ref{res-cc-300}-\ref{spec-cc-800} for
the exciton-cavity coupling strengths $g=0.3$\,meV and 0.8\,meV. The separation of the time dependence into the long-time and  short-time dynamics, using the multi-exponential fit \Eq{fit}, is attempted for these larger coupling strengths, and the results of such fits are also included in Figs.~\ref{res-cc-300}-\ref{spec-cc-800} and compared with the multi-exponential behaviour predicted by the PA, now demonstrating a clear deviation from the latter. In the $L$N, the actual Rabi splittings of both rungs are larger than those predicted by the PA, but for $g=0.3$\,meV, the PA offers a reasonably good approximation. For $g=0.8$\,meV instead, a better estimate of the Rabi splittings is given by the bare JC model, i.e. without the polaron renormalization, but both approximations, with or without the renormalization fail to capture the full dynamics. In particular, the scaling of the splitting with the square root of rung number $n$ is no longer observed in the $L$N: The average scaling of the Rabi splitting with $\sqrt{n}$ is reduced (for the second rung) by around 13\% for $g=0.3$\,meV and 5\% for $g=0.8$\,meV, as compared to only 0.4\% reduction for $g=0.05$\,meV. One can infer from this comparison a phonon-induced modification of the JC energy level structure.

As demonstrated by \Fig{res-cc-50} in \Sec{results:smallg}, the presence of first to second rung transitions in the c-c FWM dynamics increases dramatically the timescale of the initial non-Markovian behavior as compared to the linear or x-x FWM polarization, involving only ground state to first rung transitions. This effect is getting stronger as $g$ increases. Moreover, this non-Markovian timescale grows with $g$ both in the linear and FWM polarizations, as it is clear from \Figs{res-cc-300}{res-cc-800} (green lines).
Consequently, the corresponding non-Markovian contribution to the spectrum is getting narrower and higher, reducing the ZPL wights, see \Figs{spec-cc-300}{spec-cc-800}.

For $g=0.3$ meV,  the linear and the FWM response spectra in \Fig{spec-cc-300} are very different, which is not seen as clearly in the time dependence in \Fig{res-cc-300}. However, it is clear from \Fig{res-cc-300} that the initial deviation from a multi-exponential behavior is more prominent in the FWM dynamics, same as in \Fig{res-cc-50} but with longer non-Markovian times both for the linear and FWM polarizations.
We note that the overall spectrum in \Fig{spec-cc-300} (b) appears as a triplet. A tendency of nonlinearities of the JC model to manifest as triplet structures in the presence of pure dephasing, has been previously reported in Ref~\cite{Gonzalez-Tudela2010}, but in the photoluminiscence spectra of the system subject to continuous incoherent pumping.
The spectra in \Fig{spec-cc-800} for $g=0.8$\,meV show an increased asymmetry (which is discussed in \Sec{results:changeT}) and higher deviation from the PA, while the temporal dynamics demonstrates a further increase of the non-Markovian times for both linear and FWM polarizations.

To understand the enhancement with $g$ of the non-Markovian component in the linear and FWM c-c polarizations, let us recall that within the IB model, the exciton-phonon interaction results in a non-Markovian pure dephasing on the timescale $\tau_{\rm IB}$, which is the memory time of the phonon bath and the time of the polaron formation around the QD in the absence of the cavity or for a small coupling
strength $g$ between the cavity and the QD. This pure dephasing manifests itself as a quick non-exponential drop of the excitonic polarization, see e.g. the inset in \Fig{res-xx-50}. For small $g$, i.e. in the limit $\tau_{\rm JC} \gg \tau_{\rm IB}$, this quick non-Markovian dynamics of the polaron formation is spread over the much slower JC evolution, so that on any time interval $\Delta t \sim\tau_{\rm IB}$, only a very little change of lattice deformations around the QD occurs, and the whole process of the polaron formation or disappearance (due to the JC Rabi rotations) can be treat as adiabatically slow. In other words, the non-Markovian part of the overall dynamics of the whole system, which would manifest itself as a non-exponential behaviour, is negligible, apart from a very short period of time at the beginning during which the systems reaches a ``steady state'' in this adiabatic process. This can be clearly seen in the quality of the multi-exponential fit (green curves in \Fig{res-cc-50}). As $g$ increases and the timescales $\tau_{\rm JC}$ and $\tau_{\rm IB}$ become comparable, this period of time to reach the steady state is getting longer, as it is clear from \Figs{res-cc-300}{res-cc-800} (green curves), and the quasi-periodic process of the polaron formation and disappearance can no longer be treated as adiabatic. Finally, in the c-c FWM polarization for $g=0.8$\,meV, this steady state cannot be reached at all, see \Fig{res-cc-800}(b). In fact, even though the quality of the multi-exponential fit looks reasonable for sufficiently long times ($t>50$\,ps), the fit is not unique in this case and does not bring much value to the understanding of the spectrum, see the green line and the inset in \Fig{spec-cc-800}(b). In this regime, the quantum dynamics observed in the FWM and even in the linear polarization becomes essentially non-Markovian, with a much stronger effect in the FWM, due to a larger number of quantum transitions involved and owing to the quantum nonlinearity of the JC system.

\subsection{Temperature effects and spectral asymmetry}
\label{results:changeT}

\begin{figure}[t]
\centering
\raisebox{4.5cm}{(a)}{\includegraphics[width=0.95\columnwidth]{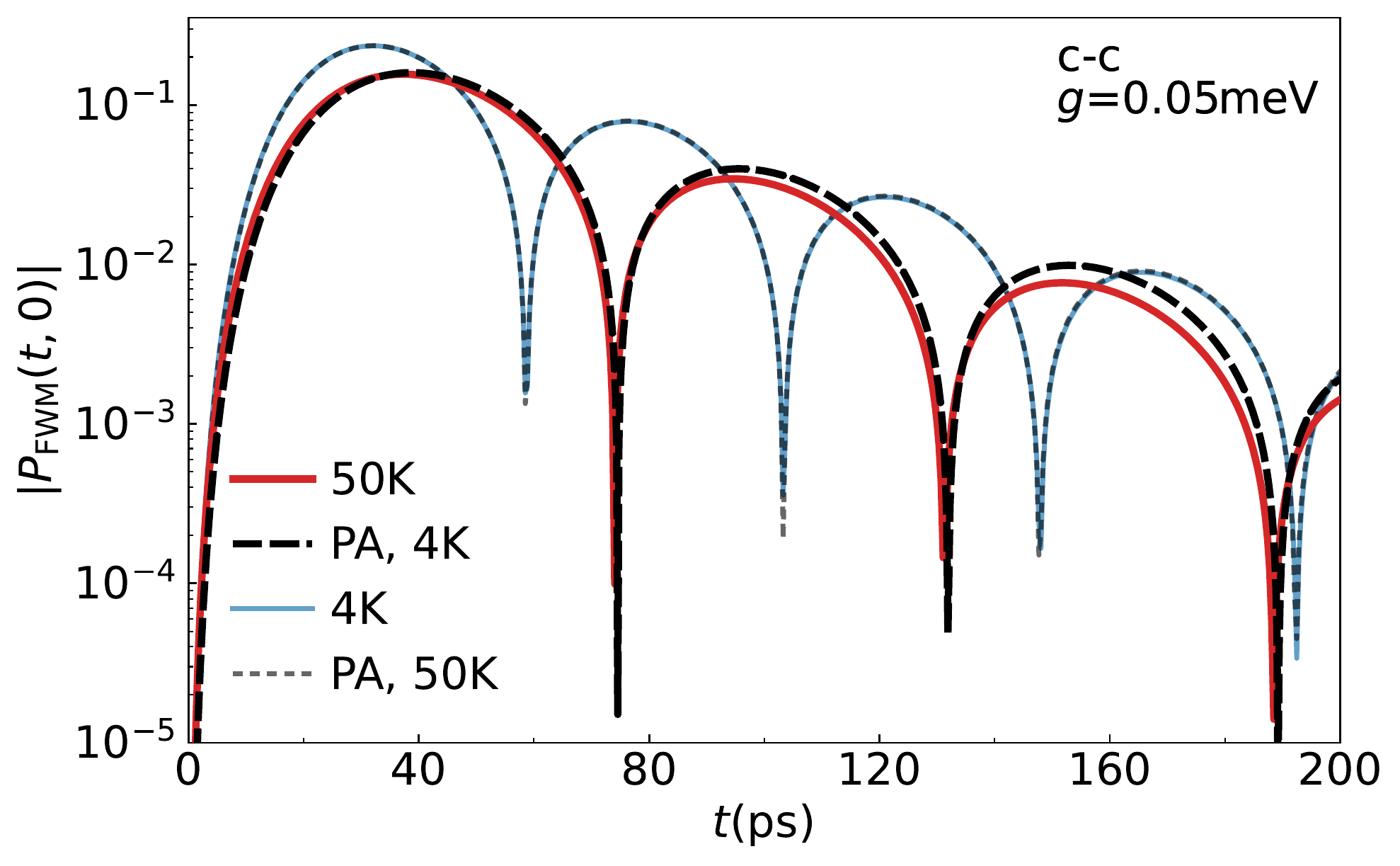}}
\raisebox{4.5cm}{(b)}{\includegraphics[width=0.95\columnwidth]{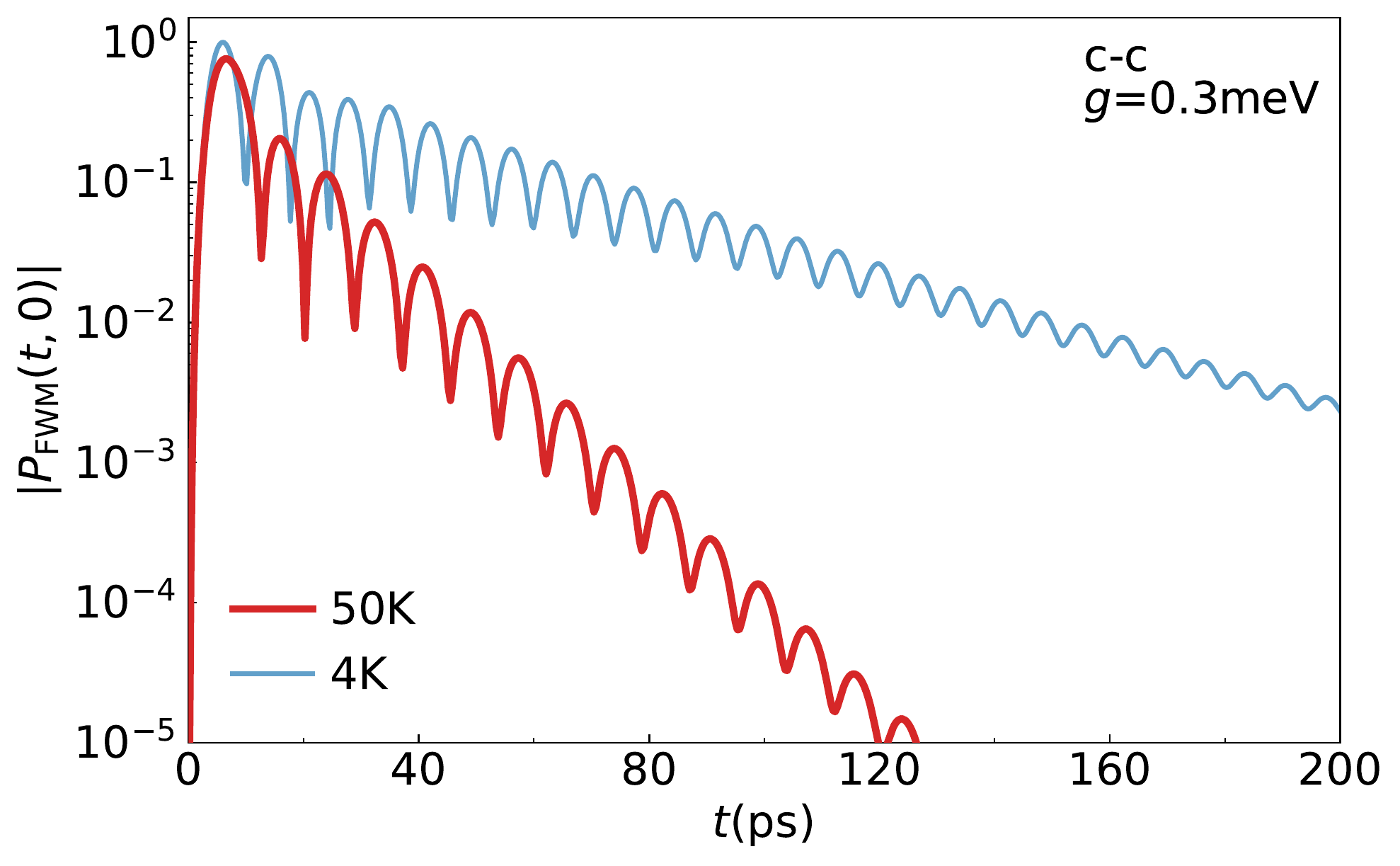}}
\raisebox{4.5cm}{(c)}{\includegraphics[width=0.95\columnwidth]{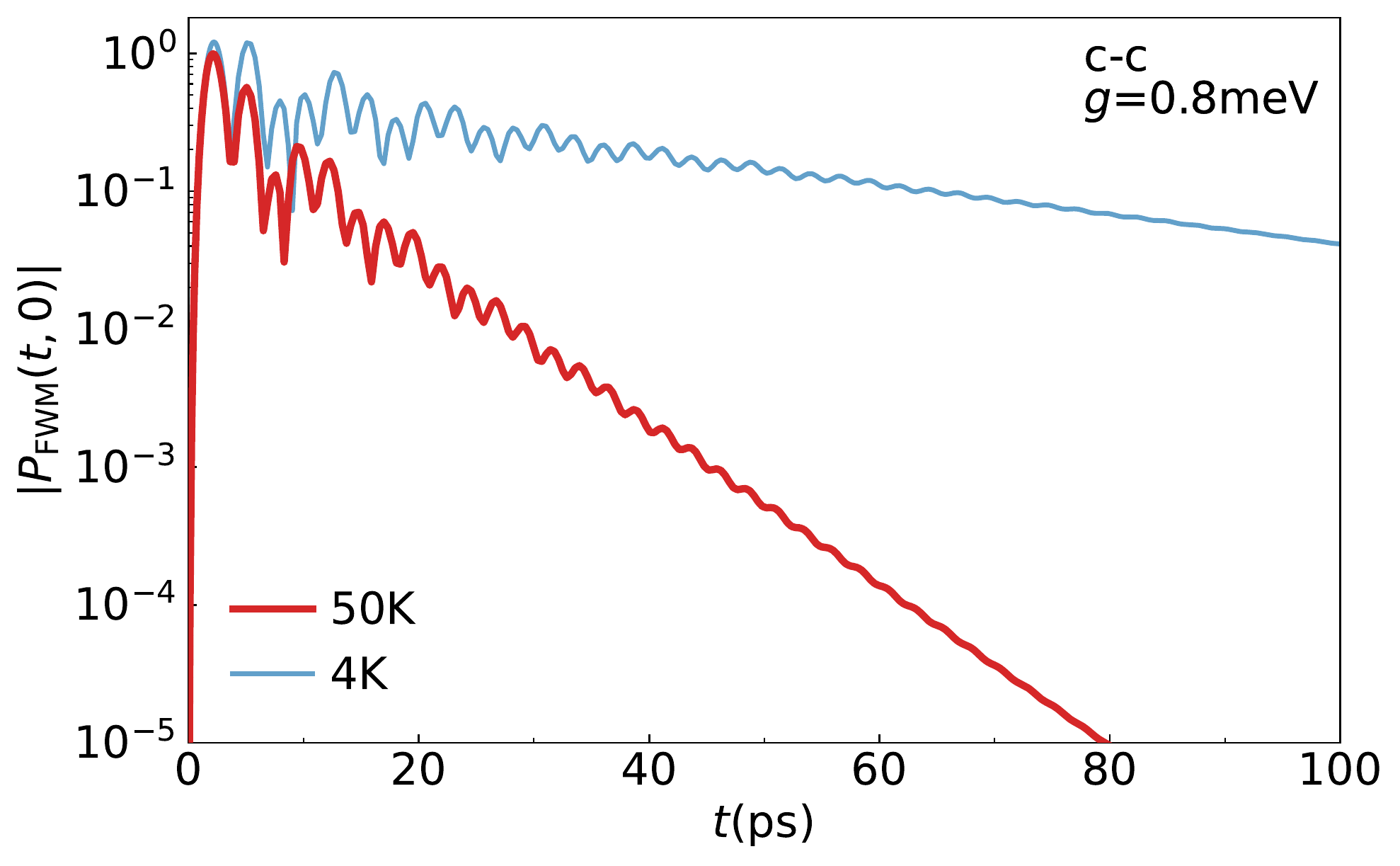}}
\caption{ Zero-delay c-c FWM polarization signals for (a) $g=0.05$\,meV, (b) $g=0.3$\,meV, and (c) $g=0.8$\,meV, calculated in the $L$N approach (solid lines), for $T=4$\,K (blue) and $T=50$\,K (red) (blue). The PA is also shown in (a) by black dashed lines.
}
\label{res-temp}
\end{figure}

\begin{figure}[t]
\centering
\raisebox{4.5cm}{(a)}{\includegraphics[width=0.95\columnwidth]{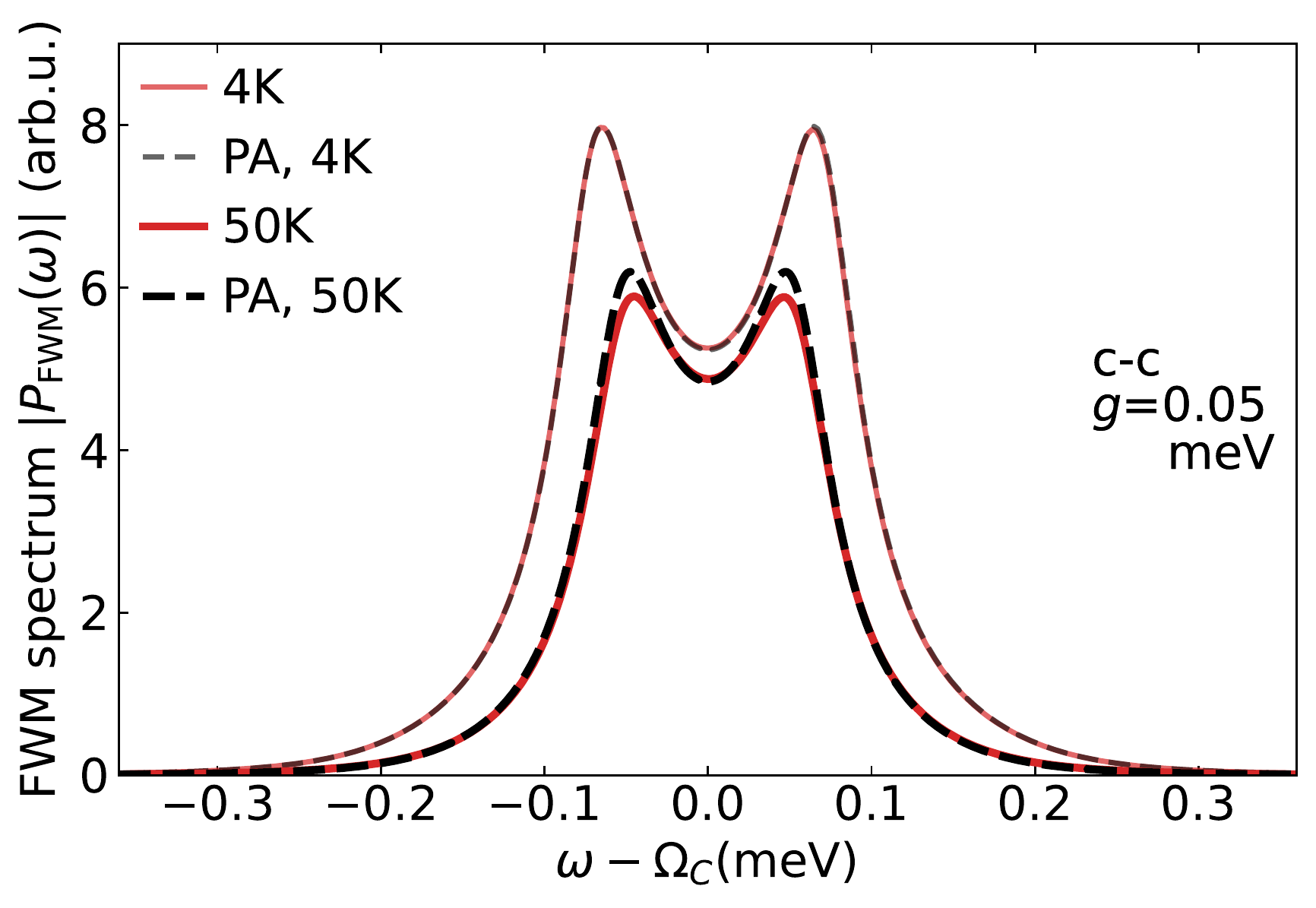}}
\raisebox{4.5cm}{(b)}{\includegraphics[width=0.95\columnwidth]{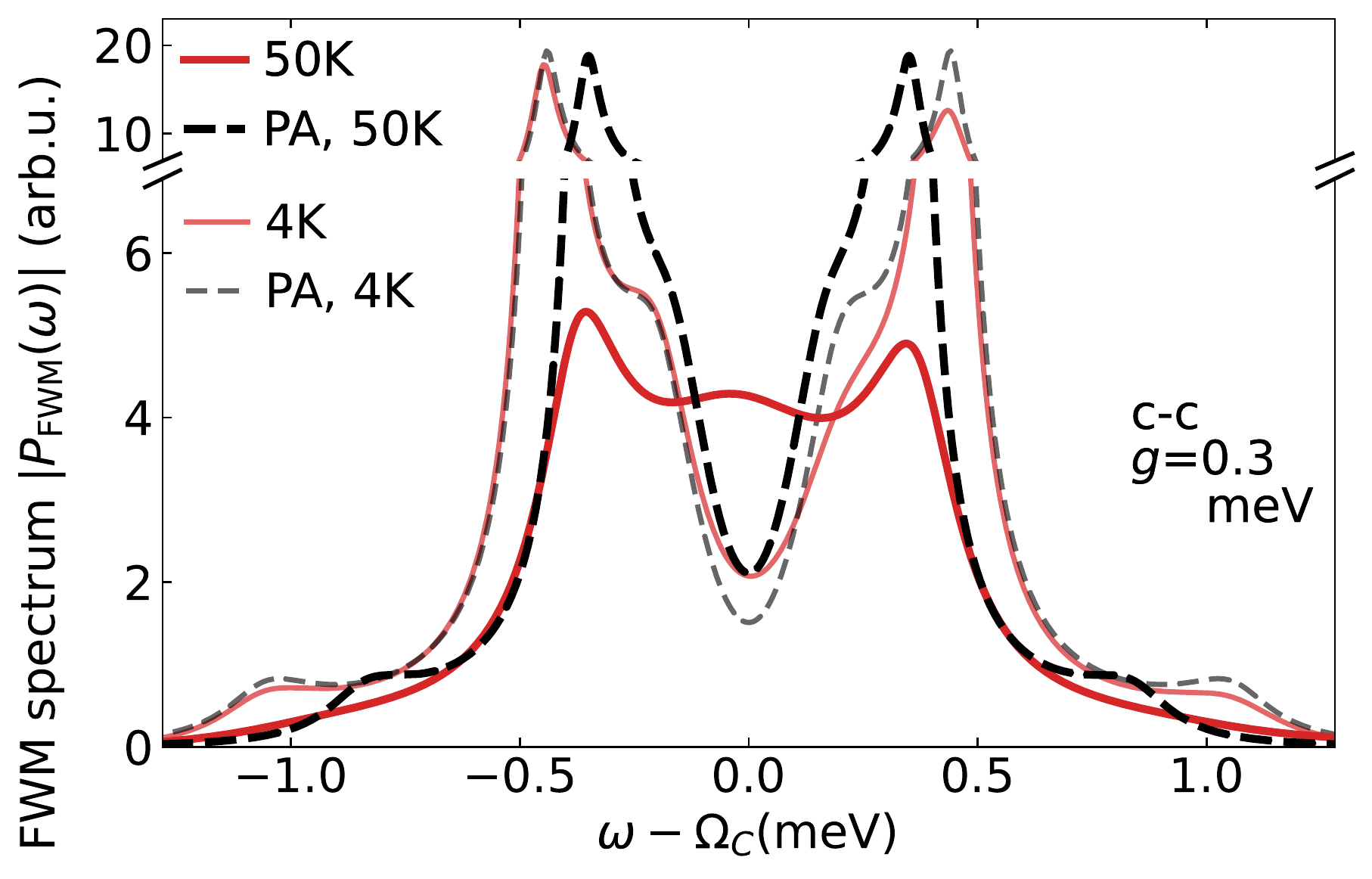}}
\raisebox{4.5cm}{(c)}{\includegraphics[width=0.95\columnwidth]{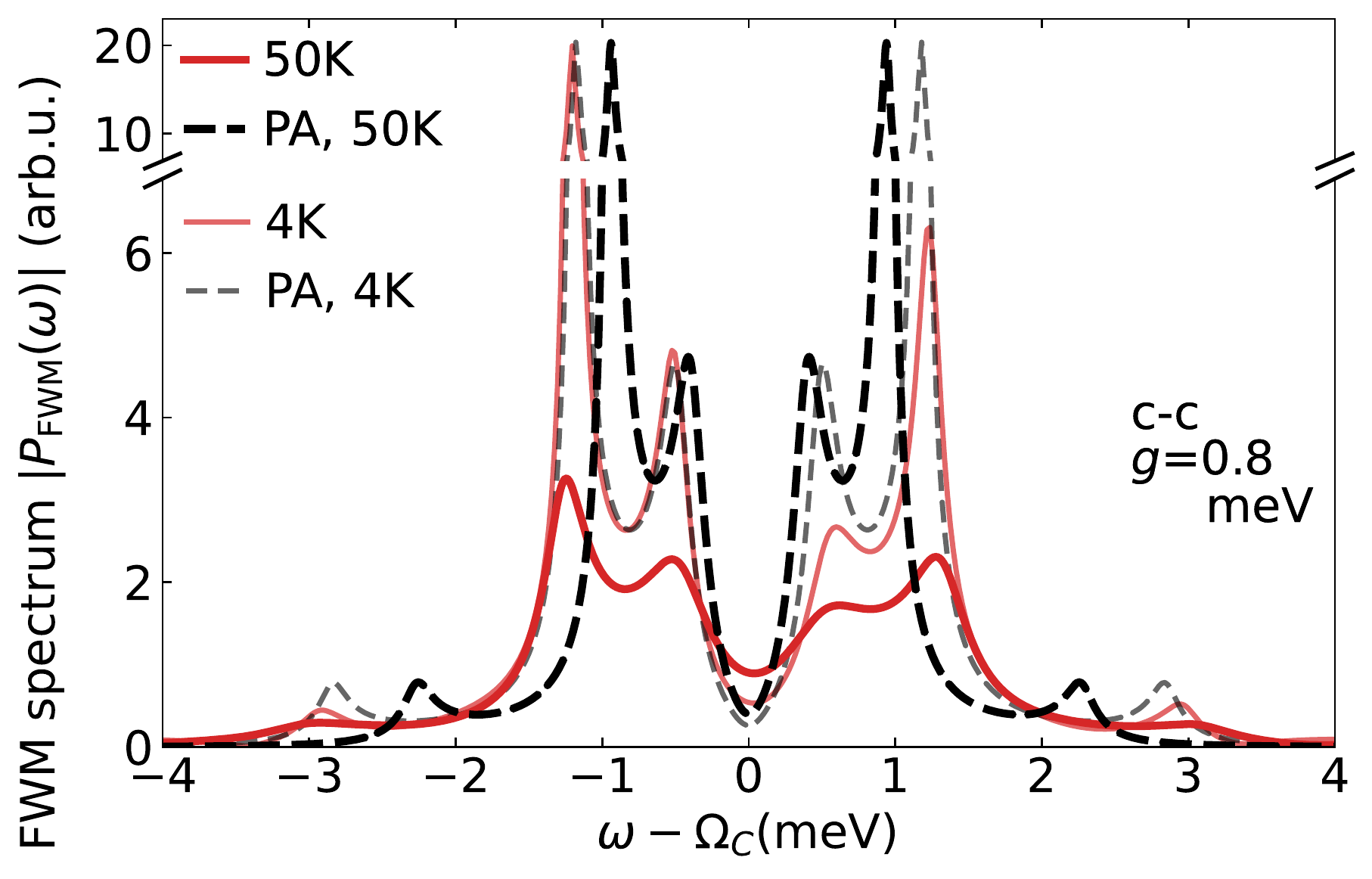}}
\caption{ Zero-delay c-c FWM spectrum for (a) $g=0.05$\,meV, (b) $g=0.3$\,meV, and (c) $g=0.8$\,meV, the Fourier transform of the time dependent polarization in \Fig{res-temp}, calculated in the $L$N approach (red solid lines) and PA (black dashed lines), for $T=4$\,K (thin) and $T=50$\,K (think lines).
}
\label{spec-temp}
\end{figure}

Here we concentrate on the temperature effects in the FWM polarization.
Figures \ref{res-temp} and \ref{spec-temp} show, respectively, the zero-delay FWM c-c polarization and its spectrum, for two different temperatures ($T=4$\,K and $T=50$\,K), and for three different coupling strengths ($g=0.05$\,meV, 0.3\,meV, and 0.8\,meV), in each case comparing $L$N results with the PA.  For small $g$, the $L$N shows good agreement with the PA, as previously discussed (see \Sec{results:smallg}), and the agreement is better at low temperatures. Note a significant change with temperature of the doublet splitting in the spectrum in \Fig{spec-temp}(a), corresponding to the change in the beating frequency in \Fig{res-temp}(a), in agreement with the PA, see \Eqs{Polaron}{Huang-Rhys}. In fact, the coupling constant $g$ is renormalized (reduced) stronger at larger temperatures, as dictated by the temperature dependence of the Huag-Rhys factor $S$.

For an intermediate exciton-cavity coupling strength of $g=0.3$\,meV [panels (b) in \Figs{res-temp}{spec-temp}], the impact of phonons on the FWM is much stronger. The exciton-cavity dynamics occurs faster (in accordance with a larger Rabi splitting), and phonons do not have enough time to adapt to the changes in the QD-cavity system. In this regime, the behavior is more complex and in general requires a non-Markovian treatment, which is provided by the $L$N solution. For both temperatures, the spectrum for
$g=0.3$\,meV is asymmetric due to phonon-assisted transitions between the polariton states~\cite{morreau2019phonon}. Surprisingly, for $T=4K$, the PA, despite its fundamental simplicity, still gives a reasonable agreement with the $L$N result but fails to predict the spectral asymmetry. At $T=50K$ the spectrum shows some drastic changes in shape as compared to the PA: The frequencies of all six transitions and their linewidths are significantly modified by the presence of the phonon environment.

At even larger exciton-cavity coupling strength of $g=0.8$\,meV [panels (c) in \Figs{res-temp}{spec-temp}] the deviation between the PA and $L$N approaches becomes dramatic.  The exciton-cavity dynamics now occurs on a faster timescale than the exciton-phonon dynamics. Note that comparable timescales $\tau_{\rm JC}\approx\tau_{\rm IB}$ are observed in the linear polarization at $g\approx0.6$\,meV (see Fig.\,3 in Ref.~\cite{morreau2019phonon}). The spectral asymmetry becomes more pronounced for both temperatures.
For $T=5$\,K, the PA still agrees with the $L$N results surprisingly well, apart from the asymmetry. This can be understood from the fact that at low temperatures, there is a very little polaron renormalization of the coupling $g$, hence the transition energies match quite well in both approaches. This is not true at higher temperatures for large $g$, see Fig.\,3 in Ref.~\cite{morreau2019phonon} demonstrating a strong deviation of the actual Rabi splitting in the first rung of the JC ladder from that predicted by the PA. This effect becomes even more pronounced in the second rung involved in the FWM dynamics.

\section{Conclusions}
\label{section:conclusions}

We have presented an asymptotically exact solution for the FWM response of a QD embedded in a photonic cavity, taking into account the LA phonon contribution in full. We have implemented a real-time path integral-based approach, combining Trotter's decomposition and linked-cluster expansion. Our work presents an important generalization of the method recently developed for calculating the linear polarization~\cite{morreau2019phonon}. This generalization, applied here to the FWM response, clearly demonstrates how the method can be used for other elements of the density matrix and applied in principle to any observable. This semi-analytic approach unites the behavior of two exactly solvable models into a common framework, where they can influence each other by means of a memory kernel responsible for temporal correlations within the system. Our solution has a microscopic origin and is capable of addressing intrinsically quantum dynamics beyond perturbative regimes. We were thus able to explore the effect of phonons on linear and nonlinear QD-cavity dynamics with comparable exciton-cavity and exciton-phonon timescales, as well as arbitrary temperatures.

The exactly solvable JC and IB models constituting the system are characterised by the timescales $\tau_{\rm JC}$ and $\tau_{\rm IB}$, respectively. In the limit of fast phonon environment, $\tau_{\rm JC}\gg\tau_{\rm IB}$, our microscopic approach allows us to develop and evaluate the validity of two useful approximations: the nearest-neighbor approach and the fully analytic polaron approximation following from it. For $\tau_{\rm JC}\gg\tau_{\rm IB}$, both approximation  describe the dynamics of the system well but are computationally much simpler, with the nearest-neighbor approach being able to capture even the non-Markovian dynamics at the initial times ($t\lesssim\tau_{\rm IB}$).  In the polaron approximation, the effect of phonons is reduced to the polaron shift of the exciton frequency and a renormalization of the exciton-cavity coupling strength by the Huang-Rhys factor, so that the quantum dynamics is described as a superposition of exponentials. Inspired by the polaron approximation, we have applied a complex-valued multi-exponential fit to the exact solution, that allowed us to precisely quantify the individual quantum transitions contributing to the FWM signal. Moreover, the multi-exponential fit separates the full quantum dynamics into a short-time non-Markovian and a long-time multi-exponential behavior. The latter represents optical spectra as superpositions of complex Lorentzian lines, corresponding to quantum transitions between phonon-dressed states of the JC ladder, in agreement with the polaron approximation. We have demonstrated this by focusing on the excitation of the optical nonlinearity via the cavity mode, in which case the second-rung states of JC ladder are directly excited and a more complex dynamics (compared to the linear polarization) arises. We have also studied other excitation and measurement channels corresponding to the third-order optical nonlinearity.

We have studied the regime of $\tau_{\rm JC}\gg\tau_{\rm IB}$ for the exciton-cavity coupling strength of $g=0.05$\,meV which is close to the one measured in Ref.~\cite{kasprzak2010up}. For this coupling strength, the calculated FWM spectra are qualitatively similar to those simulated in Ref.~\cite{kasprzak2010up} for $g=0.035$\,meV without phonons. In this regime, phonons around the QD can quickly adapt to any changes caused by the Rabi rotation. For larger values of the coupling strength, $g=0.3$\,meV (close to the one measured in Ref.~\cite{dory2016complete}) and $g=0.8$\,meV, the similarity of the timescales $\tau_{\rm JC}$ and $\tau_{\rm IB}$ opens up a possibility of phonon-assisted transitions between different polariton states, previously studied in the linear polarization~\cite{morreau2019phonon}. These transitions cause spectral asymmetry which we have also demonstrated in FWM spectra.

In the regime of comparable system and environment timescales,   $\tau_{\rm JC}\sim \tau_{\rm IB}$, we have found an increased deviation from the multi-exponential behavior, clearly demonstrating a non-Markovian character of the exciton-cavity FWM dynamics, not seen in the linear polarization. The phonon cloud around the QD is unable to adiabatically adapt to a varying optical state that results in non-Markovian effects seen in the quantum dynamics. We have found a significant deviation from the polaron approximation and multi-exponential fit, which becomes even more pronounced at elevated temperatures.

In a model of increased complexity, where linear and nonlinear effects, as well as environmental interactions are combined, there is a richer variety of interesting physical behavior that can be seen. While the dynamics of the full system exhibits characteristics of its individual parts, when they are combined the actual behavior is different that results in emerging phenomena, such as those demonstrate above. These phenomena can only be captured in a fully microscopic treatment, which is presented in this work.

\section*{Acknowledgments}

We thank Wolfgang Langbein and Amy Morreau for useful discussions.
L.S. acknowledges support from the EPSRC under grant no. EP/R513003/1.

\appendix

\section{$L$N approach to the finite-delay FWM polarization}
\label{appendix:delay}

Here we consider the general case of the FWM with an arbitrary delay time $\tau \geqslant 0$ between pulses.
The regions of delay time $\tau$ and the observation time $t$ are split into discrete numbers of time steps, $N_\mathrm{I}$ and $N_\mathrm{II}$, respectively. For simplicity and for brevity of notations, we assume equidistant grids, so $\tau=N_\mathrm{I}\Delta \tau$ and $t=N_\mathrm{II}\Delta t$, with generally different   time steps $\Delta \tau$ and $\Delta t$. Note, however, that the formalism presented below can be adapted to non-equidistant grids. To distinguish regions I and II in the time domain, we add to the operators ${Y}_{i_n}$ defined by \Eq{Y-vector}, respectively, indices (I) and (II), and take into account  the corresponding basis sizes of the DM, JC Liouvillian matrix, and the vectors $\vec{\alpha}$ and $\vec{\beta}$.

The time evolution of the DM after the application of the first pulse, described by \Eq{firstpulse}, has the form
\be
   \vec{\rho}(0_-) =  \sum_{j_{N_\mathrm{I}-1}...j_0}
   \tilde{T} {Y}^\mathrm{(I)}_{j_{N_\mathrm{I}}}{M}^\mathrm{(I)}_{j_{N_\mathrm{I}},j_{N_\mathrm{I}-1}}\dots {Y}^\mathrm{(I)}_{j_1} {M}^\mathrm{(I)}_{j_{1},j_{0}} \vec{\rho}(-\tau)\,.
\label{rho-delay-region}
\ee
After the application of the second pulse, which is described by \Eq{secondpulse},
the DM takes the form
\be
    \vec{\rho}(t)= \sum_{i_{N_\mathrm{II}-1}...i_0} \tilde{T} {Y}^{\mathrm{(II)}}_{i_{N_\mathrm{II}}} {M}^{\mathrm{(II)}}_{i_{N_\mathrm{II}} i_{N_\mathrm{II}-1}} \dots
    {Y}^{\mathrm{(II)}}_{i_1} {M}^{\mathrm{(II)}}_{i_{1} i_{0}} \vec{\rho}(0_+)\,.
   \label{rho-obs-tim-region}
\ee
Matrices  ${M}^{\mathrm{(I)}}$ and ${M}^{\mathrm{(II)}}$ are given by
\be
    {M}^{\mathrm{(I)}} = e^{-i \mathcal{L}_{\rm JC}^{\mathrm{(I)}} \Delta\tau}\,,\quad\quad
    {M}^{\mathrm{(II)}} = e^{-i \mathcal{L}_{\rm JC}^{\mathrm{(II)}} \Delta t}.
\label{M2-matrix}
\ee

Combining these equations and using the definition of the total polarization Eq.(\ref{Pfwm}),
where the vector $\vec{O}$ has the meaning of the annihilation operator of the observation channel,
the FWM polarization becomes:
\begin{widetext}
\begin{align}
P(t,\tau) =&\sum_{i_{N_\mathrm{II}}} O^{\mathrm{(II)}}_{i_{N_\mathrm{II}}} \sum_{i_{N_\mathrm{II}-1}\dots i_1} {M}^{\mathrm{(II)}}_{i_{N_\mathrm{II}} i_{N_\mathrm{II}-1}}\dots {M}^{\mathrm{(II)}}_{i_2 i_1} \sum_{i_0} {M}^{\mathrm{(II)}}_{i_1 i_{0}} \sum_{j_{N_\mathrm{I}}} Q^{\mathrm{(II)}}_{i_0 j_{N_\mathrm{I}}} \nonumber \\
     & \times \sum_{j_{N_\mathrm{I}-1}\dots j_1} {M}^{\mathrm{(I)}}_{j_{N_\mathrm{I}},j_{N_\mathrm{I}-1}}\dots {M}^{\mathrm{(I)}}_{j_2 j_1}
     \sum_{j_{0}} {M}^{\mathrm{(I)}}_{j_1,j_{0}}{Q_{j_{0}}^{\mathrm{(I)}}}
    \Big\langle \tilde{T}
   {Y}^{\mathrm{(II)}}_{i_{N_\mathrm{II}}} \dots {Y}^{\mathrm{(II)}}_{i_2} {Y}^{\mathrm{(II)}}_{i_1} {Y}^{\mathrm{(I)}}_{j_{N_\mathrm{I}}}\dots
    {Y}^{\mathrm{(I)}}_{j_2} {Y}^{\mathrm{(I)}}_{j_1}
     \Big\rangle\,.
     \label{P-before-LCE-delay}
\end{align}
\end{widetext}
The form of $\vec{Q}^{\mathrm{(I)}}$, $Q^{\mathrm{(II)}}$, and $\vec{O}^\mathrm{(II)}$ depends on the excitation and measurement channels, and is given, respectively, by \Eqsss{Q1-x-c}{Q2-x-c}{O2-x-c}.

Similar to the derivation in \Sec{theory:linkedcluster}, the application of the linked-cluster expansion to the expectation value of the $\tilde{T}$-ordered product in \Eq{P-before-LCE-delay} gives
\begin{align}
\Big\langle \tilde{T}
   {Y}^{\mathrm{(II)}}_{i_{N}} \dots  {Y}^{\mathrm{(II)}}_{i_{N_\mathrm{I}+1}} &{Y}^{\mathrm{(I)}}_{i_{N_\mathrm{I}}}\dots
    {Y}^{\mathrm{(I)}}_{i_1}
     \Big\rangle
\nonumber \\
     =&
\exp{\left(\sum_{m=1}^N \sum_{n=1}^N \tilde{\mathcal{K}}_{i_m i_n}(m,n)\right)},
\label{linked-cluster-expansion-delay}
\end{align}
where $N=N_\mathrm{I}+N_\mathrm{II}$, and a unified index $i_n$, having different dimensions in regions I and II, is introduced.
The cumulant elements in \Eq{linked-cluster-expansion-delay} are given by
\be
  \tilde{\mathcal{K}}_{i_m i_n}(m,n) =-\frac{1}{2} \int_{t_{m-1}}^{t_m} d\tau_1 \int_{t_{n-1}}^{t_n} d\tau_2  \left\langle \Tilde{T} \tilde{V}_{i_m}(\tau_1) \tilde{V}_{i_n}(\tau_2) \right\rangle
\label{barK}
\ee
where
\be
\tilde{V}_{i_n}(\tau')= \alpha_{i_n}^{(\zeta_n)} V^{(+)}(\tau')-\beta_{i_n}^{(\zeta_n)} V^{(-)}(\tau') \label{Vtilde-left-right-vec-delay}
\ee
with $V^{(\pm)}(\tau')$ introduced in \Eq{Vtilde-left-right-vec}. Here the operator $\tilde{V}_{i_n}(\tau')$, however, has a more general meaning than in \Eq{Vtilde-left-right-vec}, as it differentiates the two time regions by means of an extra index $\zeta_n$ defined as
\be
\zeta_n = \begin{cases}
\mathrm{I} \quad   &\phantom{N_\mathrm{I}+\,\,\,}1\leqslant n \leqslant N_\mathrm{I}\,, \\
\mathrm{II} \quad   &N_\mathrm{I}+1\leqslant n \leqslant N\,.
\end{cases}
\ee
The time steps  $t_n$ used in \Eq{barK} and below are defined in the following way:
\be
t_n = \begin{cases}
n\Delta \tau  -\tau  \quad   &\phantom{N_\mathrm{I}+\,\,\,}0\leqslant n \leqslant N_\mathrm{I}\,, \\
(n-N_\mathrm{I})\Delta t \quad   &N_\mathrm{I}+1\leqslant n \leqslant N\,.
\end{cases}
\ee

Now, using the definition of the time-ordering operator $\tilde{T}$ and the interaction $\tilde{V}_i(\tau')$, introduced in \Eqs{evolve-Tordering}{Vtilde-left-right-vec-delay}, respectively, \Eq{barK} can be written more explicitly as
\begin{align}
  \tilde{\mathcal{K}}_{i_m i_n}(m,n) =-\frac{1}{2} &\int_{t_{m-1}}^{t_m} d\tau_1 \int_{t_{n-1}}^{t_n} d\tau_2
\label{cumulant-Voriginal}
\\
 \times\Big\{   & \alpha^{(\zeta_m)}_{i_m}\alpha^{(\zeta_n)}_{i_n} \left\langle T V(\tau_1) V(\tau_2) \right\rangle \nonumber\\
 & - \alpha^{(\zeta_m)}_{i_m}\beta^{(\zeta_n)}_{i_n} \left\langle  V(\tau_2) V(\tau_1) \right\rangle \nonumber \\
    &  -\beta^{(\zeta_m)}_{i_m}\alpha^{(\zeta_n)}_{i_n} \left\langle  V(\tau_1) V(\tau_2) \right\rangle \nonumber \\
 & + \beta^{(\zeta_m)}_{i_m}\beta^{(\zeta_n)}_{i_n} \left\langle T_{\mathrm{inv}} V(\tau_1) V(\tau_2) \right\rangle \Big\}\,,
   \nonumber
\end{align}
from what immediately follows its symmetry, \Eq{Ksym}.
To evaluate it for $m\geqslant n$, we  use the phonon propagator (also known as autocorrelation function) with normal and inverse time ordering
\begin{align}
     \left\langle T V(\tau_1) V(\tau_2) \right\rangle= & {D} (\tau_1-\tau_2)\,,
\nonumber\\
     \left\langle T_{\mathrm{inv}} V(\tau_1) V(\tau_2) \right\rangle= & {D}^\ast (\tau_1-\tau_2)\,,
\end{align}
where $D(t)$ is given explicitly in \Eq{D} and in \App{appendix:Propagator}. Then we determine the form of each of the four terms in \Eq{cumulant-Voriginal}, by using the definition of $V$ given in \Eq{V}:
\begin{align}
    &\left\langle  V(\tau_1) V(\tau_2) \right\rangle =  \sum_{\q,\q'} \lambda_\q  \lambda_{\q'}   \Big\langle \Big(b_\q (\tau_1) +{b_{-\q}}^\dagger (\tau_1)\Big) \nonumber\\
    & \hspace{3.5cm} \times \Big(b_{\q'}(\tau_2) +{b_{-\q'}}^\dagger (\tau_2)\Big) \Big\rangle\ \nonumber \\
     &= \sum_\q | \lambda_\q |^2 \Big\{ \left\langle  b_\q (\tau_1) {b_\q}^\dagger (\tau_2) \right\rangle + \left\langle  {b_\q}^\dagger (\tau_1) b_\q (\tau_2) \right\rangle \Big\} \nonumber \\
    &=  \sum_\q | \lambda_\q |^2 \left[ (N_q+1) e^{-i \omega_q (\tau_1-\tau_2)} + N_q e^{i \omega_q (\tau_1-\tau_2)}  \right]  \nonumber \\
&=  \begin{cases}
    {D}(\tau_1-\tau_2)  & \tau_1>\tau_2\,, \\
    {D}^\ast(\tau_2-\tau_1) &\tau_1<\tau_2\,,  \\
    \end{cases}
    \label{PhGreensFn}
\end{align}
where $b_\q (t)= {b_{\q}}   e^{-i \omega_q t}$, $\lambda^\ast_{\q}=\lambda_{-\q}$, and the phonon occupation number $N_q$ is defined by \Eq{Bose}. Then \Eq{cumulant-Voriginal} simplifies to
\be
\tilde{\mathcal{K}}_{i_m i_n}(m,n) =    (\alpha^{(\zeta_m)}_{i_m}-\beta^{(\zeta_m)}_{i_m})( \alpha^{(\zeta_n)}_{i_n} {K}_{mn} - \beta^{(\zeta_n)}_{i_n} {K}_{mn}^\ast )
\label{cumulant-general-delay}
\ee
for $m\geqslant n$, using the definition of ${K}_{mn}$ given by \Eq{cumulant-element-defn}. The vectors $\vec{\alpha}^{(\zeta)}$ and $\vec{\beta}^{(\zeta)}$ in the above formula are given by  \Eqs{alphabeta1}{alphabeta2} for $\zeta = \mathrm{I}$ and II, respectively.

\begin{figure}[htbp]
\resizebox{9cm}{9cm}{%
 \begin{tikzpicture}
\pgfdeclarelayer{Llayer}
\pgfsetlayers{main,Llayer}
\definecolor{lightyellow}{rgb}{1.0, 0.86, 0.35}

\foreach \x in {0,...,10} {
      \draw[fill=lightgray,fill opacity=0.03,draw=none] (0,0) rectangle ++(4.5+0.05*\x,4.5+0.05*\x);
      \draw[fill=almond,fill opacity=0.08,draw=none] (0,0) rectangle ++(-4-0.05*\x,-4-0.05*\x);
      \draw[fill=lightyellow,fill opacity=0.01,draw=none] (0,0) rectangle ++(-4-0.05*\x,4.4+0.05*\x);
      \draw[fill=lightyellow,fill opacity=0.01,draw=none] (0,0) rectangle ++(4.5+0.05*\x,-4-0.05*\x);
}
\draw node at (-3.5,-4.3) {(I)};
\draw node at (0.7,4.8) {(II)};
\draw node at (-3.5,4.8) {(I-II)};
\draw node at (0.7,-4.3) {(I-II)};

\foreach \x [
evaluate=\x as \xb using int(4-\x),
evaluate=\x as \xp using int(3-\x),
evaluate=\x as \xg using int(2-\x),
evaluate=\x as \xy using int(1+\x),
evaluate=\x as \xr using int(2+\x),
evaluate=\x as \xn using int(3+\x)
] in {0,...,3} {
 \Ks{-\x-1}{-\x-1}{1}{1}{}{cyan};
\ifthenelse{\x<3}{
 \Ks{-\x-2}{-\x-1}{1}{1}{}{Purple};	
 \Ks{-\x-1}{-\x-2}{1}{1}{}{Purple};	
 \Ks{\x*1.5}{\x*1.5}{1.5}{1.5}{}{yellow}
}{}
\ifthenelse{\x<2}{
 \Ks{1.5*\x}{1.5*\x+1.5}{1.5}{1.5}{}{red};
 \Ks{1.5*\x+1.5}{1.5*\x}{1.5}{1.5}{}{red};
 \Ks{1*\x-2}{1.5*\x}{1}{1.5}{}{Lavender}
 \Ks{1.5*\x}{1*\x-2}{1.5}{1}{}{Lavender}
 \Ks{-1*\x-3}{-1*\x-1}{1}{1}{}{green!60!yellow}
 \Ks{-1*\x-1}{-1*\x-3}{1}{1}{}{green!60!yellow}
 \Lshape{black}{1}{\x-4}{3};
}{}
\ifthenelse{\x<1}{
 \Ks{\x-1}{1.5*\x}{1}{1.5}{}{Gray}
 \Ks{1.5*\x}{\x-1}{1.5}{1}{}{Gray}
 \Ks{1.5*\x}{1.5*\x+3}{1.5}{1.5}{}{Blue};
 \Ks{1.5*\x+3}{1.5*\x}{1.5}{1.5}{}{Blue};
\Lshape{black}{1.5}{\x}{3};
 \Lshape{black}{1}{\x-1}{4};
 \Lshape{black}{1.5}{\x+3}{1};
  \Lshape{black}{1.5}{\x+1.5}{2};
   \Lshape{black}{1}{\x-2}{3.5};
}{}
}

\draw[<->] (5.1,3) -- (5.1,4.5) node at (5.4,3.75) {$\Delta t$};
\draw[<-] (4.8,0) -- (4.8,2.1) node at (4.9,2.25) {$\tau_{\rm IB}$};
\draw[->] (4.8,2.4) -- (4.8,4.5);


  \draw[style=help lines,step=1.5] (0,0) grid (5,5);
   \draw[style=help lines,step=1] (-4.5,-4.5) grid (0,0);
 \draw[help lines,xstep=1,ystep=1.5] (-4.5,0)  grid (0,5);
  \draw[help lines,xstep=1.5,ystep=1] (0,-4.5)  grid (5,0);

\draw [black,thick] (0,0) circle [radius=0.1];
\draw[->] (-4.5,0) -- (5,0) node[right] {};
\draw[->] (0,-4.5) -- (0,5) node[above] {};
\draw node at (5.25,0.0) {$t_1$};
\draw node at (0.0,5.25) {$t_2$};
\draw node at (-4,-0.2) {$-\tau$};
\draw node at (-0.3,-4) {$-\tau$};
\draw node at (4.5,-0.2) {$t$};

\draw[-] (-3,-0.25) -- (-3.5,0.25) node[above] {(a)};
\draw[-] (-3.5,-1) -- (-3.5,0.25);
\draw[-] (-2,0.75) -- (-2.5,1.75) node[above] {(b)};
\draw[-] (-1,2.25) -- (-1.5,3.25) node[above] {(c)};
\draw[-] (-0,3.75) -- (-0.7,3.9) node[above] {(d)};

\foreach [count=\i] \x in {$i_4$, $i_3$, $i_2$, $i_1$} {\node (\i) at (-\i+0.5,-4.7) {\x};}
\foreach [count=\i] \x in {$i_5$, $i_6$, $i_7$} {\node (\i) at (1.5*\i-0.75,-4.7) {\x};}
\foreach [count=\i] \x in {$i_4$, $i_3$, $i_2$, $i_1$} {\node (\i) at (-4.7,-\i+0.5) {\x};}
\foreach [count=\i] \x in {$i_5$, $i_6$, $i_7$} {\node (\i) at (-4.7,1.5*\i-0.75) {\x};}
\end{tikzpicture}
}
\caption{
As \Fig{NNkr} (with $N_\mathrm{I}=4$ and  $N_\mathrm{II}=3$) but for $L=2$. The L-shaped features belonging to four different regions (a,b,c,d) discussed at the end of Appendix~\ref{appendix:delay} are indicated. The entire first quadrant belongs to the region (d).
}
\label{LNkrr2}
\end{figure}
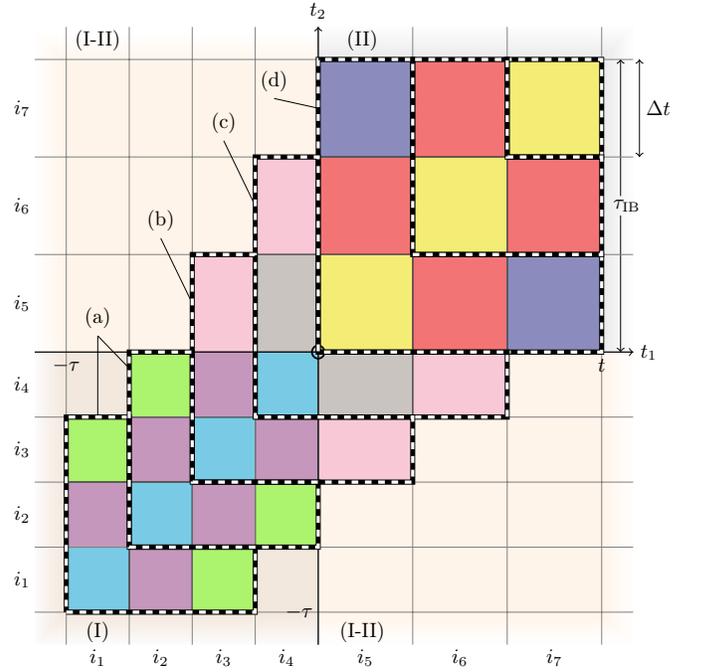

To derive more explicit expressions for $\tilde{\mathcal{K}}_{i_n i_m}(n,m)$, we start by distinguishing three cases for the cumulant elements $K_{mn}$:
\be
K_{mn} = \begin{cases}
R_l^{\mathrm{(I)}} \quad   &{\rm (i)\ \ if}\ \ m,n \leqslant N_\mathrm{I}\,, \\
R_l^{\mathrm{(II)}} \quad   &{\rm (ii)\ \ if}\ \ m,n > N_\mathrm{I}\,, \\
R_{ll'}^{\mathrm{(I-II)}} \quad   &{\rm (iii)\ \ otherwise}\,,
\end{cases}
\ee
where $l=|m-n|$ in the first two cases, while in the last case, $l=m-N_{\mathrm{I}}$ and $l'=N_{\mathrm{I}}+1-n$ for $m>n$ (for the opposite condition, $m<n$, one can simply use the symmetry $K_{nm}=K_{mn}$, which always holds). As in \Eq{Rl}, $R_l^{\mathrm{(I)}}$ and $R_l^{\mathrm{(II)}}$ can be found recursively:
\begin{align}
2R^{\mathrm{(I)}}_{l-1}&= K(l\Delta \tau)-lR^{\mathrm{(I)}}_0 -2 \sum_{k=1}^{l-2} (l-k) R^{\mathrm{(I)}}_k\,,
\label{RlI}
\\
2R^{\mathrm{(II)}}_{l-1}&= K(l\Delta t)-lR^{\mathrm{(II)}}_0 -2 \sum_{k=1}^{l-2} (l-k) R^{\mathrm{(II)}}_k\,,
\label{RlII}
\end{align}
starting from $l=2$ and using, respectively, $R^{\mathrm{(I)}}_0=K(\Delta \tau)$ and $R^{\mathrm{(II)}}_0=K(\Delta t)$, with $K(t)$ defined in \Eq{diag-cumulant-element-fn}. $R_{ll'}^{\mathrm{(I-II)}}$ can be found in a similar way:
\begin{align}
2R_{ll'}^{\mathrm{(I-II)}}=&K(l\Delta t + l'\Delta \tau) - lR^{\mathrm{(II)}}_0 -l'R^{\mathrm{(I)}}_0
\nonumber\\
&-2 \sum_{k=1}^{l-1} (l-k) R^{\mathrm{(II)}}_k -2 \sum_{k=1}^{l'-1} (l'-k) R^{\mathrm{(I)}}_k
\nonumber\\
&
-2 \sum_{k=1}^{l} \sum_{k'=1}^{l'} R^{\mathrm{(I-II)}}_{kk'} (1-\delta_{kl}\delta_{k'l'})
\,.
\label{Rli-II}
\end{align}
Then for (i) $n\leqslant m\leqslant N_{\mathrm{I}}$,
\begin{align}
\tilde{\mathcal{K}}_{i_m i_n}(m,n)=&
\mathcal{K}^{\mathrm{(I)}}_{i_{n+l}i_n}(l)
\label{KlI}\\
\equiv&(\alpha_{i_{n+l}}^{\mathrm{(I)}}-\beta_{i_{n+l}}^{\mathrm{(I)}})\left(\alpha_{i_n}^{\mathrm{(I)}} R_l^{\mathrm{(I)}}-\beta_{i_n}^{\mathrm{(I)}} {R_l^{\mathrm{(I)}}}^\ast\right),
\nonumber
\end{align}
and for (ii) $m\geqslant n>N_{\mathrm{I}}$,
\begin{align}
\tilde{\mathcal{K}}_{i_m i_n}(m,n)=&
\mathcal{K}^{\mathrm{(II)}}_{i_{n+l}i_n}(l)
\label{KlII}\\
\equiv&(\alpha_{i_{n+l}}^{\mathrm{(II)}}-\beta_{i_{n+l}}^{\mathrm{(II)}})\left(\alpha_{i_n}^{\mathrm{(II)}} R_l^{\mathrm{(II)}}-\beta_{i_n}^{\mathrm{(II)}} {R_l^{\mathrm{(II)}}}^\ast\right),
\nonumber
\end{align}
in both cases depending on the difference $l=m-n\geqslant0$.
Finally, for (iii) $m>N_\mathrm{I}\geqslant n$,
\be
\tilde{\mathcal{K}}_{i_m i_{n}}(m,n)=(\alpha_{i_{m}}^{\mathrm{(II)}}-\beta_{i_{m}}^{\mathrm{(II)}})\left(\alpha_{i_n}^{\mathrm{(I)}} R_{ll'}^{\mathrm{(I-II)}}-\beta_{i_n}^{\mathrm{(I)}} {R_{ll'}^{\mathrm{(I-II)}}}^\ast\right),
\label{KI-II}
\ee
where $l=m-N_{\mathrm{I}}$ and $l'=N_{\mathrm{I}}+1-n$. If $m<n$, one can just use the symmetry \Eq{Ksym} to evaluate $\tilde{\mathcal{K}}_{i_m i_{n}}(m,n)$ in all three cases (i)-(iii).  Note also that \Eqs{KlI}{KlII} have the same form as \Eq{cumulant-equidistant}, with the upper indices (I or II) present explicitly in \Eqs{KlI}{KlII} and implicitly in \Eq{cumulant-equidistant}. These upper indices indicate the relevant regions of the two-dimensional time grid shown in \Fig{NNkr} for the NN approximation ($L=1$) and in \Fig{LNkrr2} for $L=2$. The region (I-II) in \Figs{NNkr}{LNkrr2}, consisting of two parts lying in the second and forth quadrants of the plane, correspond to case (iii) with the cumulant elements given by \Eq{KI-II}.

Now, reducing the number of time steps in the memory segments to $L$, \Eqs{P-before-LCE-delay}{linked-cluster-expansion-delay} can be written as a sequence of tensor products,
\begin{align}
\mathcal{F}_{i_{L}...i_{1}}^{(1)}=&\sum_{j=1}^{h_{\mathrm{I}}} {M}^{\mathrm{(I)}}_{i_1 j} Q^{\mathrm{(I)}}_{j}\,,
 \label{F1}
 \\
\mathcal{F}^{(n+1)}_{i_{L}...i_{1}}=&\sum_{j=1}^h \tilde{\mathcal{G}}^{(n)}_{i_{L}...i_{1}j} \mathcal{F}^{(n)}_{i_{L-1}...i_{1}j} ,
\label{Fn}
\\
P(t,\tau)=& \sum_{j=1}^{h_{\mathrm{II}}} e^{\mathcal{K}^{\mathrm{(II)}}_{jj}(0)} O^{\mathrm{(II)}}_j \mathcal{F}^{(N)}_{p\dots p j}\,.
\label{Pttau}
\end{align}
where $p=2$ or 6,  $h_\mathrm{I}=2$, $h_\mathrm{II}=6$, and $h=h_\mathrm{I}$ ($h=h_\mathrm{II}$) for $n<N_\mathrm{I}$ ($n\geqslant N_\mathrm{I}$).
Unlike zero-delay FWM, the tensor $\tilde{\mathcal{G}}^{(n)}$ for non-zero delay is generally $n$-dependent and has different definitions in four different regions. In two regions, (a) $n\leqslant N_\mathrm{I}-L$ and (d) $n> N_\mathrm{I}$, it is independent of $n$ and has essentially the same form as in \Eq{G-tensor}:
\be
\tilde{\mathcal{G}}^{(n)}_{i_{L}...i_{1}j}=\mathcal{G}^{\mathrm{(\zeta)}}_{i_{L}...i_{1}j}={M}^{\mathrm{(\zeta)}}_{i_{1} j}
e^{\mathcal{K}^{\mathrm{(\zeta)}}_{jj}(0) +2\mathcal{K}^{\mathrm{(\zeta)}}_{i_1 j}(1)+\dots +2\mathcal{K}^{\mathrm{(\zeta)}}_{i_L j}(L)}\,,
\label{G-zeta}
\ee
where $\zeta= \mathrm{I}$ for region (a)  and $\zeta= \mathrm{II}$ for region (d), while  $\mathcal{K}^{\mathrm{(\zeta)}}_{i_{n+l}i_n}(l)$ are given by \Eqs{KlI}{KlII}, respectively.
In the other two regions, it is $n$-dependent. In region (b), where $N_\mathrm{I}-L<n< N_\mathrm{I}$,
\be
\tilde{\mathcal{G}}^{(n)}_{i_{L}...i_{1}j}={M}^{\mathrm{(I)}}_{i_{1} j}
e^{\tilde{\mathcal{K}}_{jj}(n,n) +2\tilde{\mathcal{K}}_{i_1 j}(n+1,n)+\dots +2\tilde{\mathcal{K}}_{i_L j}(n+L,n)}\,,
\label{Gn3}
\ee
and in region (c), consisting of a single element $n=N_\mathrm{I}$,
\begin{align}
\tilde{\mathcal{G}}^{(n)}_{i_{L}...i_{1}j}=&[{M}^{\mathrm{(II)}}Q^{\mathrm{(II)}}]_{i_{1} j}
\label{Gn4}\\
&\times
e^{\tilde{\mathcal{K}}_{jj}(n,n) +2\tilde{\mathcal{K}}_{i_1 j}(n+1,n)+\dots +2\tilde{\mathcal{K}}_{i_L j}(n+L,n)}\,,
\nonumber
\end{align}
where $\tilde{\mathcal{K}}_{i j}(m,n)$  are given by \Eq{KlI}, (\ref{KlII}) or (\ref{KI-II}), depending on the numbers $m$ and $n$ used.

In the NN approximation, corresponding to $L=1$, $\mathcal{F}^{(n)}$ and $\tilde{\mathcal{G}}^{(n)}$ become, respectively, vectors and matrices, region (b) disappears, and in region (c) $\tilde{\mathcal{G}}^{(n)}$ becomes  $\mathcal{G}^{(\mathrm{I-II})}$ defined in \Eq{G1-2-mat}, so \Eqsss{F1}{Fn}{Pttau} transform into \Eq{P-NN-delay}. In fact, within \Eq{Gn4}, $\tilde{\mathcal{K}}_{jj}(n,n)$ and $\tilde{\mathcal{K}}_{i_1 j}(n+1,n)$ are the only remaining cumulants which become, respectively, $\mathcal{K}^\mathrm{(I)}_{jj}(0)$  and $\mathcal{K}^{\mathrm{(I-II)}}_{i_1j}(1)$, the latter being defined in \Eq{cumulant-1-2-mat}, in agreement with \Eq{KI-II}.

\section{QD phonon propagator}
\label{appendix:Propagator}

A quantum state of the phonon subsystem is described by a set of excitations in different modes, which are distinguished by their wave numbers $\q$. The coupling of exciton to the acoustic phonon mode $\q$ is determined by the coupling matrix element
\begin{align}
\lambda_\q &=\frac{q \mathcal{D}(\q)}{\sqrt{2 \rho_M v_s q \mathbb{V}}}\,,
\label{ex-ph-coupling}
\end{align}
where $\rho_M$ is the mass density, $v_s$ is the sound velocity, and $\mathbb{V}$ is the sample volume. The form-factor $\mathcal{D}(\q)$ is given by
\begin{align}
\mathcal{D}(\q) &=\int d\mathbf{r} \left\{ D_c|\psi_e(\mathbf{r})|^2-D_v|\psi_h(\mathbf{r})|^2\right\} e^{i \mathbf{q} \cdot \mathbf{r}}\,,
\label{form-factor}
\end{align}
where  $\psi_e(\mathbf{r})$ and $\psi_h(\mathbf{r})$  are, respectively, the electron and hole confined wave functions,  and $D_c$ and $D_v$ are, respectively, the conduction- and valence-band deformation potentials.
For a symmetric QD, when the wave function has mirror symmetry, $\lambda_\q$ is real and $\lambda_{-\q}=\lambda_{\q}$.

In the phonon propagator $D(t)$, given by \Eq{D}, the phonon spectral density $J(\omega)=\sum_\q |\lambda_\q|^2 \delta(\omega-\omega_q)$ can be used to convert a discrete sum into an integral:
 \begin{align}
    D(t)=\int_0^\infty J(\omega) & \Big[  (N(\omega)+1) e^{-i \omega |t|}
+ N(\omega) e^{i \omega |t|}   \Big] d\omega,
    \label{PhGreensFn-int}
\end{align}
where  $N(\omega)=\{\exp[\omega/(k_b T)]-1\}^{-1}$ is the Bose function.
For a QD with spherically symmetric parabolic confinement potential, the detailed expression for $J(\omega)$ is given in Appendix~F of Ref.~\cite{morreau2019phonon}.

\section{Characteristic timescales}
\label{appendix:timescales}

\begin{figure}[htbp]
\centering
\includegraphics[width=0.4\textwidth]{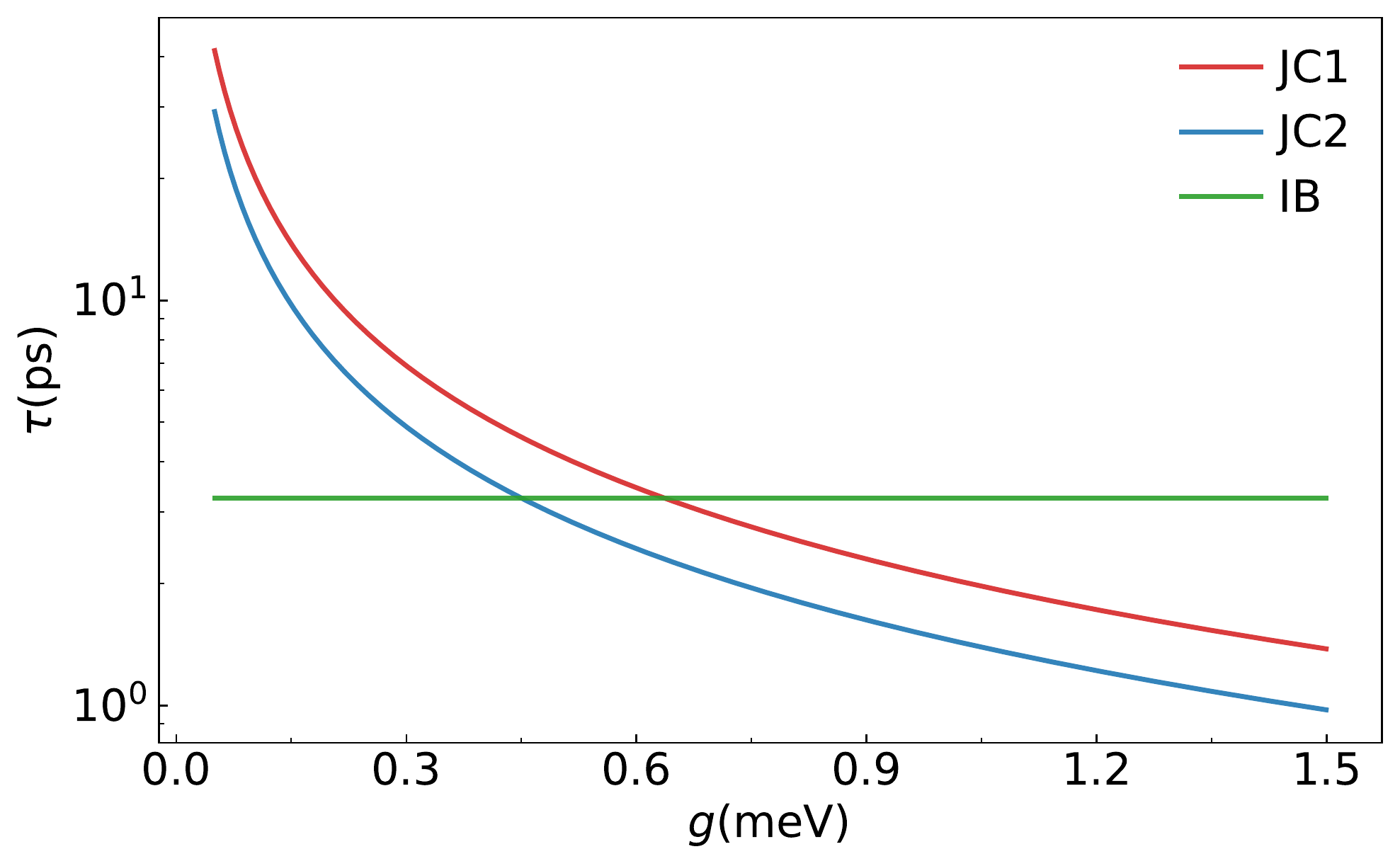}
\caption{Characteristic timescales for the exciton-phonon dynamics (green) and the exciton-cavity dynamics of the first (red) and the second (blue) rungs as functions of the coupling strength $g$.
}
\label{timescales}
\end{figure}

The characteristic timescales determining the full quantum dynamics of the system are introduced, following Ref.\,\cite{morreau2019phonon}:

\begin{align}
\tau_{\rm IB} &=  \sqrt{2} \pi l/ v_s\,,\\
\tau_{\rm JC1} &= \pi/g\,,\\
\tau_{\rm JC2} &=  \tau_{\rm JC1}/\sqrt{2}\,,
\end{align}
where $l$ denotes the exciton confinement radius. The timescale of the IB model depends on temperature (saturating at large $T$) and the parameters of the QD.
The above expression for $\tau_{\rm IB}$ is an estimate of the temporal decay of the phonon autocorrelation function \Eq{PhGreensFn-int}. Note that in our numerical calculations, we use slightly different values of $\tau_{\rm IB}$ which are determined by the condition $D(\tau_{\rm IB})-D(t\rightarrow \infty)=D(\tau_{\rm IB})+i\Omega_p t+S \approx 10^{-4}$. As for the above given timescales $\tau_{\rm JC1}$ and $\tau_{\rm JC2}$, they express the exact values of the periods of the Rabi rotations in the first and second rungs of JC ladder at zero detuning and in the absence of phonons, taking into account the $\sqrt{n}$ increase of the Rabi splitting with the rung number $n$.

Figure~\ref{timescales} shows all three timescales as functions of $g$.
For the linear polarization, the maximum phonon-induced dephasing is observed in the regime of comparable JC1 and IB timescales, which is at $g\approx 0.6$ meV \cite{morreau2019phonon}. We have also observed similar behavior of the first rung transitions in the FWM.

\section{FWM results for other excitation and measurement channels}
\label{appendix:channels}

\begin{figure}[htpb]
\centering
\raisebox{3.5cm}{(a)}{\includegraphics[width=0.45\textwidth]{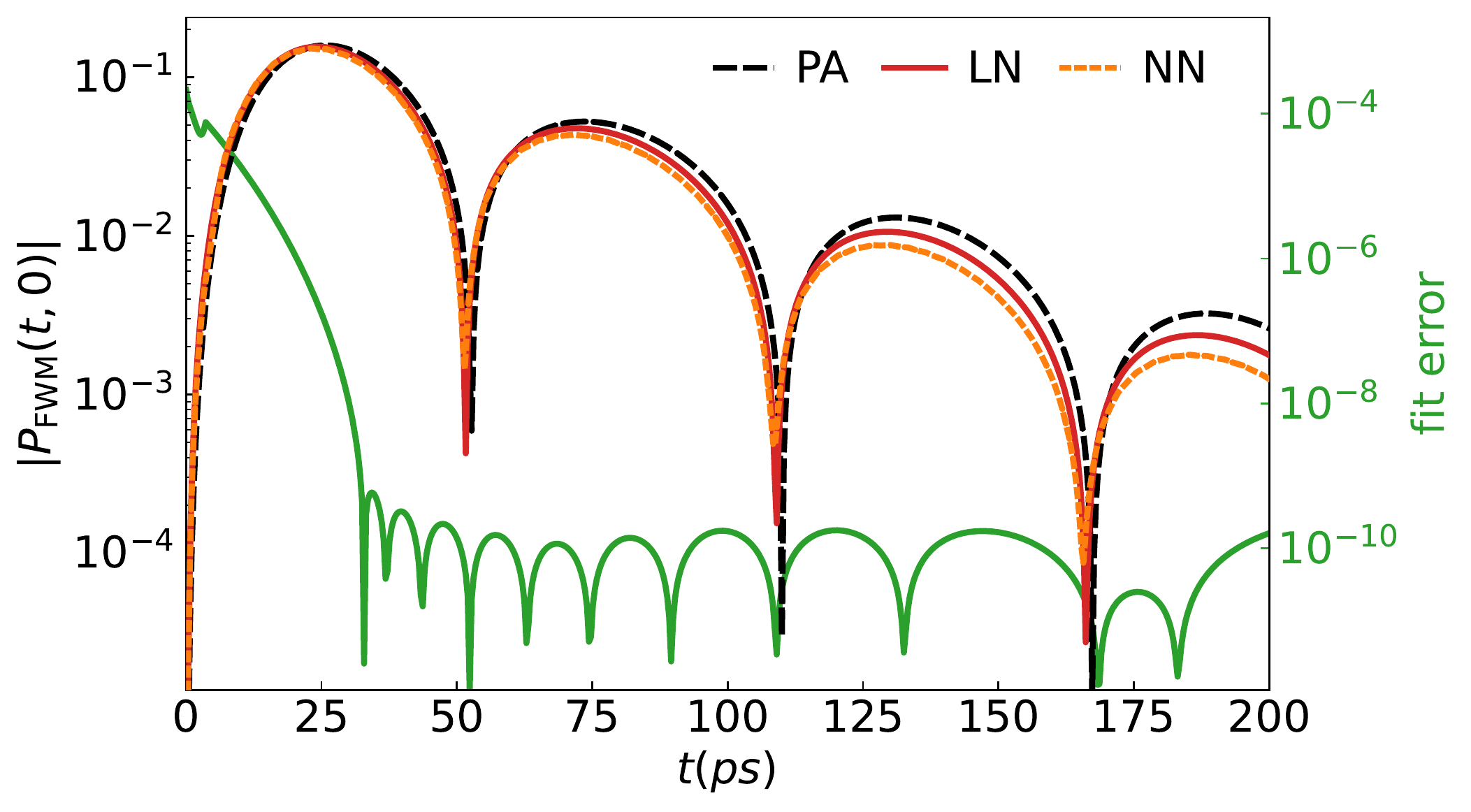}}
\raisebox{3.5cm}{(b)}{\includegraphics[width=0.45\textwidth]{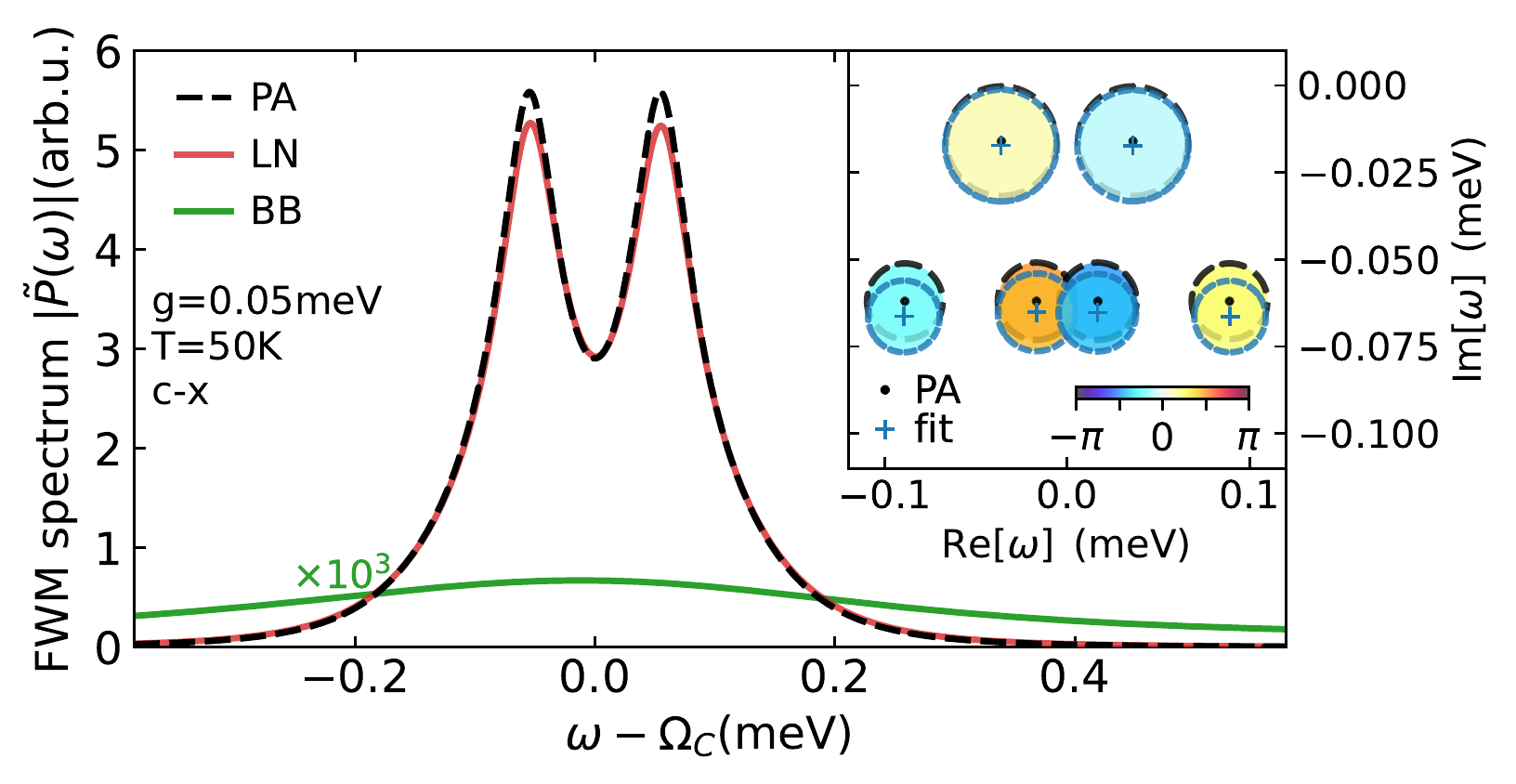}}
\caption{
(a) c-x FWM polarization at $\tau=0$, $g=0.05$\,meV, and $T=50$\,K, calculated in the $L$N ($L=9$, red solid), NN (red dotted), and PA (black dashed line). The $L$N result is fitted with six complex exponentials [see \Eq{fit}], to separate the long-time from the short-time behavior, here shown as a relative error of the fit (green curve).
(b) Fourier transform of c-x FWM polarization, calculated in the $L$N and PA approaches and shown in (a). The full numeric $L$N result (red), is separated into ZPLs (the inset) and a phonon broad band (green line), corresponding, respectively, to the long-time and short-time parts of the signal. The inset shows the complex frequencies of the multi-exponential fit (blue crosses) and  PA (black dots) along with the corresponding amplitudes $|A_i|$ given by the circle area and their phases color coded.
}
\label{rs-cx-50}
\end{figure}

\begin{figure}[htpb]
\centering
\raisebox{3.5cm}{(a)}{\includegraphics[width=0.45\textwidth]{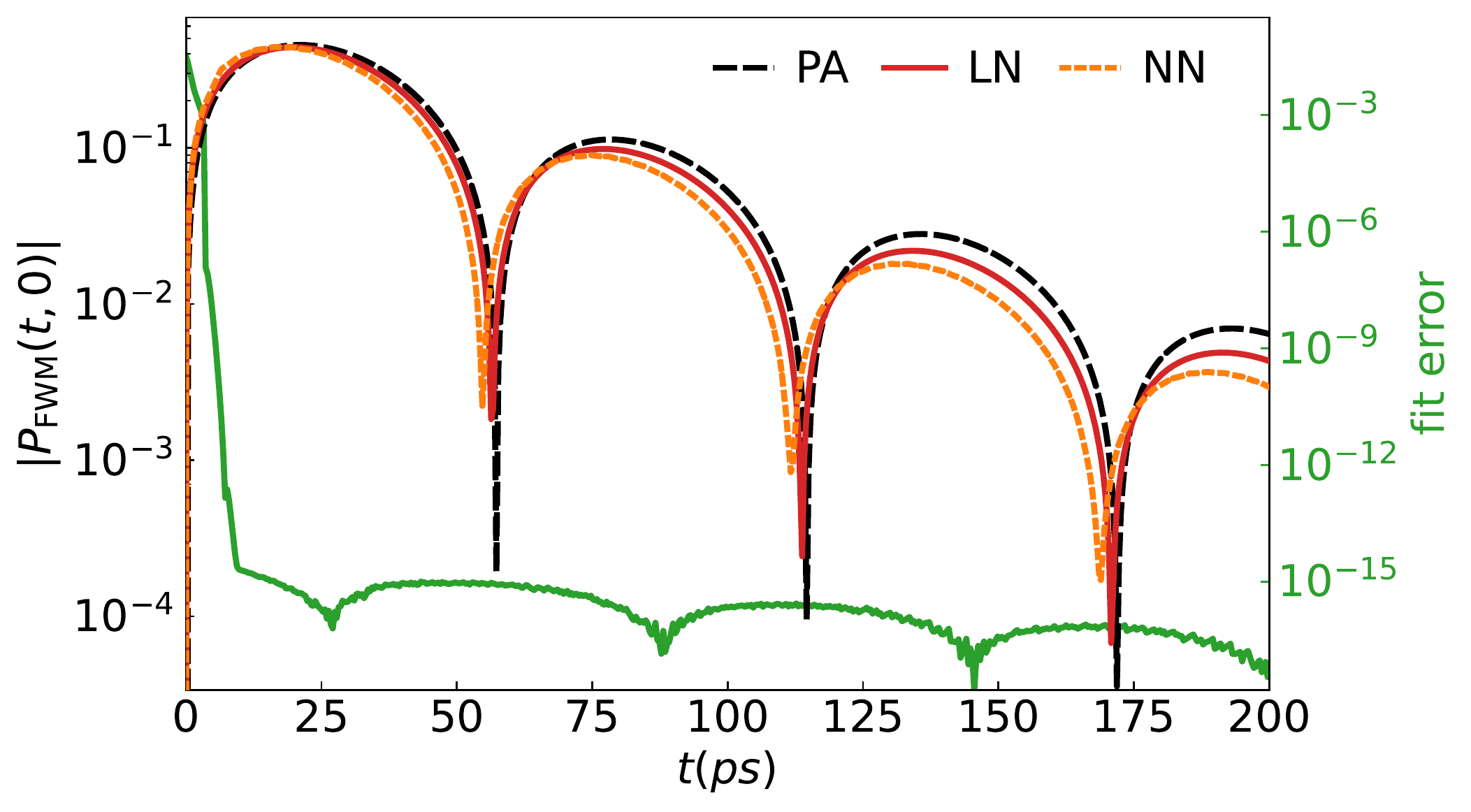}}
\raisebox{3.5cm}{(b)}{\includegraphics[width=0.45\textwidth]{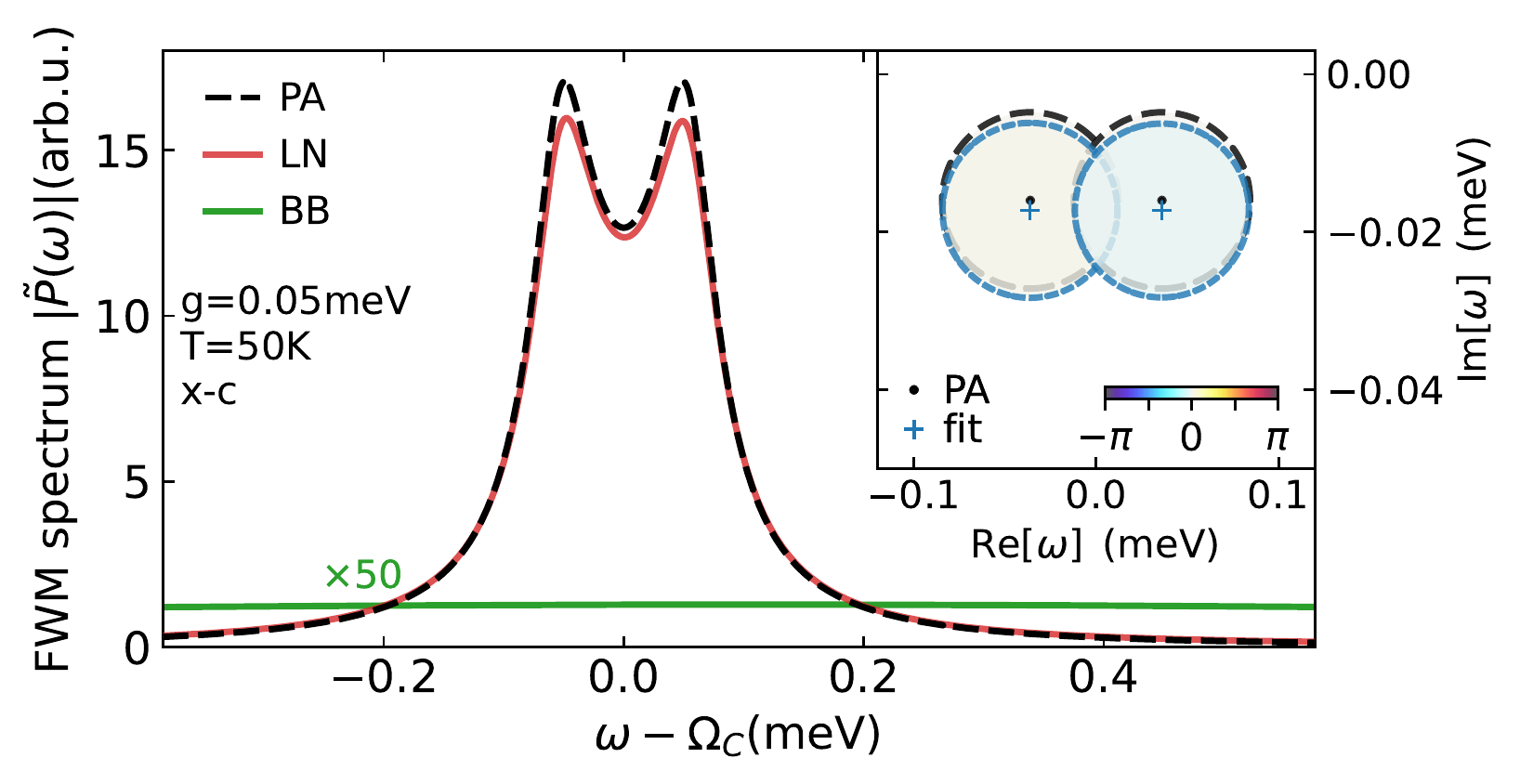}}
\caption{
As \Fig{rs-cx-50} but for the x-c FWM channel.
}
\label{rs-xc-50}
\end{figure}

In Section \ref{section:results} we have shown the FWM results when both the excitation and the measurement are done in the same mode. Here we consider the other possibilities, when the system is excited via the cavity, while the response is measured via the exciton channel (c-x FWM, see \Fig{rs-cx-50}) and vice versa (x-c FWM, see \Fig{rs-xc-50}). The former is similar to the c-c polarization shown in Figs.\,\ref{res-cc-50}(b) and \ref{spec-cc-50}(b), although the results are quantitatively different. In particular, a faster rise time of the signal is seen in \Fig{rs-cx-50}, since the cavity excitation needs to be converted to the excitonic polarization in order to be observed in this nonlinearity channel. This takes a shorter time as compared to a further conversion of the excitation back to the cavity, which is required in the c-c FWM, see a discussion at the end of \Sec{results:smallg}. 
Another feature is a larger relative phase difference between the ``inner transitions''~\cite{allcock2022quantum}, compare the insets in Figs.\,\ref{spec-cc-50}(b) and \ref{rs-cx-50}(b). Also, an increased contribution of the second-rung transitions is seen in \Fig{rs-cx-50}(b).  All of these features are well reproduced in the PA. The x-c FWM signal in \Fig{rs-xc-50}(a) demonstrates an even shorter rise time, and its deviation for the bi-exponential fit is similar to the x-x FWM polarization, shown in \Fig{res-cc-50}. The spectrum also consists of only two transitions, but this time with very similar phases.

\begin{figure}[htpb]
\centering
\raisebox{3.5cm}{(a)}{\includegraphics[width=0.45\textwidth]{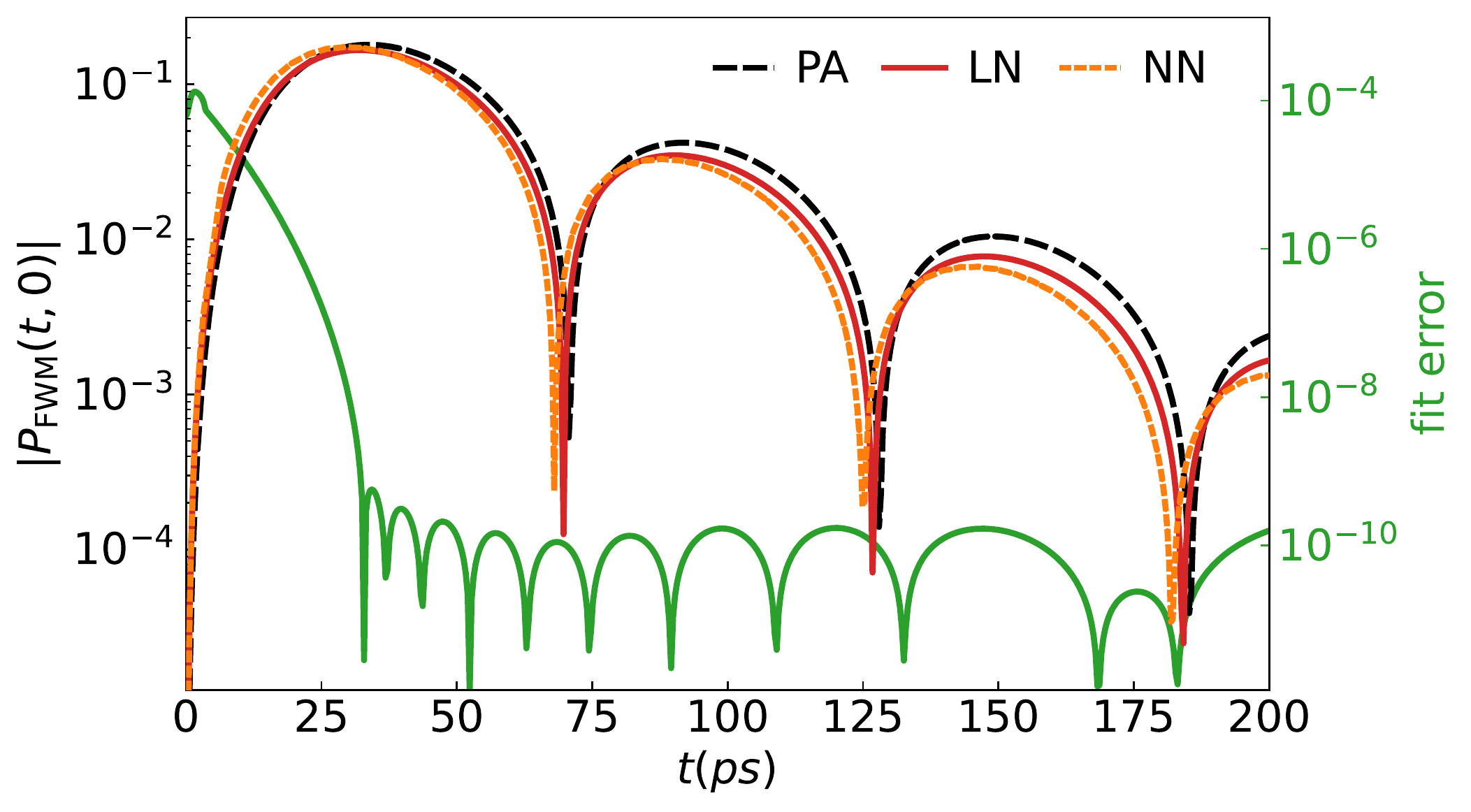}}
\raisebox{3.5cm}{(b)}{\includegraphics[width=0.45\textwidth]{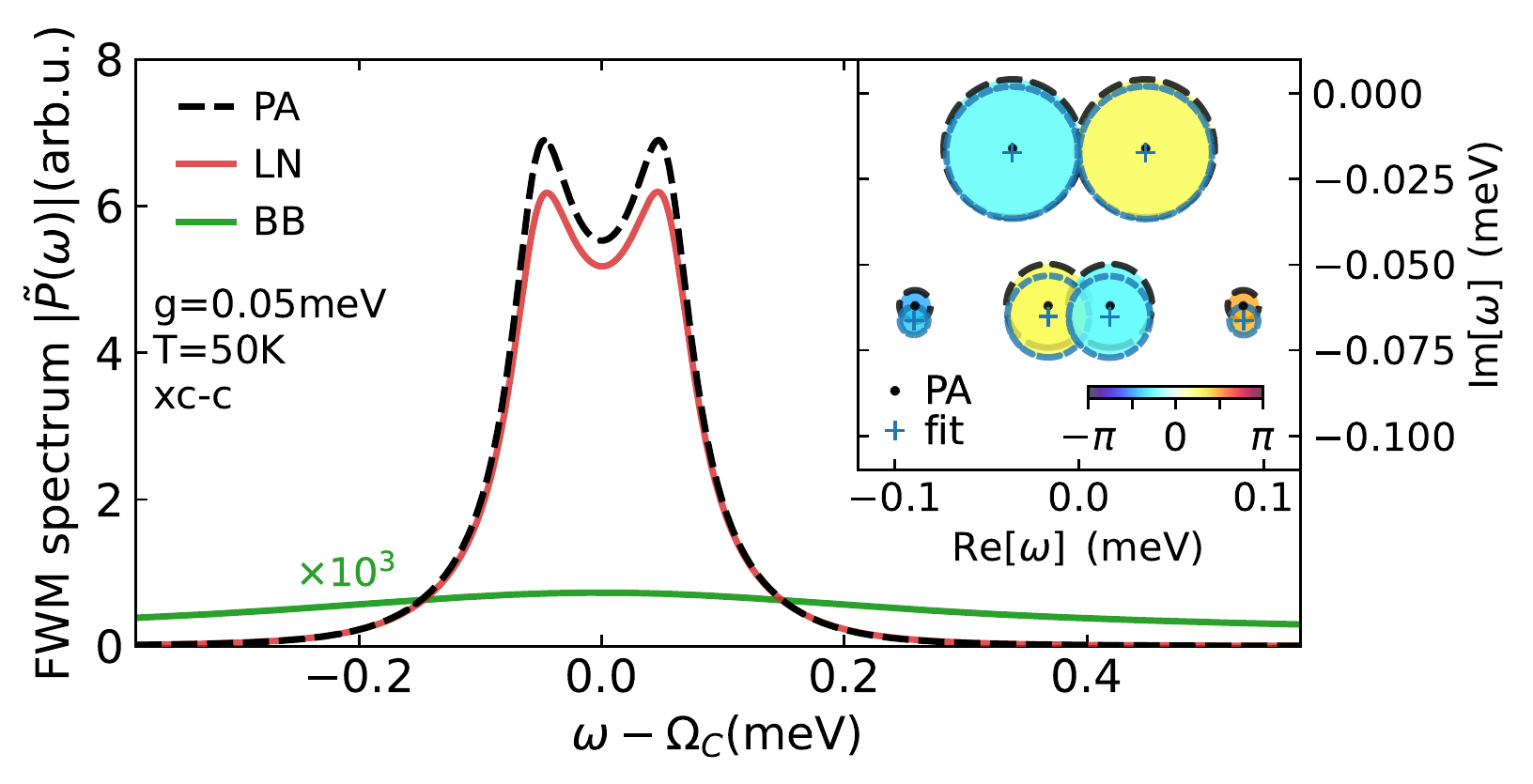}}
\caption{
As \Fig{rs-cx-50} but for the xc-c FWM channel, in which the system is excited by $\vec{Q}^{\mathrm{(I)}}_x$ and ${Q}^{\mathrm{(II)}}_c$.
}
\label{rs-xcc-50}
\end{figure}

\begin{figure}[htpb]
\centering
\raisebox{3.5cm}{(a)}{\includegraphics[width=0.45\textwidth]{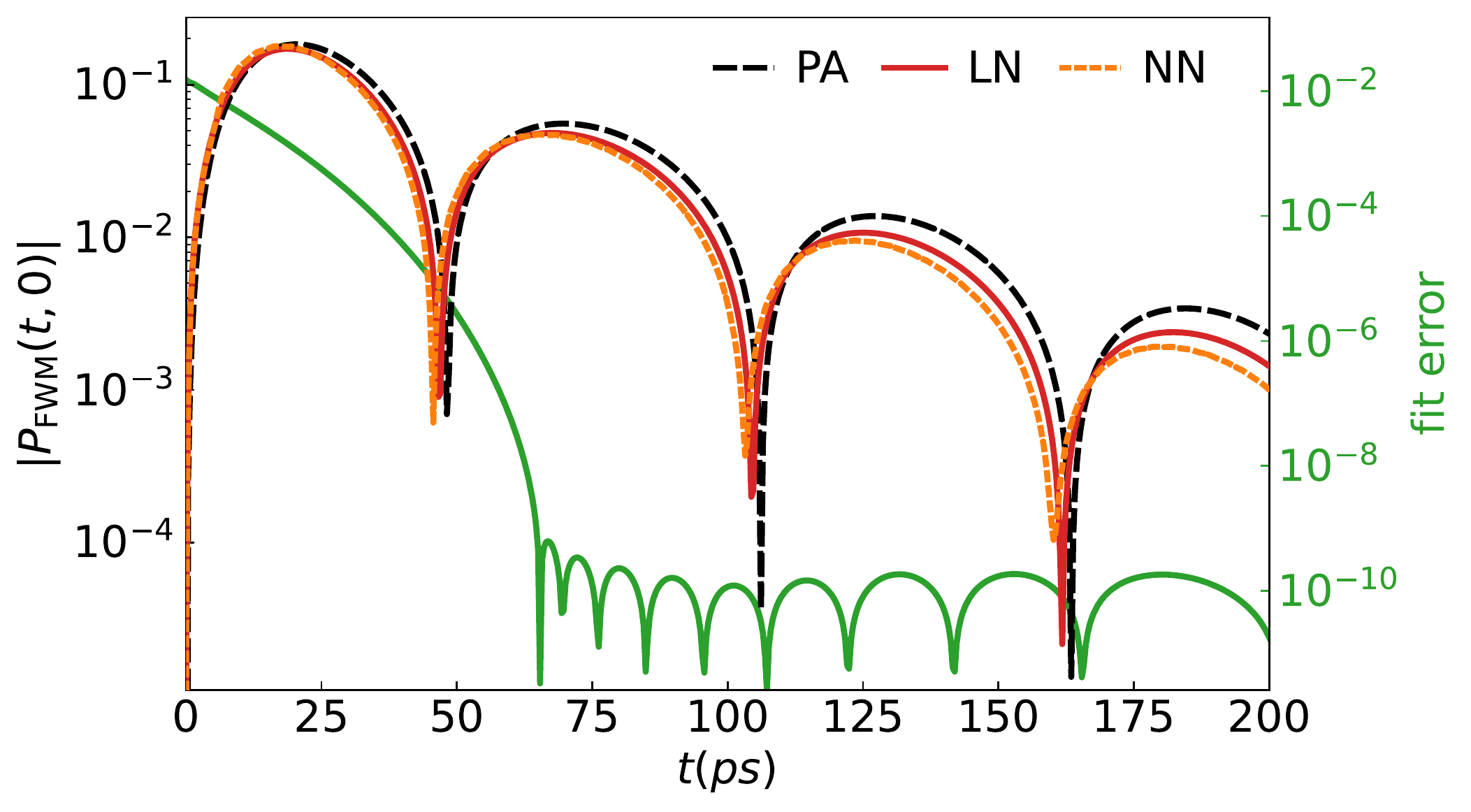}}
\raisebox{3.5cm}{(b)}{\includegraphics[width=0.45\textwidth]{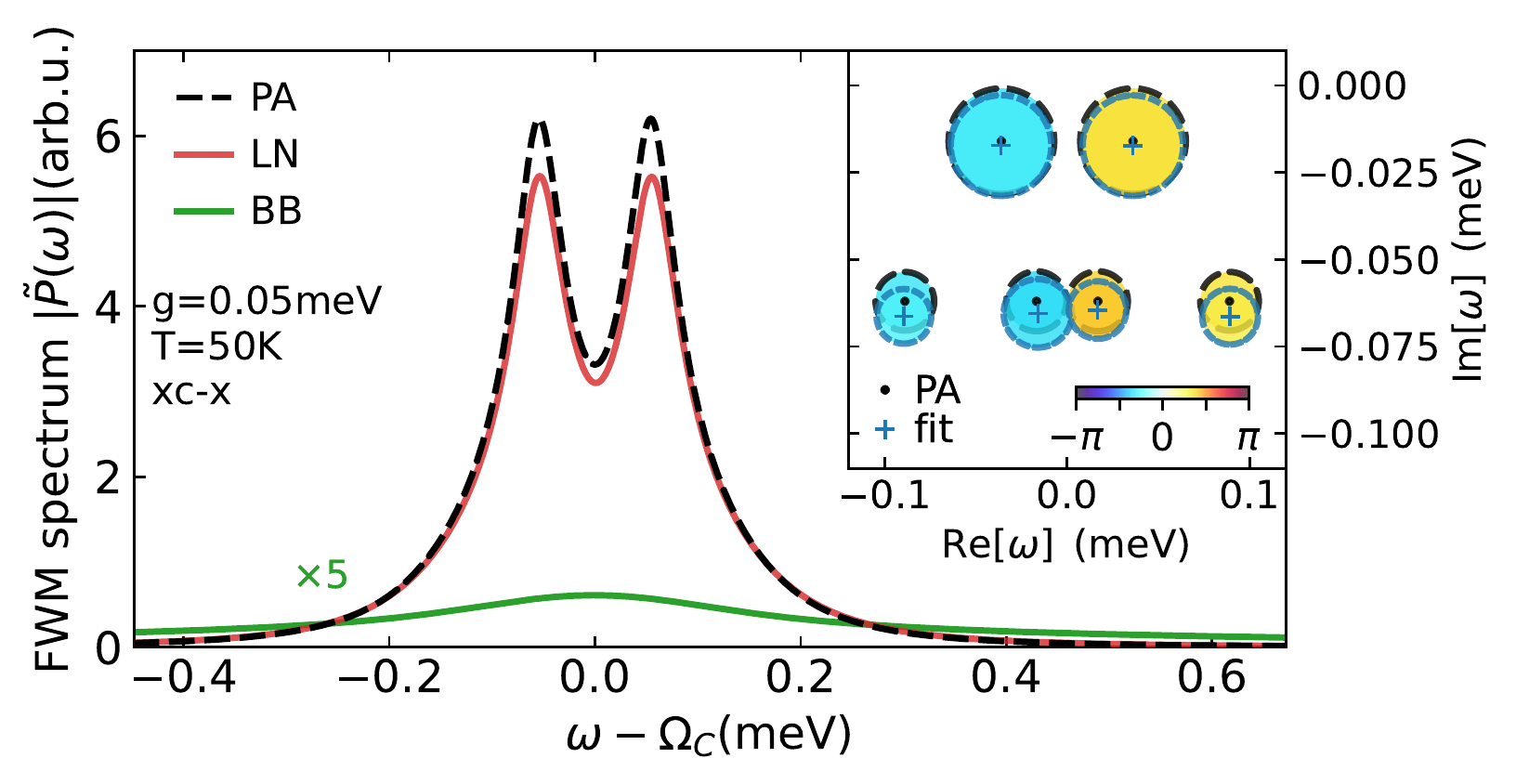}}
\caption{
As \Fig{rs-xcc-50} but for the xc-x FWM channel.
}
\label{rs-xcx-50}
\end{figure}

\begin{figure}[t]
\centering
\raisebox{4cm}{(a)}{\includegraphics[width=0.4\textwidth]{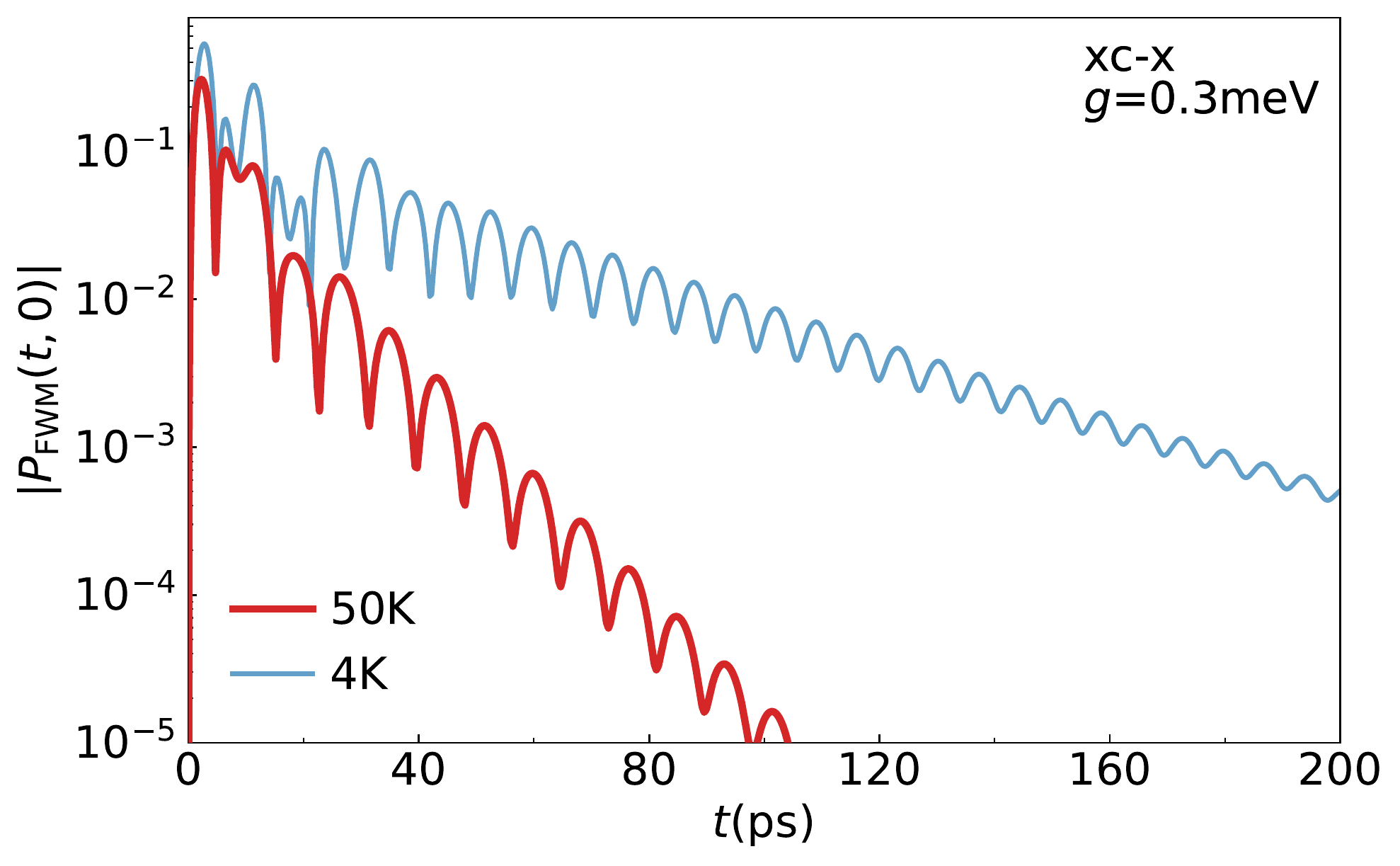}}
\raisebox{4cm}{(b)}{\includegraphics[width=0.4\textwidth]{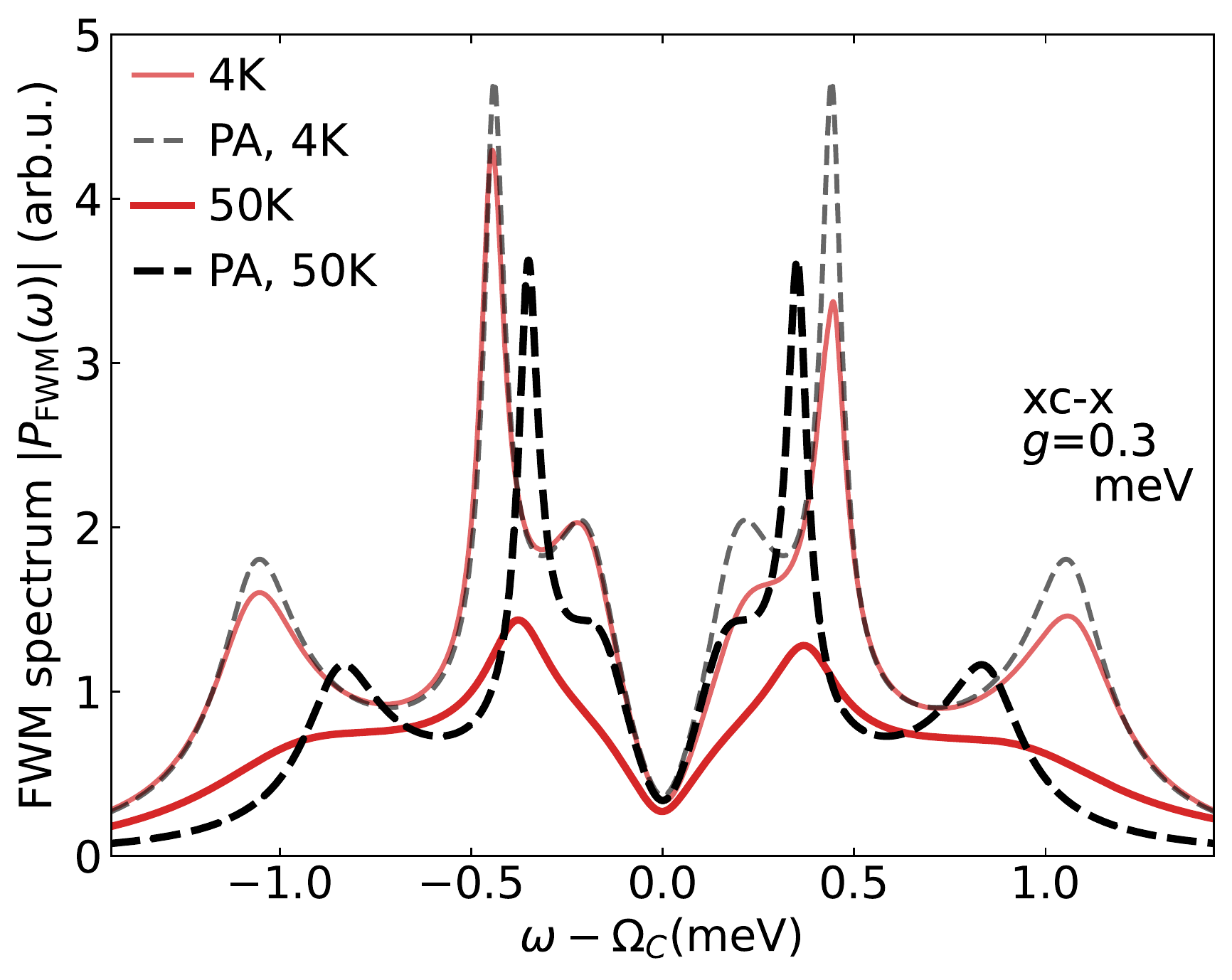}}
\caption{
As \Fig{rs-xcx-50} but for $g=0.3$\,meV and two different temperatures, $T=4$\,K and $50$\,K, showing only the $L$N results in (a) and both the $L$N (red solid lines) and PA (black dashed lines) in (b), with the inset removed.
}
\label{rst-xcx-300}
\end{figure}

So far, we have considered the situation, where both excitation pulses $\vec{Q}^{\mathrm{(I)}}$ and ${Q}^{\mathrm{(II)}}$ correspond to the same mode. In general, this is not always the case, and we give here examples of the xc-c and xc-x FWM polarizations, for which the first (second) pulsed excitation $\vec{Q}^{\mathrm{(I)}}_x$ (${Q}^{\mathrm{(II)}}_c$) occurs in the exciton (cavity) mode and the measurement is done, respectively, in the cavity and exciton mode.  Figures \ref{rs-xcc-50} and \ref{rs-xcx-50} show, respectively, the xc-c and xc-x FWM polarizations and the corresponding spectra for $g=0.05$\,meV.
In this regime, the xc-c and c-c FWM polarizations are very similar, compare \Fig{rs-xcc-50} with
Figs.\,\ref{res-cc-50}(b) and \ref{spec-cc-50}(b). The xc-x FWM is much different from all other nonlinear channels considered so far. In fact, its deviation from the multi-exponential behaviour is stronger, taking place at a longer timescale, which leads to a more pronounced BB in the spectrum, see the green lines in \Fig{rs-xcx-50}. Because of the stronger BB and the presence of all six transitions at the same time, below we explore this FWM channel further for different values of $g$ and $T$, see \Figs{rst-xcx-300}{rst-xcx-800}.

\begin{figure}[htpb]
\centering
\raisebox{4cm}{(a)}{\includegraphics[width=0.4\textwidth]{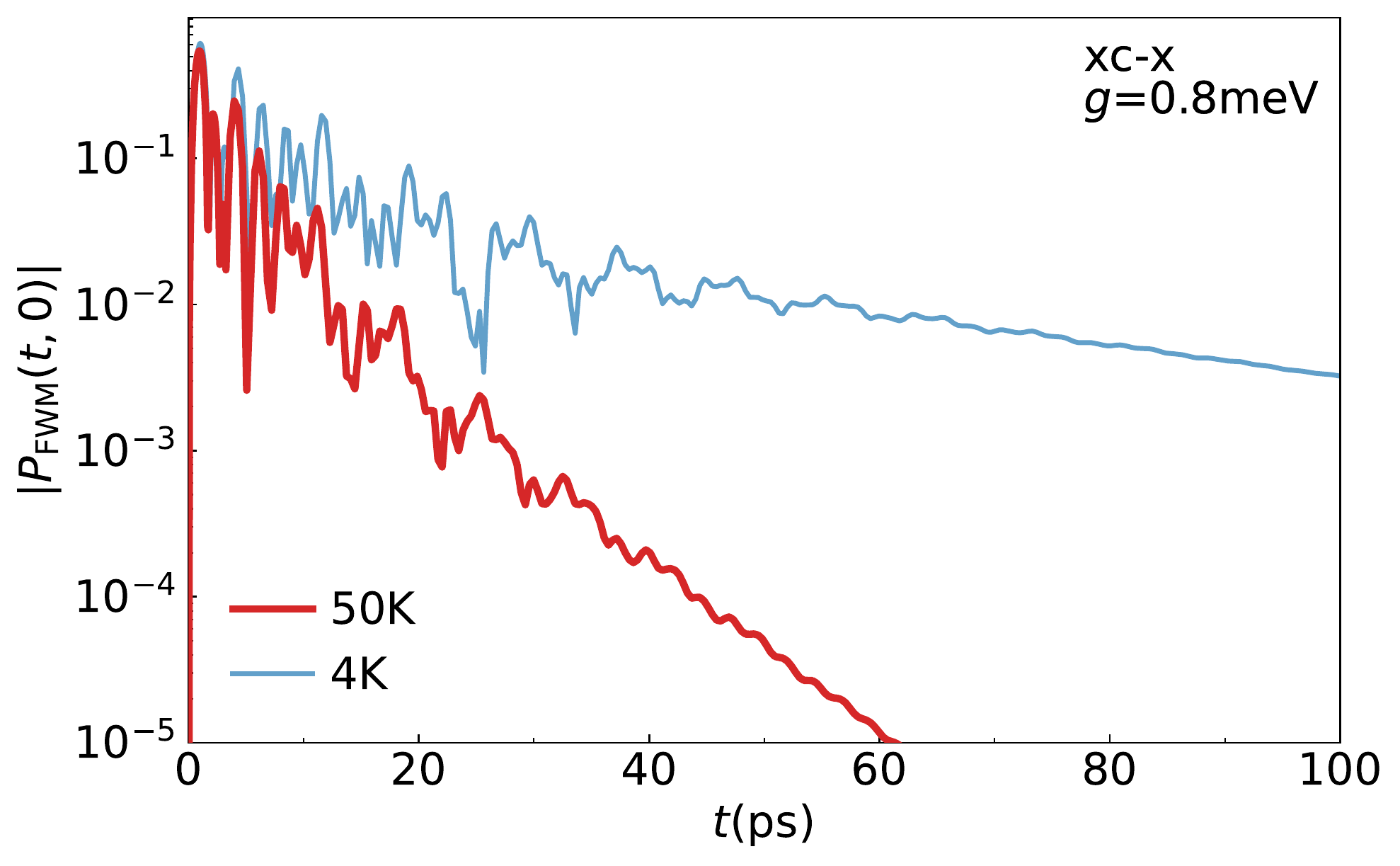}}
\raisebox{4cm}{(b)}{\includegraphics[width=0.4\textwidth]{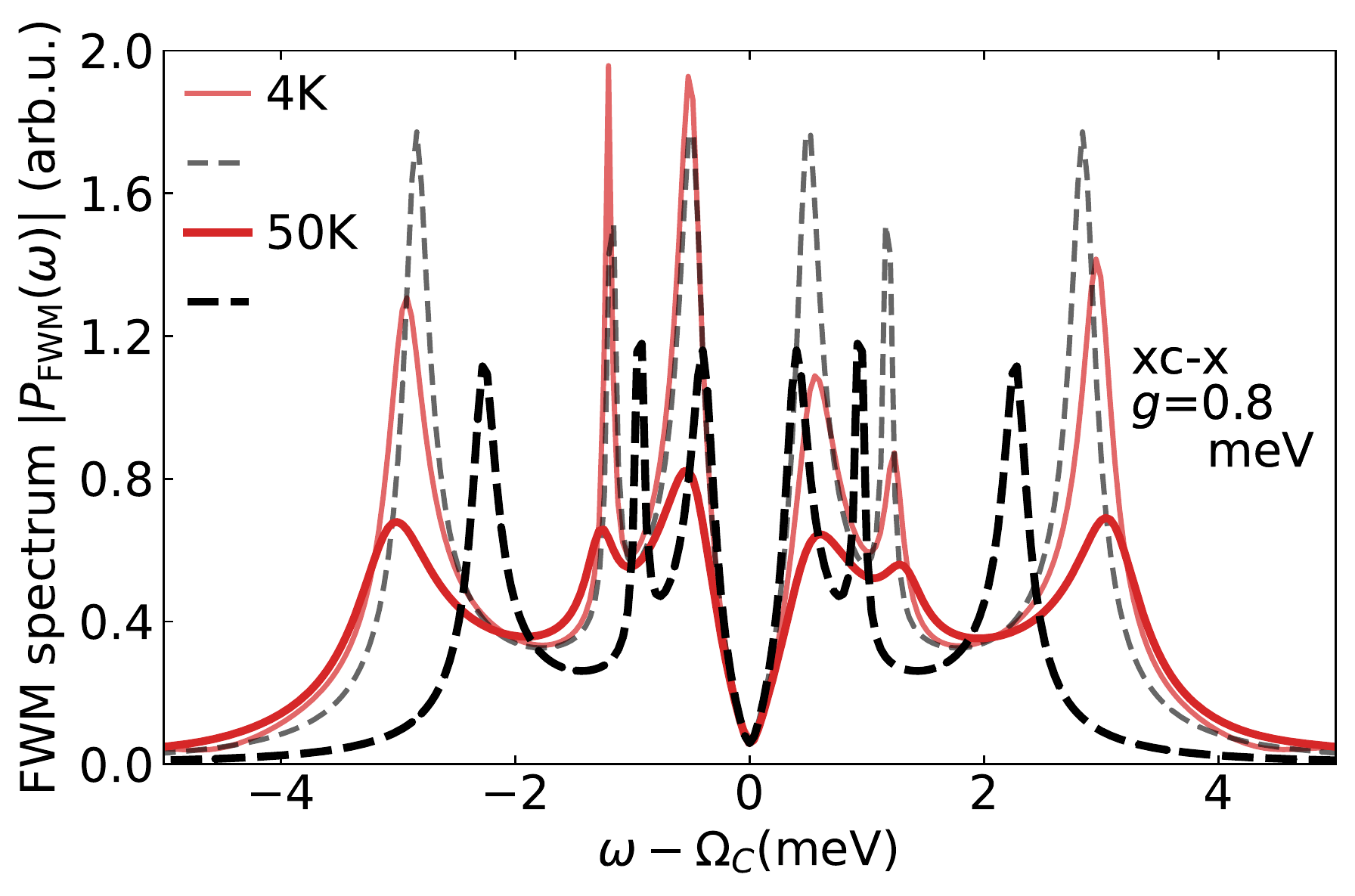}}
\caption{
As \Fig{rst-xcx-300} but for $g=0.8$\,meV.
}
\label{rst-xcx-800}
\end{figure}

For $g=0.3$\,meV and $T=4$\,K, the xc-x FWM spectrum is wider, and the individual transitions are already well resolved as opposed to the c-c FWM, compare Figs.\,\ref{rst-xcx-300}(b) and \ref{spec-temp}(b). At $T=50$\,K  there is still a considerable overlap of the transitions in the spectrum and only a small spectral asymmetry.
For $g=0.8$ meV (\Fig{rst-xcx-800}) there is a deviation from a regular periodic behavior in the time domain. The deviation from a multi-exponential behavior appears to be also very significant, so these results cannot be fitted at all.
The transitions are spectrally better resolved, having less overlap and rather similar weights, both in the $L$N and PA. The $L$N spectrum has an interesting profile, with more uneven distribution of weights caused by phonon assisted transitions.

Another possibilities of excitation and measurement channels are cx-x and cx-c, produced by $\vec{Q}^{\mathrm{(I)}}_c$ and ${Q}^{\mathrm{(II)}}_x$. However, no signal is produced in these channel for zero delay between the pulses, since ${Q}^{\mathrm{(II)}}_x\vec{Q}^{\mathrm{(I)}}_c=0$, as can be seen from \Eqs{Q1-x-c}{Q2-x-c}. Any other excitation and measurement conditions can be expressed as linear combinations of those discussed here and in the main text.

\section{Convergence of FWM results with $L$}
\label{appendix:convergence}

We explore convergence of both the linear and the FWM numerical results towards the exact solution. For the linear polarization, it is easier to assess the convergence, as we have access to larger $L$. For both the linear and the FWM polarizations, we compare our numeric results to those obtained using the TEMPO algorithm~\cite{strathearn2018efficient,strathearn2020modelling}.

\begin{figure}[t]
\raisebox{4.5cm}{(a)}{\includegraphics[width=0.45\textwidth]{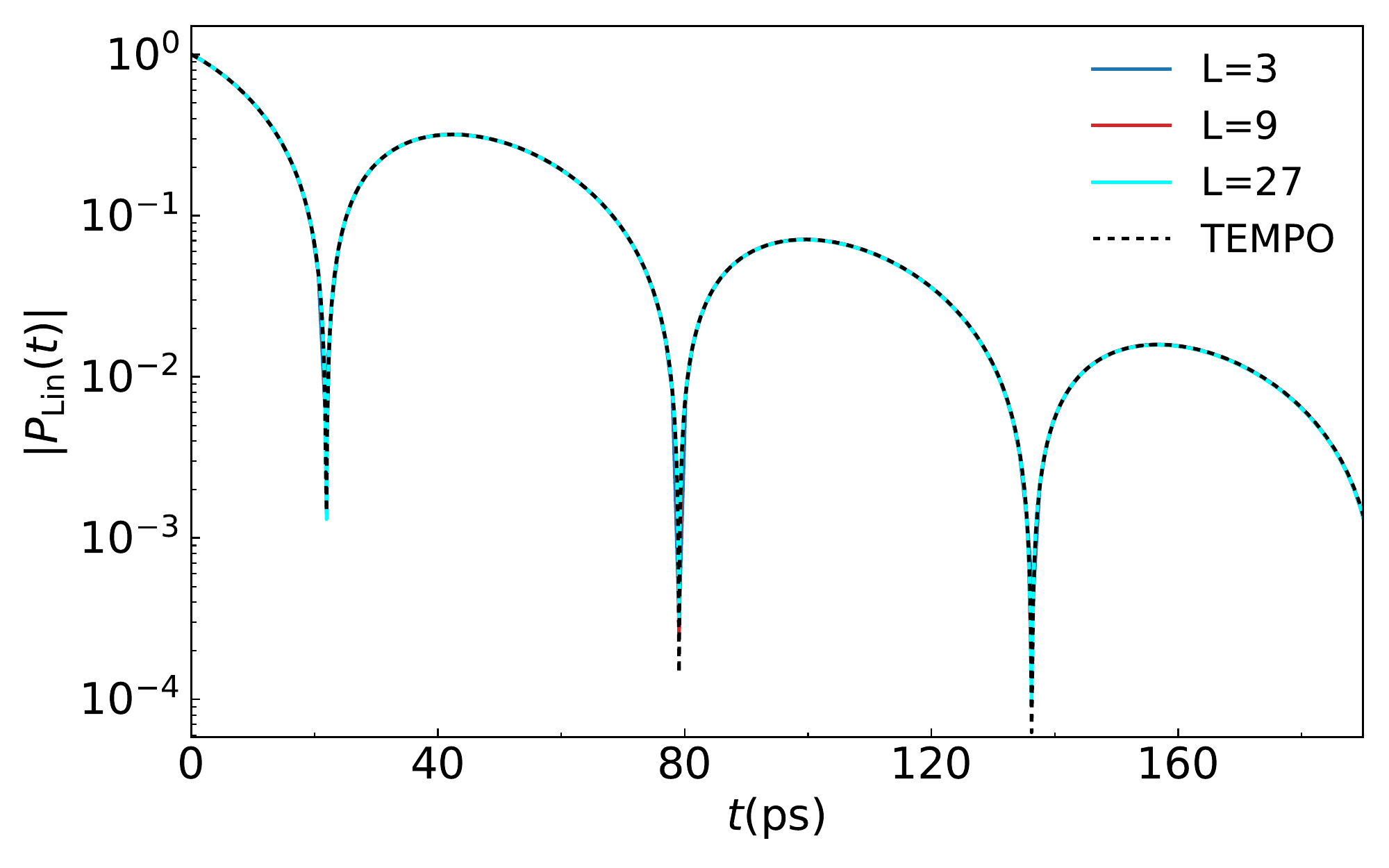}}

\raisebox{4.5cm}{(b)}{\includegraphics[width=0.45\textwidth]{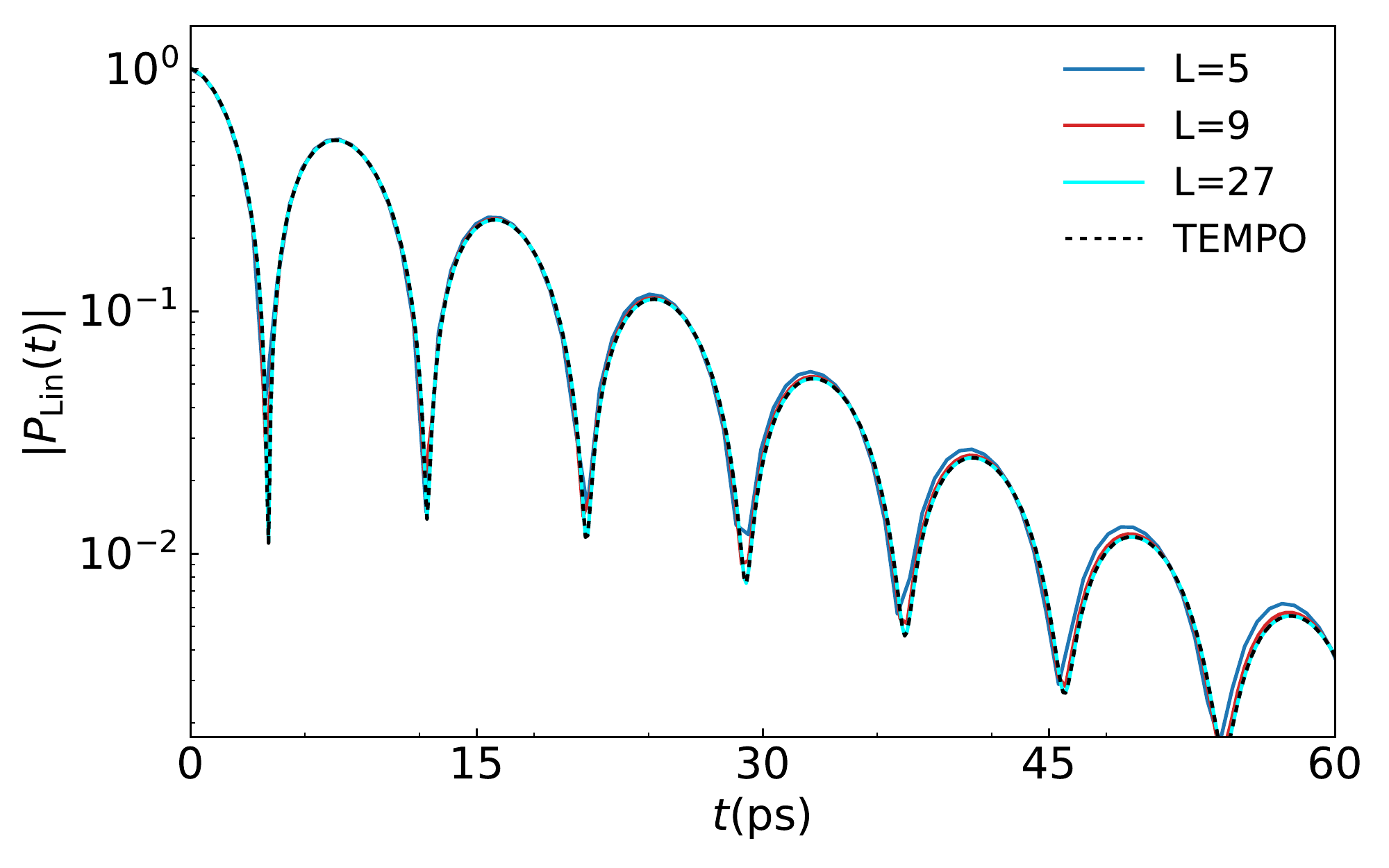}}

\raisebox{4.5cm}{(c)}{\includegraphics[width=0.45\textwidth]{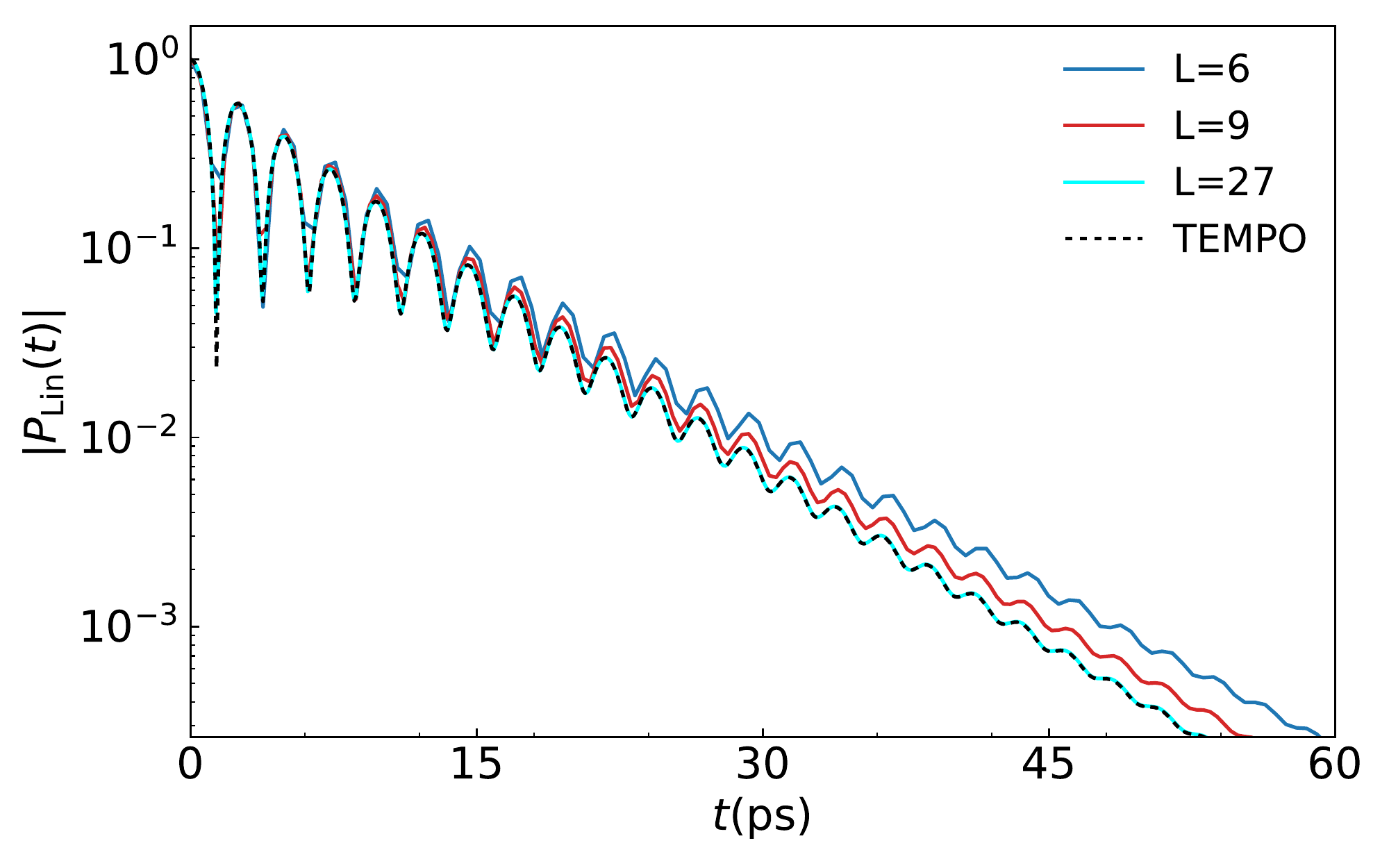}}
\caption{
Linear polarization, showing small L (blue), L=9 (red), L=27 (cyan) and TEMPO (black) results for $T=50$\,K, c-c channel and (a) $g=0.05$\,meV, (b) $g=0.3$\,meV, (c) $g=0.8$\,meV. }
\label{LPtempo3}
\end{figure}

Fig.~\ref{LPtempo3} shows, the convergence of c-c linear polarization with $L$, at $T=50$\,K and for different values of $g$ (the same as used in Figs.~\ref{res-temp}-\ref{spec-temp}). For the linear polarization we have considered up to $L=27$. For $L=27$, it takes around two seconds to propagate the system by one time step $\Delta t$ on a standard PC.

\begin{figure}[t]
\includegraphics[width=0.45\textwidth]{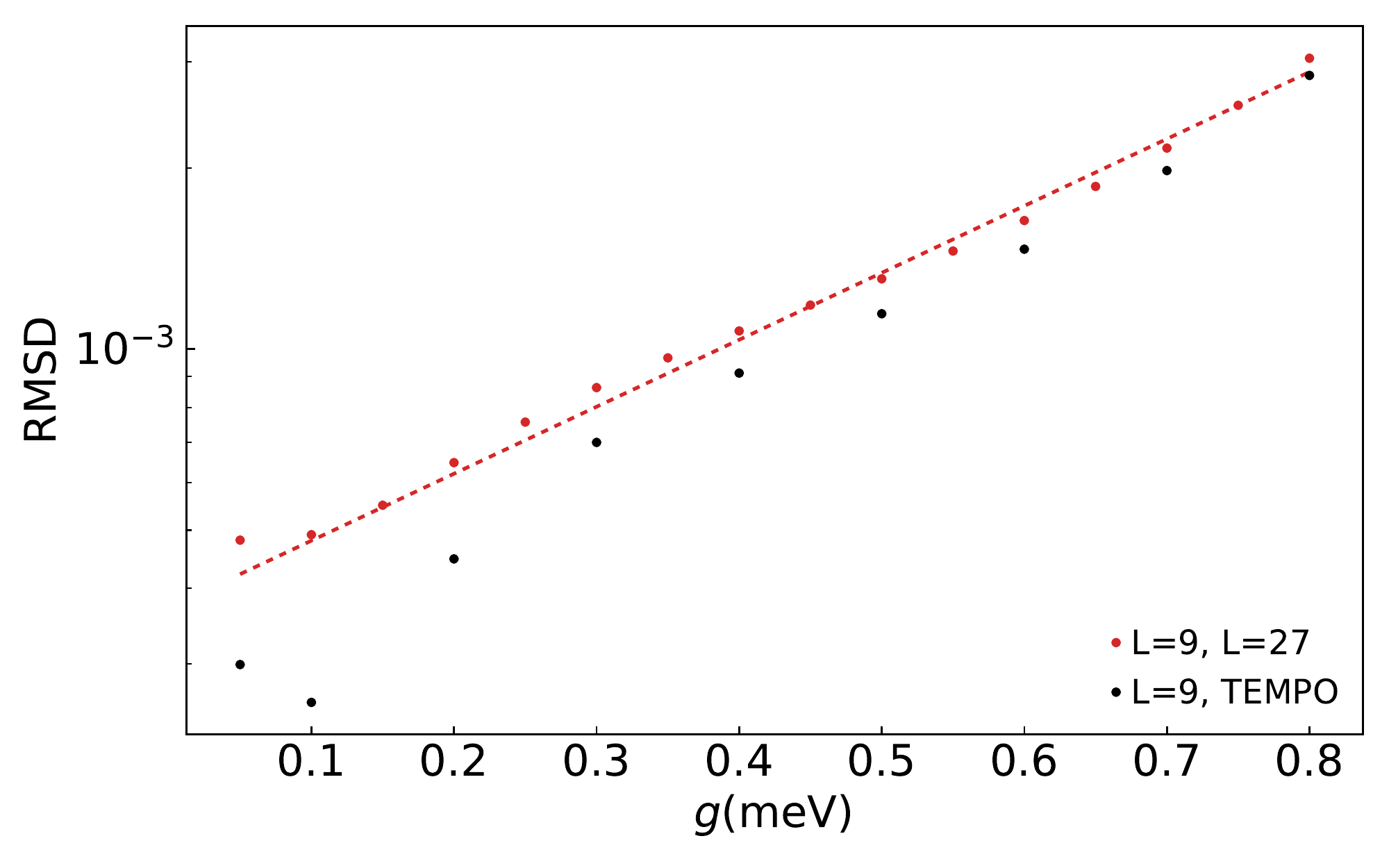}
\caption{RMSD for the linear polarization, given by Eq.(\ref{eqer2}), between $L=9$ and $L=27$ results (red). The dashed line shows the corresponding fit with $0.000686 e^{2.50 g}$, where $g$ is measured in meV.}
\label{rmsd}
\end{figure}

Figure \ref{rmsd} shows the root mean square deviation (RMSD) defined by
\begin{align}
\mathrm{RMSD}=\sqrt{\frac{1}{N+1}  \sum_{n=0}^N \Big| P_{ex}(t_n)-P_{9}(t_n) \Big|^2},
\label{eqer2}
\end{align}
for the linear polarization vs. $g$. A comparison of the $L=9$ result with an exact solution $P_{ex}$ is considered, taking $L=27$ or TEMPO as $P_{ex}$. We had to manually tune the convergence parameters in TEMPO each time $g$ was varied.
Having observed the RMSD in the range 0.03-0.3\%, we believe that the convergence we have reached is sufficient for demonstrating the key physical behavior across the entire range of $g$ here explored.

\begin{figure}[t]
\raisebox{4.5cm}{(a)}{\includegraphics[width=0.45\textwidth]{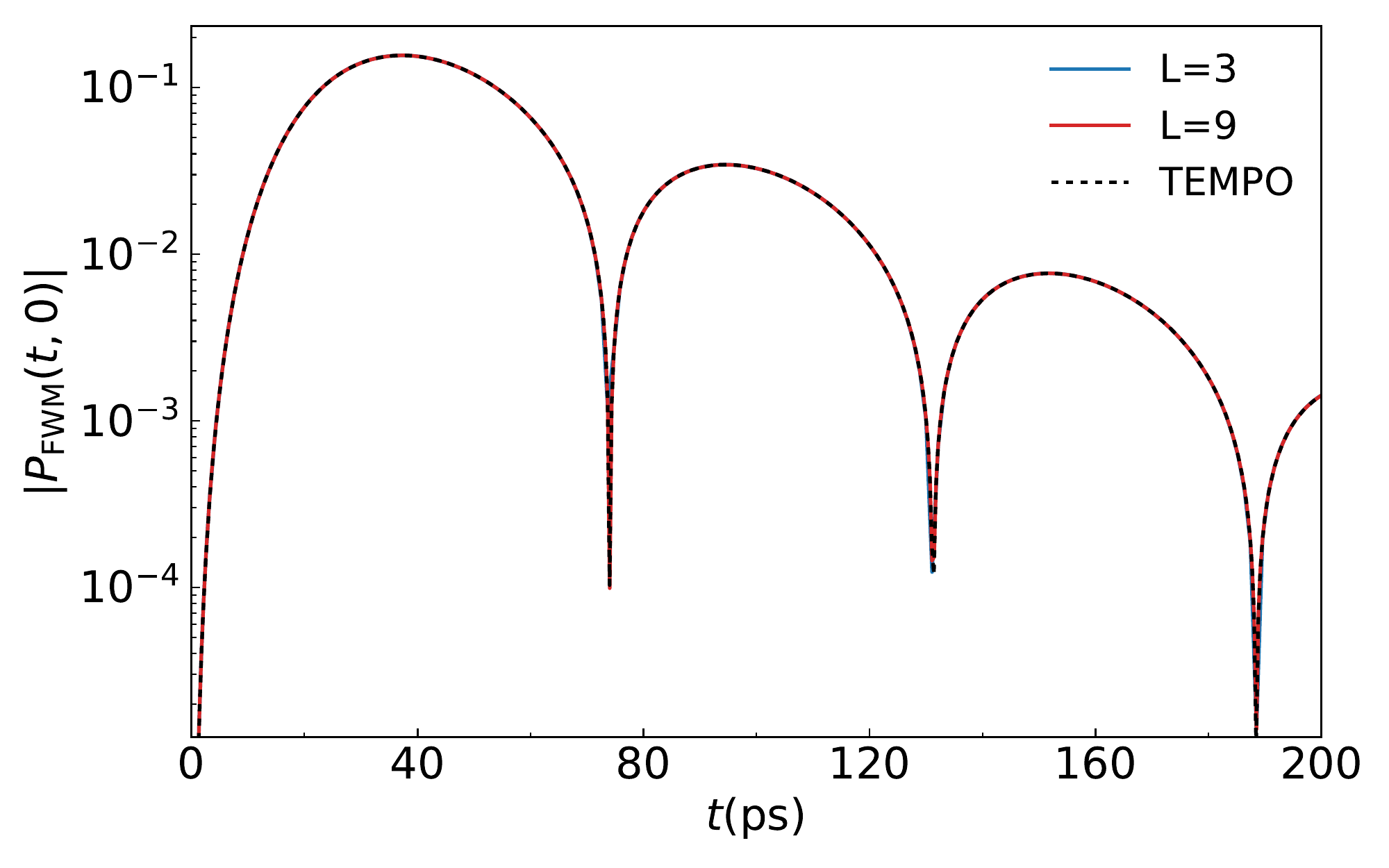}}

\raisebox{4.5cm}{(b)}{\includegraphics[width=0.45\textwidth]{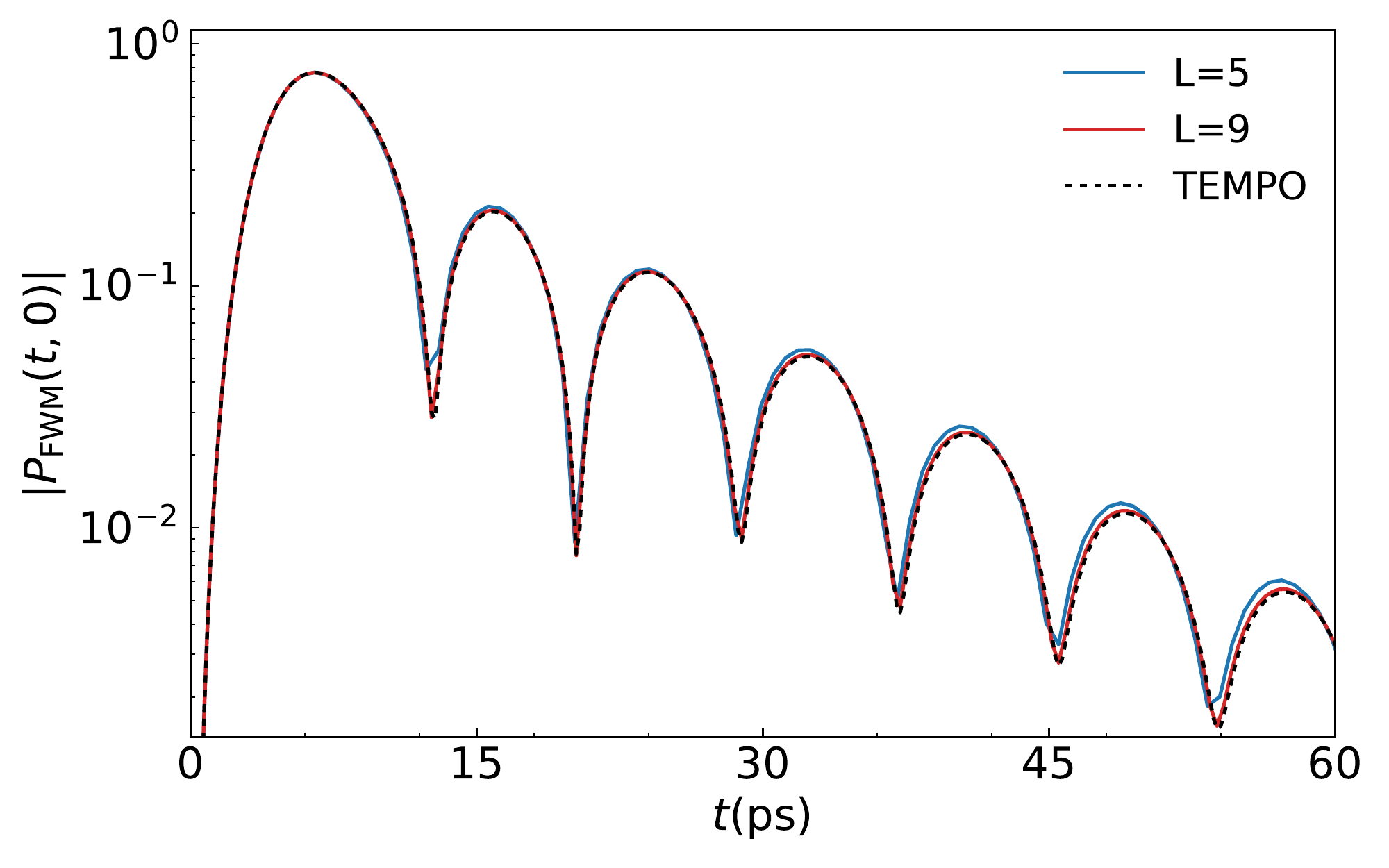}}

\raisebox{4.5cm}{(c)}{\includegraphics[width=0.45\textwidth]{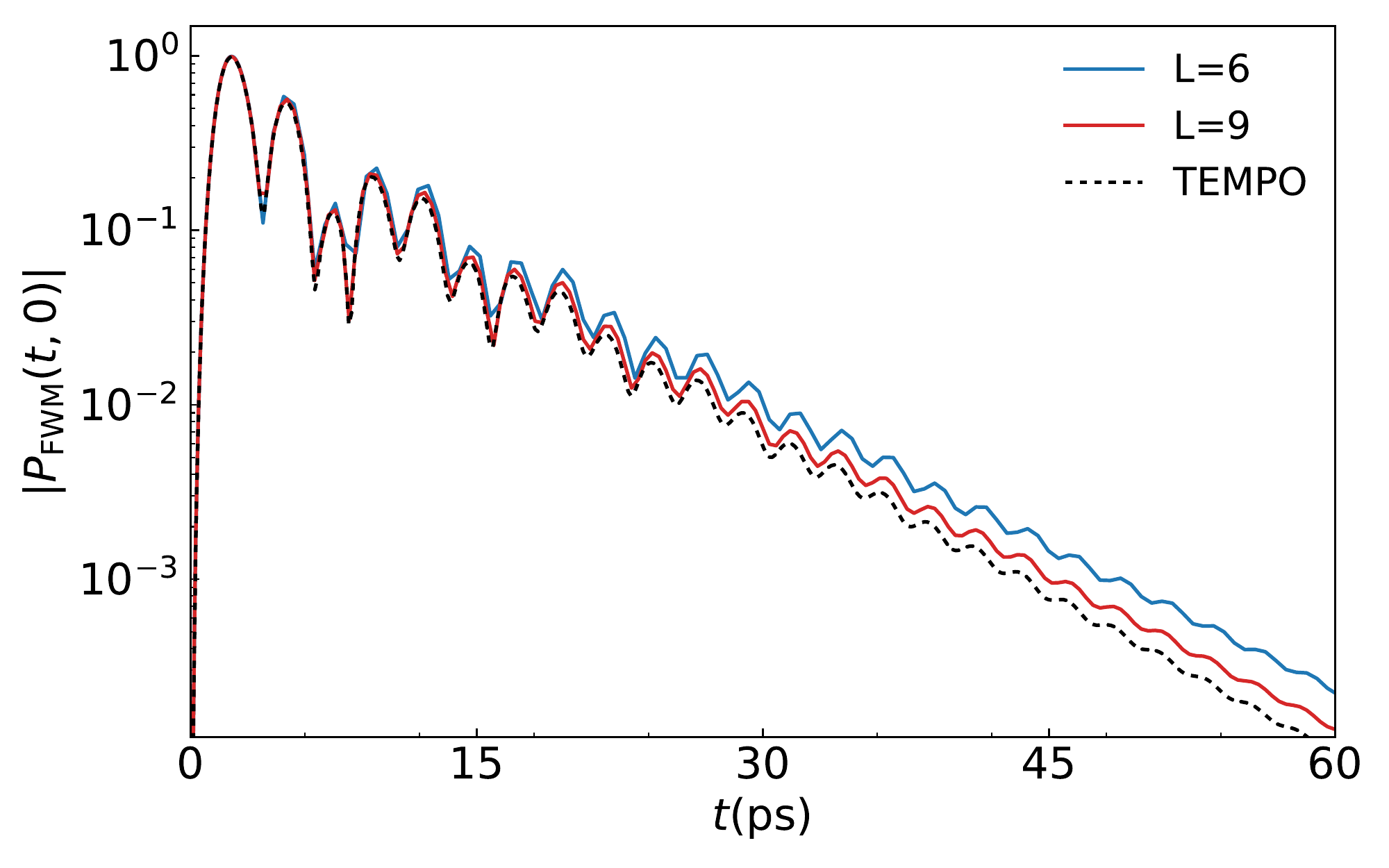}}
\caption{FWM polarization, showing small L (blue), L=9 (red) and TEMPO (black) results for $T=50$\,K, c-c channel and (a) $g=0.05$\,meV, (b) $g=0.3$\,meV, (c) $g=0.8$\,meV. }
\label{tempo3}
\end{figure}

Fig.~\ref{tempo3} shows the convergence of c-c FWM results with $L$, using the same parameters as in Fig.~\ref{LPtempo3}. 
The difference between $L=9$ and TEMPO FWM results in Fig.~\ref{tempo3} is similar to the difference we observed between $L=9$ and the exact results ($L=27$, TEMPO) for the linear polarization in Fig.~\ref{LPtempo3}. We therefore expect that the convergence of the FWM signal with $L$ is similar to that of a linear signal. The $L=9$ result presented in this work is well converged to the exact result for $g=0.05$\,meV and for $g=0.3$\,meV. Although the situation looks worse for $g=0.8$\,meV, the signal itself is fast decaying and any significant difference between $L=9$ and the exact result becomes visible only on a log scale, when the signal is small.

We have experienced difficulties in obtaining FWM results with good precision using TEMPO. The number of combinations of different convergence parameters we could sample was limited by a long computational time.
Therefore we only show a visual comparison of the FWM results with TEMPO in Fig.~\ref{tempo3}.
As stated in the main text, it is sufficient to consider six elements of the density matrix in order to calculate the FWM response.
We note that in the TEMPO results shown here, we have used the full exciton-cavity density matrix for the first two rungs of the JC ladder (36 elements), which may have resulted in a poorer convergence.

\FloatBarrier

\bibliography{refferences}

\end{document}